\documentclass[twocolumn,trackchanges]{aastex61}

\usepackage{graphicx}	
\usepackage{amsmath}	
\usepackage{amssymb}	
\usepackage{enumitem}

\defcitealias{lehner13}{L13}
\defcitealias{lehner18}{Paper~I}
\defcitealias{wotta19}{Paper~II}
\defcitealias{wotta16}{W16}
\defcitealias{haardt96}{HM05}
\defcitealias{haardt12}{HM12}

\newcommand{\ssz}{223}
\newcommand{\logU}{\ensuremath{\log U}}

\def\nodata{ ~$\cdots$~ }

\newcommand{\zabs}{$z_{\rm abs}$}
\newcommand{\mzabs}{z_{\rm abs}}

\newcommand{\cmm}{cm$^{-2}$}
\newcommand{\cmmm}{cm$^{-3}$}

\newcommand{\lya}{Ly$\alpha$}

\newcommand{\lyb}{Ly$\beta$}

\newcommand{\nhi}{$N_{\rm H\,I}$}
\newcommand{\nh}{$N_{\rm H}$}
\newcommand{\nnh}{$n_{\rm H}$}

\newcommand{\novi}{$N_{\rm O\,VI}$}

\newcommand{\mnnh}{n_{\rm H}}
\newcommand{\mnh}{N_{\rm H}}
\newcommand{\mnhi}{N_{\rm H\,I}}
\newcommand{\mnhii}{N_{\rm H\,II}}
\newcommand{\lnhi}{$\log N_{\rm H\,I}$}

\newcommand{\mlnhi}{\log N_{\rm H\,I}}
\newcommand{\mlnh}{\log N_{\rm H}}
\newcommand{\mlnnh}{\log n_{\rm H}}

\newcommand{\km}{${\rm km\,s}^{-1}$}

\newcommand{\hst}{{\em HST}}
\newcommand{\fuse}{{\em FUSE}}

\newcommand{\xh}{\ensuremath{{\rm [X/H]}}}
\newcommand{\ah}{\ensuremath{{\rm [\alpha/H]}}}
\newcommand{\ca}{\ensuremath{{\rm [C/\alpha]}}}

\newcommand{\hi}{\ion{H}{1}}
\newcommand{\hii}{\ion{H}{2}}

\newcommand{\cii}{\ion{C}{2}}
\newcommand{\ciii}{\ion{C}{3}}

\newcommand{\oi}{\ion{O}{1}}
\newcommand{\oii}{\ion{O}{2}}
\newcommand{\oiii}{\ion{O}{3}}
\newcommand{\oiv}{\ion{O}{4}}

\newcommand{\ovi}{\ion{O}{6}}

\newcommand{\siii}{\ion{Si}{2}}
\newcommand{\siiii}{\ion{Si}{3}}

\newcommand{\mgii}{\ion{Mg}{2}}

\newcommand{\feii}{\ion{Fe}{2}}

\shortauthors{Lehner et al.}
\shorttitle{CCC. III: Metallicity and Physical Properties of the Cool $\lowercase{z} \la 1$ CGM}

\begin{document}

\title{The COS CGM Compendium. III: Metallicity and Physical Properties of the Cool Circumgalactic Medium at $\lowercase{z} \la 1$}

\author{Nicolas Lehner}
\affiliation{Department of Physics, University of Notre Dame, Notre Dame, IN 46556, USA}
\author{Christopher B. Wotta}
\affiliation{Department of Physics, University of Notre Dame, Notre Dame, IN 46556, USA}
\author{J. Christopher Howk}
\affiliation{Department of Physics, University of Notre Dame, Notre Dame, IN 46556, USA}
\author{John M. O'Meara}
\affiliation{W.M. Keck Observatory 65-1120 Mamalahoa Highway Kamuela, HI 96743, USA}
\author{Benjamin D. Oppenheimer}
\affiliation{CASA, Department of Astrophysical and Planetary Sciences, University of Colorado, Boulder, CO 80309, USA}
\author{Kathy L.~Cooksey}
\affiliation{Department of Physics and Astronomy, University of Hawai`i at Hilo, HI 96720, USA}

\begin{abstract}
We characterize the metallicities and physical properties of cool, photoionized gas in a sample of 152 $z\la 1$ strong \lya\ forest systems (SLFSs, absorbers with $15<\mlnhi < 16.2$). The sample is drawn from our COS circumgalactic medium (CGM) compendium (CCC), an ultraviolet survey of \hi-selected circumgalactic gas around  $z\la 1$ galaxies that targets 262 absorbers with $15< \mlnhi < 19$. We show that the metallicity probability distribution function of the SLFSs at $z\la 1$ is  unimodal, skewed to low metallicities with a mean and median of  $\xh = -1.47$ and $-1.18$ dex.  Very metal-poor gas with $\xh < -1.4$ represents about half of the population of absorbers with $15 <\mlnhi \la 18$. Thus, there are important reservoirs of primitive (though not pristine) gas around $z\la 1$ galaxies. The photoionized gas around $z\la 1$ galaxies is highly inhomogeneous based on the wide range of metallicities observed ($-3 \la \xh \la +0.4$) and that there are large metallicity variations (factors of 2 to 25) for most of the closely-spaced absorbers ($\Delta v \la 300$ \km) along the same sightlines. These absorbers show a complex evolution with redshift and \hi\ column density, and we identify subtle cosmic evolution effects that affect the interpretation of metallicity distributions and comparison with other of absorbers samples. We discuss the physical conditions and cosmic baryon and metal budgets of the CCC absorbers. Finally, we compare the CCC results to recent cosmological zoom simulations and explore the origins of the $15< \mlnhi < 19$ absorbers within the EAGLE high-resolution simulations.
\end{abstract}


\section{Introduction}\label{s-intro}
The distribution of metals in the circumgalactic medium (CGM) of galaxies, and how the distribution changes with time and density, constrains the processes that drive galaxy evolution. Typically, metallicities are best studied in the densest gas -- which gives rise to high column density systems -- due to the diminished ability to detect metals at lower densities. One approach to studying the metal distribution in the rare, high column density CGM gas is to compile an absorption-selected sample from a blind search of observed sight lines, e.g., selecting on \hi\ column density. This is complementary to galaxy-selected studies that target QSOs behind specific galaxies (e.g., COS-Halos---\citealt{tumlinson13,werk14}; COS-GASS---\citealt{borthakur13,borthakur16}, luminous red galaxy surveys---\citealt{chen18,berg19}). The advantage of the \hi\ selection over a metal-line selection (e.g., \mgii) is that it does not bias the study toward high (or low) metallicities.  Based on several studies on galaxy-absorber relationships, low-redshift absorbers with $\mlnhi >15$ are known to probe predominantly the gas surrounding galaxies (e.g., \citealt{lanzetta95,chen00,chen01a,penton02,bowen02,prochaska11c,lehner13,tejos14}). These absorbers probe the denser regions of the CGM about galaxies and the phenomena occurring therein. Thus, they probe CGM material produced by tidally stripped material, feedback-expelled gas, previously ejected matter that is recycling onto the central galaxy, and accretion of intergalactic gas. Understanding the origins, chemical abundances, and physics of these absorbers is thus critical to understanding the mixture of these phenomena and their influence on galaxy evolution. 

The chemical abundances of gas and stars in galaxies have been characterized for several decades (see, e.g., recent review by \citealt{maiolino19}). Studies of the metallicities of gas around galaxies have been much more limited and have necessarily focused on the strongest \hi\ column density absorbers that typically probe the gas in the vicinity ($\la 50$ kpc) of galaxies and on gas at high redshifts where ground-based telescopes can be used to study the rest-frame ultraviolet absorption (e.g., \citealt{prochaska00,prochaska03,kulkarni05,peroux06,peroux08,rafelski12}). Prior to the installation of the Cosmic Origins Spectrograph (COS) on the {\it Hubble Space Telescope} (\hst), there were only $\sim 6$ metallicity estimates of strong \hi\ absorbers with $15 < \mlnhi<19$ at $z\la 1$ \citep[e.g.,][]{zonak04,jenkins05,prochaska05,cooksey08,lehner09}. Thanks to the sensitivity of COS, the archive of ultraviolet observations of QSOs has grown to a sizeable level. It is now possible to assemble a sample of high column density \hi\ absorbers at $z\la 1$ that is sufficiently large to not only robustly characterize the metallicity distribution  of high \nhi\ absorbers but also determine if and how it evolves with \nhi\ and $z$. When these datasets are compared with complementary studies of high \nhi\ absorbers at $z>2$ (e.g., \citealt{cooper15,fumagalli16,lehner16,glidden16}), we can study the cosmic evolution of dense gas about galaxies in the universe over several billions of years. 

With the COS CGM Compendium (CCC), we have undertaken an archival ultraviolet spectroscopic survey of \hi-selected absorbers with $15<\mlnhi < 19$ at $z\la 1$. Our main aim has been to characterize the properties---the metallicities in particular---of the cool gas probed by these absorbers and observed in absorption \citep[see ][hereafter \citetalias{lehner18}]{lehner18}. In our first paper describing CCC \citepalias{lehner18}, we built the largest sample to date of \hi-selected absorbers with $15<\mlnhi < 19$ at $z\la 1$. The sample presented in \citetalias{lehner18} is based on COS G130M and/or G160M spectra, drawn largely from the first data release of the \hst\ Spectroscopic Legacy Archive (HSLA, \citealt{peeples17}) available at the \textit{Barbara A. Mikulski Archive for Space Telescopes} (MAST)  with some complementary in-house data reduction (see \citetalias{lehner18}).\footnote{It is useful to consider the \hi\ absorbers in specific \nhi\ ranges (e.g., to differentiate predominantly neutral from ionized gas or using \nhi\ as a proxy of  densities), and  we adopt here the  \citetalias{lehner18} definition for the CCC absorbers (see \citetalias{lehner18} for the rational behind these different \hi\ column density intervals): Systems with $15< \mlnhi <16.2$, which are the focus of this paper, are defined as the strong \lya\ forest systems (SLFSs). The pLLSs and LLSs have \hi\ column densities  $16.2 \la \mlnhi <17.2 $ and $17.2 \le \mlnhi <19$, respectively. The super LLSs (SLLSs) have $19.0 \le \mlnhi <20.3$ and are intermediate between the LLSs and the damped \lya\ absorbers (DLAs) that have $\mlnhi \ge 20.3$.}

In the second paper \citep[see ][hereafter \citetalias{wotta19}]{wotta19},  we describe the overall methodology to derive the metallicities of the absorbers in CCC, which requires large ionization correction since the gas is nearly fully ionized at \hi\ column density $15<\mlnhi < 19$. In \citetalias{wotta19}, we focus on characterizing in detail the metallicity distribution of the pLLSs and LLSs at $z\la 1$. These results strengthen the findings from our previous surveys (\citealt{lehner13,wotta16}, hereafter \citetalias{lehner13} and \citetalias{wotta16}, respectively). With a larger sample, our CCC survey confirms the metallicity distribution of the pLLSs is bimodal, with a lack of gas at $\sim$10\% solar metallicity. However, CCC also reveals that the functional form of the metallicity distribution is redshift dependent since at $z\la 0.45$ the distribution is consistent with a unimodal distribution and only at $z>0.45$, the distribution shows two significant peaks at $\xh \simeq -1.7$ and $-0.4$ and a dip at $\xh \simeq -1$.\footnote{We use the standard notation $\xh \equiv \log N_{\rm X}/N_{\rm H} - \log {\rm X/H}_\sun$ where in CCC X is typically an $\alpha$-element, unless otherwise stated.}  CCC confirms a large fraction of the CGM is metal enriched, as demonstrated by other studies (e.g., \citealt{stocke13,liang14,werk14,keeney17}). However, what was not appreciated prior to our surveys is that the pLLS and LLSs at $z\la 1$ trace a very wide range of metallicities from chemically-primitive gas $\xh <-2.8$ to super-solar metallicity gas with $\xh \simeq +0.4$. This is in remarkable contrast to the DLAs and SLLSs that mostly have metallicities in the range $-1.4 \la \xh \la 0$ over a similar redshift interval. While  "very metal-poor" gas with $\xh <-1.4$ is rarely seen in DLAs or SLLSs at $z\la 1$, it is as common in absorbers with $16.2 \le \mlnhi \la 18$ as metal-rich gas with $\xh \ga -1$. The metallicities in these very metal-poor absorbers at $z\la 1$ are similar to pLLSs and LLSs and even the Ly$\alpha$ forest observed at $z\sim 2.3$--3.5 \citep{lehner16,fumagalli16}. Thus, a subset of gas in this column density range has experienced relatively little net chemical enrichment over several billions of years, despite the fact that it is found in the CGM of galaxies, which we expect to have been polluted continuously by galaxy feedback. This also implies that the metallicity distribution of CGM gas at $z\la 1$ is a strong function of \nhi\ for absorbers in the range  $16.2\le \mlnhi \la 22$.

In this third paper of our \hst\ Legacy CCC program, we now explore for the first time how the metallicities are distributed in lower \nhi\ gas, down to about $\mlnhi \simeq 15.1$. Although we would have been interested in determining the metallicities in even lower \nhi\ absorbers in order to be able to  connect the strong \hi\ absorbers with \lya\ forest absorbers that probe the intergalactic medium (IGM), the signal-to-noise ratios (SNRs) of the COS (or STIS) spectra are not high enough (SNR\,$\sim 5$--30) to provide an unbiased census of the metallicities for absorbers with $\mlnhi \la 15.1$. That is, there is a bias against measuring abundances in low-metallicity absorbers with $\xh <-1$ gas when the \hi\ column density is $\mlnhi \la 15.1$ given the weakness of the metal lines \citepalias{lehner18}. 

To derive the metallicity of the SLFSs, we follow the methodology developed in \citetalias{wotta19} (and see also  \citealt{crighton15,fumagalli16}), which combines a Bayesian formalism and Markov Chain Monte Carlo (MCMC) techniques with grids of photoionization models to determine the ionization corrections necessary to derive gas-phase metal abundance. We use the same extreme-ultraviolet background (EUVB)  so that our analysis is consistent across the entire range of \nhi.\footnote{We adopt variations in the Haardt-Madau EUVB radiation field from quasars and galaxies---HM05 and HM12 \citep[see][]{haardt96,haardt12}. The fiducial ionization models in the CCC surveys employ the HM05 field, though we consider the implications of adopting the HM12 field in \citetalias{wotta19} and in this work.} As demonstrated in \citetalias{wotta19} for the absorbers in CCC, while the statistical errors can be small (depending on the observational constraints), the systematic error resulting from the use of different radiation fields is at the factor 1.5--2.5 level (0.2--0.4 dex) on average. This is a limitation from having to derive the metallicities in largely ionized gas. However, it is accurate enough, e.g., to separate low from high metallicities or to enable a determination of the metallicity PDF of these absorbers. A systematic error at the factor 1.5--2.5 level on the metallicity of the absorbers is in fact comparable to the uncertainties in chemical abundances determined from emission lines in galaxy spectra (e.g., \citealt{berg16,maiolino19}). In this paper, we also show that the metallicity is not too sensitive to the ionization conditions compared to other physical parameters (such as the hydrogen density, length-scale, or total column density of hydrogen), a finding also inferred by \citet{fumagalli16} in high $z$ LLSs. We further infer in this paper that the possible nature of the gas having multiple ionized phases rather than the singly ionized phase used in our modeling would not change the derived metallicities by more than 0.1--0.2 dex. 

Our paper is organized as follows. In \S\ref{s-data}, we describe the sample of SLFSs and how we determine their metallicities. We refer the reader to 1) \citetalias{lehner18} for the full description of the sample and data, the measurements of the column densities, and direct empirical properties from the observations, and 2) \citetalias{wotta19} for the full description of the ionization modeling of the CCC absorbers that probe the cool photoionized gas, the assumptions and caveats in making these models. The metallicity distribution of the SLFSs and evolution the metallicities with \nhi\ and $z$ as well as relative abundance variation of \ca\ with \xh\ are presented in \S\ref{s-cgm-prop}. In \S\ref{s-phys-prop}, we  present the physical properties of the absorbers in CCC, including their density (\nnh), total column density (\nh), temperature, and linear scale ($l\equiv \mnh/\mnnh$); we also estimate in this section the cosmic metal and baryon budgets of these absorbers. In \S\ref{s-disc} we discuss the implications of our results, including a comparison between our CCC survey results with those of the galaxy-selected COS-Halos survey (\citealt{werk13,tumlinson13}). We also discuss in that section our results in the context of cosmological and zoom simulations \citep{hafen17,rahmati18}, assessing the plausible origins of the CCC absorbers by comparison with the  Evolution and Assembly of GaLaxies and their Environments (EAGLE) simulations \citep{schaye15}. In \S\ref{s-sum}, we summarize our main conclusions and briefly describe some of the follow-up goals with CCC. 

\section{Data and Analysis}\label{s-data}
\subsection{The Sample}\label{s-sample}
\label{s-sample-slfs}
In \citetalias{lehner18}, we assemble a sample of \ssz\ \hi-selected absorbers  with $15.1\la \mlnhi < 19$ at $z\la 1$.\footnote{Although CCC nominally probes absorbers with \nhi\ ranging from $10^{15.1}$ to $10^{19}$ \cmm, our survey is not complete below $\mlnhi \la 15.3$ (i.e., it does not follow the \hi\ column density distribution function) because high SNR spectra are required to probe gas with $\xh \la -1$ at these low \hi\ column densities (see \S5.1 in \citetalias{lehner18}).} In \citetalias{wotta19}, we focus on deriving the properties of pLLS and LLS absorbers.\footnote{We have removed the strong \hi\ absorber at $z=0.278142$ toward J111507.65+023757.5 from our sample owing to a more uncertain \nhi\ than originally estimated in \citetalias{lehner18}. With newly Keck HIRES observations of \mgii\ and \feii\ and considering the velocity structure of the metal-line absorption, it is apparent that at least three components instead of two are needed to fit the Lyman series. That additional component removes any constraint from the positive velocity wing of \lyb\ and it is now not possible to constrain \nhi\ better than $\mlnhi = [18,19.6]$ (see \citealt{berg19} for more information). \label{foot-lost}} \citetalias{wotta19} discusses the origins of the samples of SLLSs and DLAs that we used for comparison, and we refer the reader to this paper for more information. The lower \hi\ column density SLFSs with $15<\mlnhi <16.2$ studied here likely probe the more diffuse CGM and the transition between the CGM and the \lya\ forest/IGM. It is the first large sample of SLFSs assembled for the purpose of studying their metallicity distribution and physical conditions.

As presented and discussed in length in \citetalias{lehner18}, the QSO spectra for the SLFS sample were retrieved from the \hst/COS G130M and/or G160M archive. Several of these QSOs were also observed from the ground with high-resolution Keck/HIRES and VLT/UVES spectra, which provide access to the strong NUV \mgii\ and \feii\ transitions. All the absorbers were uniformly analyzed to derive the column densities of \hi\ and metal atoms and ions. In total, we have a sample of 152 SLFSs spanning the redshift range $0.2 \la \mzabs \la 0.9$. The reader should refer to \citetalias{lehner18} for more details on the column density estimates and redshift distribution.

\subsection{Metallicity Determination}\label{s-met-det}
\subsubsection{General Overview}\label{s-met-det-overview}
To determine the metallicity for each absorber, we follow the same methodology and make the same assumptions described and discussed in \citetalias{lehner18} and \citetalias{wotta19}, with a notable difference is that we have now included \oiv\  in the ionization modeling when this ion is available (see below). Here we provide only a short summary of the method and the main steps, but we encourage the reader to check these two papers for all the information, assumptions and caveats. 

We reiterate that an important aspect of our analysis is where possible 1) we separate individual absorbers on the basis of our ability to cleanly resolve them from one another at the \hst/COS G130M and G160M resolution (this allows us to study metallicity variation over small redshift intervals, see \S\ref{s-pp-abs}), and 2) we estimate the \hi\ and metal-line column densities over the same velocity interval defined by the velocity width of the weak, typically unsaturated \hi\ Lyman series transitions (this is particularly important as the strength of the metal and \hi\ absorption---and hence ionization or metallicity---may drastically change with velocity, which we demonstrate it is the case in \S\ref{s-pp-abs}).

Because we are interested in deriving the metallicities of the cool ionized gas, we consider photoionization as the dominant ionization mechanism. We model the photoionization using Cloudy \citep[version C13.02, last described by][]{ferland13}, assuming a uniform slab geometry in thermal and ionization equilibrium. The slab is illuminated with a Haardt--Madau EUVB radiation field from quasars and galaxies. The absorbers in CCC are unlikely to be affected by leaking ionizing photons from local galaxies since these absorbers are typically found beyond 30--50 kpc from galaxies \citep[e.g.][]{lehner17} and hence beyond the escaping radiation from these galaxies (e.g., \citealt{fox05}). Therefore, the EUVB radiation field from quasars and galaxies likely dominate the photoionization of the absorbers.  While we consider both HM05 and HM12 EUVB fields (see \citealt{haardt96,haardt12}), we adopt HM05 as the fiducial EUVB for the CCC survey. 

As explained in \citetalias{wotta19}, the photoionization modeling is motivated by the findings in \citetalias{lehner18} and \citetalias{lehner13} where we empirically demonstrate that the properties of the low (e.g., \cii, \mgii) and intermediate (e.g., \oii, \ciii) ions and \hi\ in the SLFSs are characteristics of the gas being photoionized. As for the higher \nhi\ absorbers, we  find that the absorption profiles of the low ions and \hi\ are similar; in particular there is a good match in their peak optical depths. This also applies to the intermediate ions most of the times, but as for higher \nhi\ absorbers, sometimes the bulk of the intermediate ion absorption can be at a different velocity than that of the peak optical depth of \hi; in these cases, intermediate ions were not included to constrain the ionization models. We note, however, that owing to the lower \nhi\ in SLFSs, this was a much rarer occurrence than for the pLLSs and LLSs. Since our main goal is to assess the ionization correction needed to determine the metallicities of the cool photoionized gas, we ignore the higher ions that are not fitted by our models (e.g., \ovi), which are rarely explained with photoionization models at the same densities than that of the lower ions (\citetalias{wotta19}), implying that in many cases they can be multiple gas-phases (we will explore this in more details in future CCC papers).

To estimate the posterior metallicity PDFs of the SLFSs, we use Bayesian techniques and MCMC sampling of a grid of Cloudy photoionization models with the same strategy for the priors as in \citetalias{wotta19}. Specifically, we compare predicted metal ion column densities in a pre-computed grid of Cloudy models with observational estimates, using MCMC sampling to provide a posterior PDF for each of the parameters. From the PDF we report a central (median) value and confidence interval (CI, we adopt 80\% CI for lower/upper limits on the metallicity and 68\% CI otherwise). For each absorber, the inputs are \nhi, \zabs, and the column densities of the metal ions. For \nhi, \zabs, and metal-ion column densities that are well-constrained,  the likelihoods are calculated from Gaussian distributions centered on each of the observed ions' column densities with standard deviations based on their measurement errors.\footnote{All the \hi\ column densities in our sample are well constrained, so that we did not have to impose a flat prior value between a minimum and maximum \nhi\ values (e.g., \citealt{glidden16,prochaska17}).} Metal column densities that are lower or upper limits are modeled using a rescaled Q-function or cumulative distribution function, respectively (as described in \citetalias{wotta19} and following \citealt{fumagalli16}). 

The three main parameters in our models are the total hydrogen density (which is connected to  the ionization parameter via $U \equiv n_\gamma/n_{\rm H}=$\,H ionizing photon density/total hydrogen number density), the metallicity, $\xh$, and the carbon-to-$\alpha$ ratio, \ca\ (where $\alpha$ is an $\alpha$-element such as, e.g., O, Si, and/or Mg). We assume solar relative heavy element abundances from \citet{asplund09}, except between C and $\alpha$-elements. Iron or nitrogen does not follow necessarily the nucleosynthesis pattern of the $\alpha$-elements, especially at low metallicities \citep[e.g.,][]{welty97,lehner01a,rafelski12,jenkins17}; when they did not match the model, they were typically removed from the input list of ions used to constrain the photoionization models and hence the metallicities. 

In the best-case scenario, we assume a flat prior on the density (within the range $-4.5 \le \log n_{\rm H} \le 0$) and \ca\ ($-1 \le \ca \le 1$). However, as for the pLLSs and LLSs, the observational constraints are not always optimum, requiring additional constraints on some of the parameters, specifically $U$ and \ca. Our strategy for these absorbers follows \citetalias{wotta19} (and see also \citetalias{wotta16}). For absorbers with observations sufficient to constrain strongly the ionization models, we adopted a flat prior on $U$. For absorbers with insufficient constraints on the ionization models, we follow the approach discussed for the pLLSs and LLSs in \citetalias{wotta19} and adopt a Gaussian prior on $\log U$, with the Gaussian parameters derived from the output PDFs of $\log U$ from the well-constrained absorbers. The approach of adopting priors on \logU\ was first developed by \citet{wotta16} to determine the metallicities in 33 absorbers (pLLSs and LLSs) where only \hi\ and \mgii\ was available. The method was refined in \citetalias{wotta19}. 

In the case of the CCC absorbers observed with COS G130M and/or G160M, we are never in the situation where \mgii\ provides the only metal ion constraint. However, owing to limited wavelength coverage and/or poor SNRs, there are cases where: 1) only \ciii\ or another intermediate ion is detected (though often with non-detections of  several low and intermediate ions); 2)  only \cii\ and \ciii\ are detected (with non-detections of other ions); 3) there are solely upper and/or lower limits on metal ions; 4) only \mgii\ is detected and all other ions provide only upper limits (\ciii\ is often not available in this case owing to wavelength coverage or contamination); or 5)  \ciii\ and one or more singly-ionized species are detected. In cases 3 and 5, the Bayesian MCMC ionization models often converge with only a flat prior on $U$, but the use of a Gaussian prior on \logU\ improves significantly the results (i.e., the models with and without prior have similar median values for the metallicity, but much smaller errors are derived with the Gaussian prior). In the other cases, only limits can be derived, often not constraining (e.g., $\xh <0$), and the use of \logU\ Gaussian prior dramatically improves the results. We further discuss these priors below, first by discussing the major change from \citetalias{wotta19}  with inclusion of \oiv\ in the ionization modeling of some absorbers. 

\subsubsection{Additional Constraints}\label{s-met-det-add-const}
While \oiv\ was not considered in \citetalias{wotta19} in the ionization modeling owing to its relatively high ionizing energy range (54.9--77.4 eV), the  combination of \oii, \oiii, and \oiv\ can provide a unique constraint on the ionization conditions. Furthermore, as we explore in this paper lower \nhi\ absorbers where the gas may be more highly ionized owing to lower densities, it is important to test whether a lack of detectable metal ions is due to the absorbers being shifted to higher ionization states than those often studied. That is, \oiv\ helps us understand if we can reliably apply a Gaussian prior on \logU\ to the full \nhi\ range in CCC. Finally, with the addition of \oiv\ in the model, we can assess the impact that the possible nature of the multiple gas-phase has on the metallicity determination (such as, e.g., part of \oiii\ being in another gas-phase than the low ions). 

We therefore search for all the absorbers with a reported \oiv\ column density in \citetalias{lehner18} (including upper or lower limits). There are 17 SLFSs and 10 pLLSs with the suite of \oii, \oiii, and \oiv\ or a combination of those that includes \oiv\ (see \citetalias{lehner18}). Some of the absorbers have also coverage of \ovi\ or \oi. \oi\ is not detected in any of these absorbers; even though the upper limits on its column densities provide generally no additional constraint, they serve as a consistency check on the models. On the other hand, when \ovi\ is detected, the new models still fail to match \novi\ derived from the observations, implying the absorbers include multiple gas-phases. For the absorbers with detected \oiv\ absorption, we check the alignment of \oiv\ with lower ions and \hi\ since if there is a clear velocity shift, it would strongly imply multiple ionization processes at play or different densities (this is the same guideline for intermediate ions, like \ciii, see above). For two pLLSs, the \oiv\ is clearly shifted relative to \hi\ and low ions (in those cases, \ciii\ is also shifted and not used in the ionization models); we remove them from this sample. 

\begin{figure}[tbp]
\epsscale{1.2}
\plotone{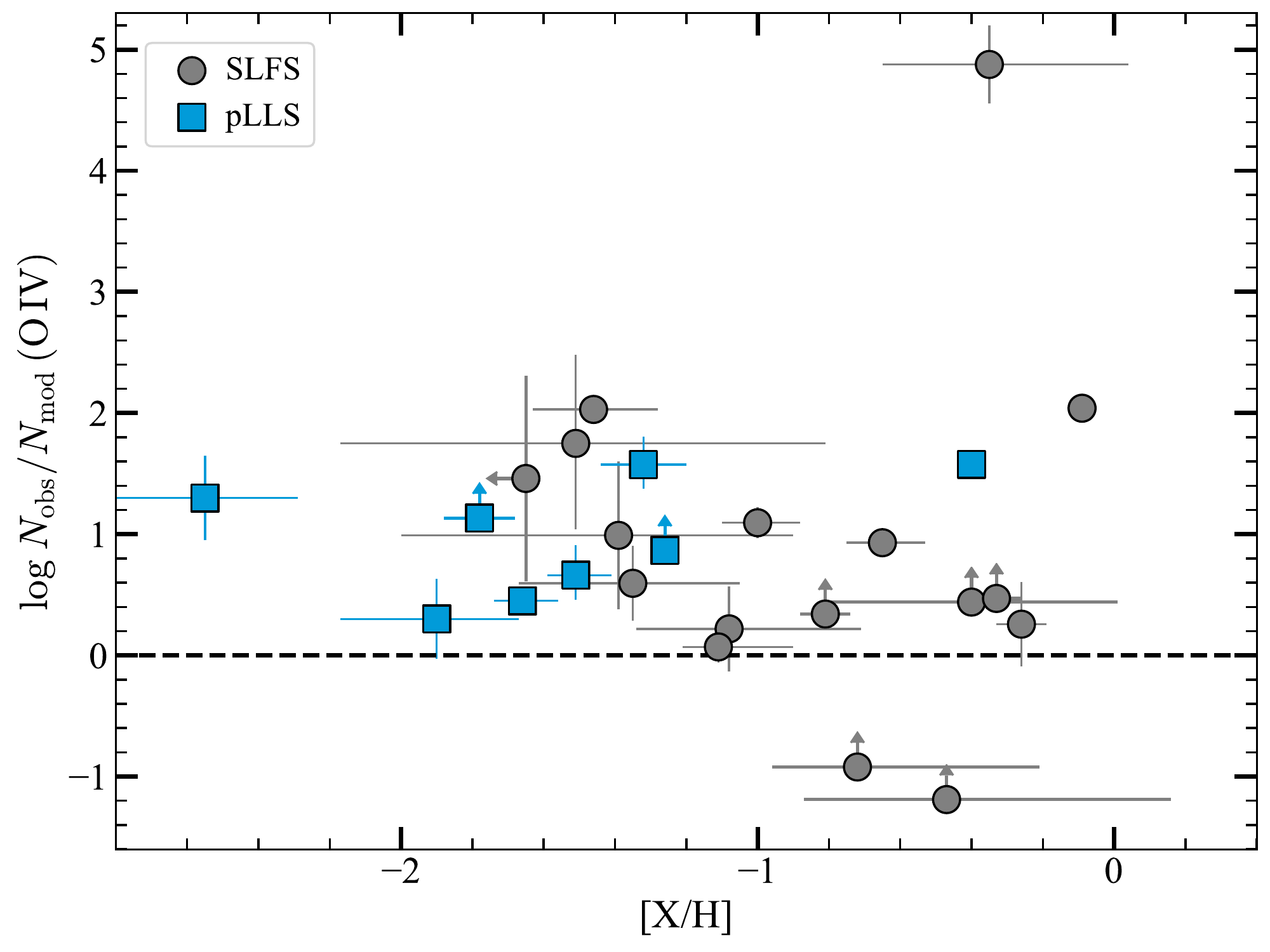}
\caption{Comparison of the observed  and predicted \oiv\ column densities as a function of the metallicity. The modeled \oiv\ densities and metallicities are determined from photoionization models derived {\it without}\ using \oiv\ in these models, and therefore any large discrepancy between the model and observation indicates that the gas has multiple ionized gas-phases. \label{f-multi}
}
\end{figure}

Prior to producing new ionization models that include \oiv, we use the results from our initial runs without \oiv\ to determine how the predicted \oiv\ column densities from the models match the columns derived from the observations. This is shown in Fig.~\ref{f-multi} where we  plot the ratio of the observed  to the predicted \oiv\ column densities for the 17 SLFSs and 10 pLLSs as a function of the metallicities.\footnote{We use the median on \logU\ and \xh\ to estimate the predicted \oiv\ column densities from the ionization models, except for the absorber with an upper limit on the metallicity where we use the $90^{\rm th}$ value on \xh.} For 5 SLFSs and 2 pLLS, the  predicted and observed \oiv\ column densities are within less than 0.1--0.4 dex from each other.  For 2 SLFSs, the observed limits converge to the predicted values. For 10/17 SLFSs and 8/10 pLLSs, the ionization models under-predict the amount of \oiv\ by more than 0.5 dex, implying that \oiv\ probes another gas-phase for these absorbers (either lower densities photoionized or collisionally ionized gas), clearly demonstrating for these absorbers the gas has multiple ionized gas-phases. As demonstrated in Fig.~\ref{f-multi} there is, however, no  visual correlation (confirmed by the Spearman rank-order test) between the difference of the observed and predicted \oiv\ column densities and the metallicity, implying that the possible multiple gas-phase nature of the gas must have a small effect on the metallicity determination as we further show below. 

Given the new information (\oiv), we assess whether we can avoid adopting a Gaussian prior on \logU\ for the six absorbers that needed this constraint before including the \oiv\ column densities. In all these cases, we have enough information when including the \oiv\ to produce robust results using a flat prior on \logU. Most importantly, the metallicities derived in these cases with a flat prior on \logU\ are within 0.1--0.3 dex of those derived previously using a Gaussian prior on \logU. To test how generally such agreement is found, we calculated the best parameters using Gaussian priors on \logU\ for  15 systems in which \oiv\ is the least underpredicted in our baseline models, then recalculated the best parameters using a flat prior. In all cases we found consistent metallicities between the cases with flat versus with Gaussian priors.  Thus, even though the best values of \logU\ may differ by a few $\sigma$ between the cases, the models in all cases converged to adequate solutions that were in good agreement.  This further demonstrates the reliability of using the Gaussian \logU\ prior to estimate the metallicities of poorly constrained absorbers. 

For 5/17 SLFSs and 7/10 pLLSs, we did not find a good match between the observations and models when \oiv\ was added, while for all these absorbers a good match was found without the inclusion of \oiv. In these cases, the models with \oiv\ converge to a solution with the reported median metallicities changing on average by $-0.2 \pm 0.2$ dex (the full range being from $-0.5$ to $+0.1$ dex). However, the predicted column densities are not in good agreement with the observed ionic column densities (while it is when \oiv\ is not included in the models). Typically, in these cases there is a strong underprediction of the singly-ionized species (e.g., \mgii, \oii) and a moderate under or overprediction of \oiii, showing that the models shift to too high ionization.\footnote{There are two cases where the opposite is observed; in both cases the error bars on the column densities of the low ions are very small, driving the solution of the models.} This discrepancy with observed column densities is good evidence that such absorbers probe multiple ionized phases. The ionization models for the low and intermediate ions provide, however, in these cases always an adequate solution. It is still plausible that a fraction of the column density of the intermediate ions (in particular \oiii\ or \ciii) in these cases could be in the same gas-phase than \oiv, but based on the change of the metallicity observed when \oiv\ is or is not included in the models, this effect must be small on the metallicity determination. In fact, we model 3 absorbers that have information from low and intermediates ions but  using only the low ions. The difference in the median metallicities between the models with just the low ions and with both the low and intermediate were consistent within 0.1--0.2 dex, further demonstrating that the possible effect of  multiple gas-phases on the derived metallicities is small. The weaker impact on the metallicity than on the ionization parameter (\nh\ and \nnh) is a result of singly ionized species (detected or constraining upper limits) being very sensitive to the metallicity and  not strongly dependent on the ionization parameter owing to  more similar ionization potentials between \hi\ and singly ionized species (in particular \mgii\ or \siii; see also Fig. 6 in \citetalias{wotta16} and Fig. 10 in \citetalias{lehner13}). 
\begin{figure}[tbp]
\epsscale{1.2}
\plotone{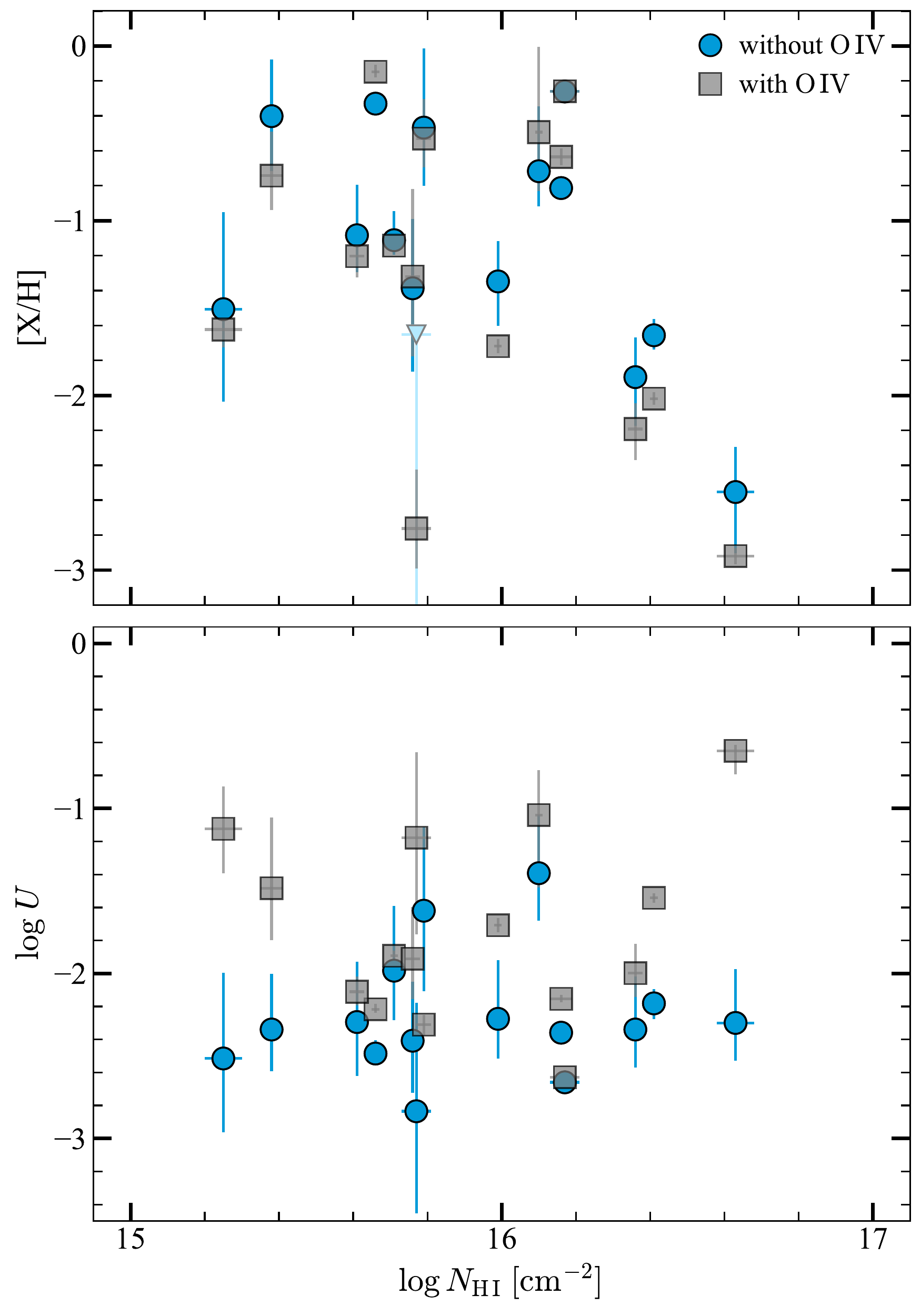}
\caption{Comparison of the metallicities ({\it top}) and ionization parameters ({\it bottom}) derived without and with using \oiv\ in the ionization modeling as a function of the \hi\ column density. The median values of posterior PDFs of the metallicity and \logU\ are adopted as the central values with 68\% CI. For the upper limit, the down triangle is placed at the $90^{\rm th}$ percentile with the light-colored lines showing the 80\% CI.
\label{f-comp-oiv}}
\end{figure}

On the other hand, for 12/17 SLFSs and 3/10 pLLSs, we find a good match between the observations and the models that include \oiv, i.e., the agreement between the observed and modeled column densities are as good as in the original models, but now with the addition of \oiv. In the top and bottom panels of Fig.~\ref{f-comp-oiv}, we show the comparison of the metallicity and \logU\ as a function of \nhi\ for these 15 absorbers with and without \oiv\ included in the ionization models. As alluded to above, the metallicities with and without \oiv\ do not drastically change even when the \logU\ values are in some cases different by more than 1 dex. The addition of \oiv\ in the models systematically reduces the magnitude of the error bars of the metallicities and \logU. However, it is important to note that while the inclusion of \oiv\ or not in the models does not change the metallicity much, other parameters such as \logU, \nh, and \nnh\ are all far more affected. The mean metallicities in this sample with and without \oiv\ are $[-1.31,-1.25]$, a $0.06$ dex difference.\footnote{This result also implies that if some unknown fraction of \oiii\ or \ciii\ arises from a higher ionization phase, it would have a very small effect on the metallicity of the gas.} However, for  \logU, \nh, and \nnh\, the mean values are $[-1.73,-2.27]$, $[-3.43,-2.90]$, $[19.70,19.10]$, implying a difference in the means of 0.53, 0.53, and 0.60 dex (a factor 3--4), respectively. This contrasts greatly compared to the metallicity estimates where the {\it largest}\ differences are at the level of a factor 2--3 (see Fig.~\ref{f-comp-oiv}). This result is not surprising (see previous paragraph), and shows that the metallicity determination at low $z$ is therefore quite robust even when large ionization corrections are applied. 

\subsubsection{Updated \logU\ Prior and Additional Considerations}\label{s-met-det-update}
\begin{figure}[tbp]
\epsscale{1.2}
\plotone{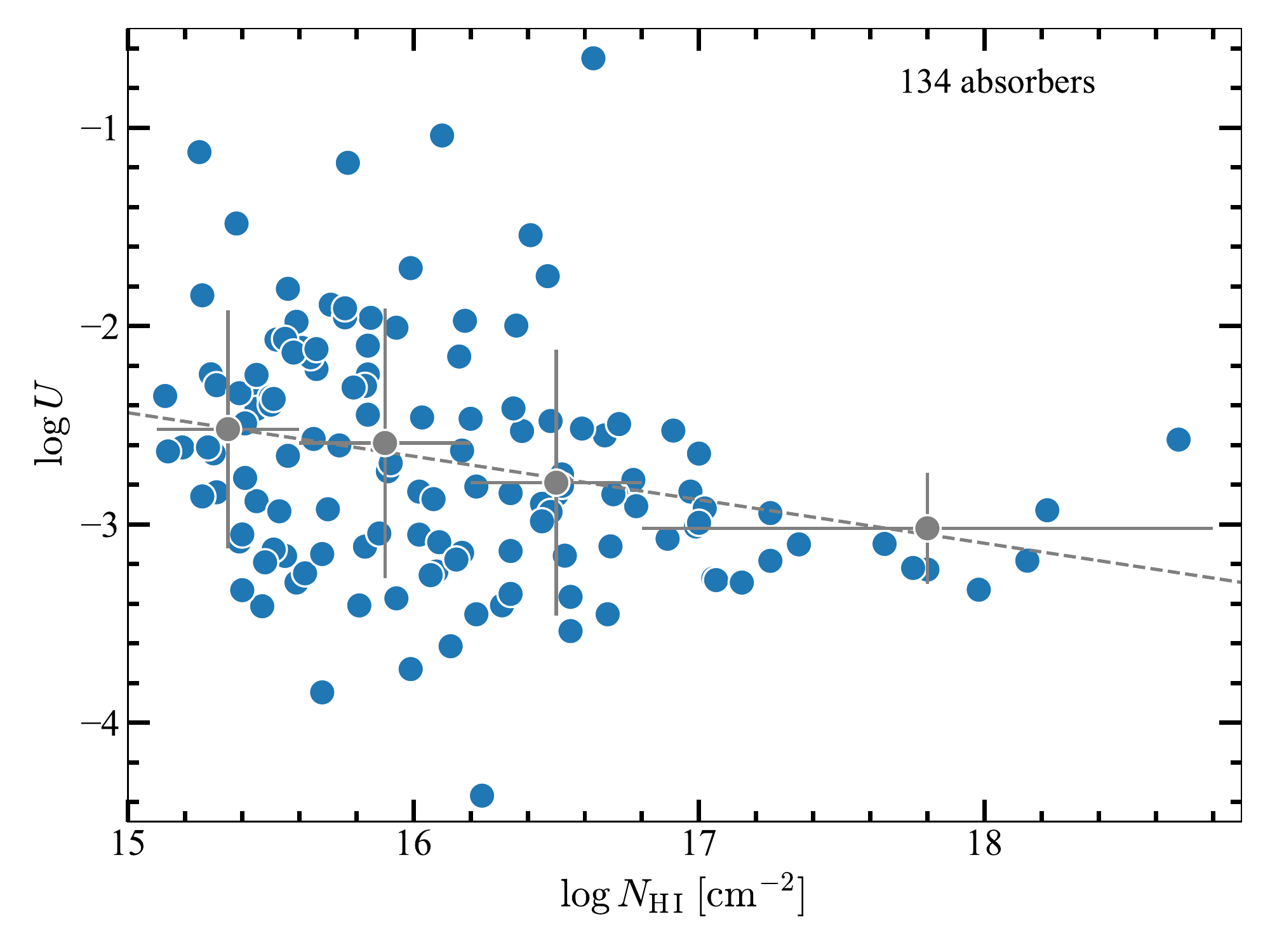}
\caption{The \logU\ median values as a function of \nhi\ for 134 $z\la 1$ absorbers in the CCC sample. The dashed-line shows a linear fit to the data (with a slope $-0.21$ and intercept $+0.85$). The crosses show the mean values of the median values in each interval \nhi\ interval indicated by the horizontal bar. 
\label{f-logU-vs-h1}}
\end{figure}

In Fig.~\ref{f-logU-vs-h1}, we show an updated figure of the \logU\ median values as a function of \nhi\ (see Fig.~2 in \citetalias{wotta19}). The same strong anti-correlation is observed between the ionization parameter and \nhi, with a change  in \logU\ that decreases by $\sim$0.5 dex in \logU\ over $\sim$3 dex in \lnhi. This implies the ionization conditions for absorbers across this regime are still comparable across the \nhi\ range probed by CCC, but the scatter in \logU\ has increased for SLFSs and pLLSs. The updated mean values in these \nhi\ intervals shown in this figure are summarized in Table~\ref{t-logu-constraint}. When applying a prior on the less-well-constrained systems, we adopt here these updated \logU\ values even for the pLLSs and LLSs.  Although the results do not quantitatively change much from \citetalias{wotta19}, we apply these updated \logU\ priors to be consistent with the SFLSs  and we therefore updated the results for the 128 absorbers in CCC that require a \logU\ Gaussian prior. 

We emphasize that the Gaussian priors on \logU\ provide adequate priors for any absorbers studied here, even for the absorbers that may be more highly ionized with just detection of intermediate ions rather than both low and intermediate ions. Indeed, we show in \S\ref{s-met-det-add-const} that the metallicity is  less sensitive to the ionization parameter variation than other parameters (\nh, \nnh). We also discuss above that using models that  did not include \oiv, the metallicities derived with the \logU\ Gaussian prior are consistent with those derived with the flat prior on \logU\ and the inclusion of \oiv\ in the models. Additionally, considering the SLFSs that could be modeled with a \logU\ flat prior, about 40\% of the well-constrained SLFSs in our sample have no low ion detected and more than one intermediate ion detected (e.g., a combination of \ciii, \siiii, \oiii).  These absorbers nearly equally populate the \logU\ distribution as absorbers with both low and intermediate ions. Therefore the  Gaussian priors on \logU\ summarized in Table~\ref{t-logu-constraint} are suitable prior for any absorbers since they were derived using absorbers include using low and intermediate ions or only intermediate ions. 

\begin{figure}
\epsscale{1.2}
\plotone{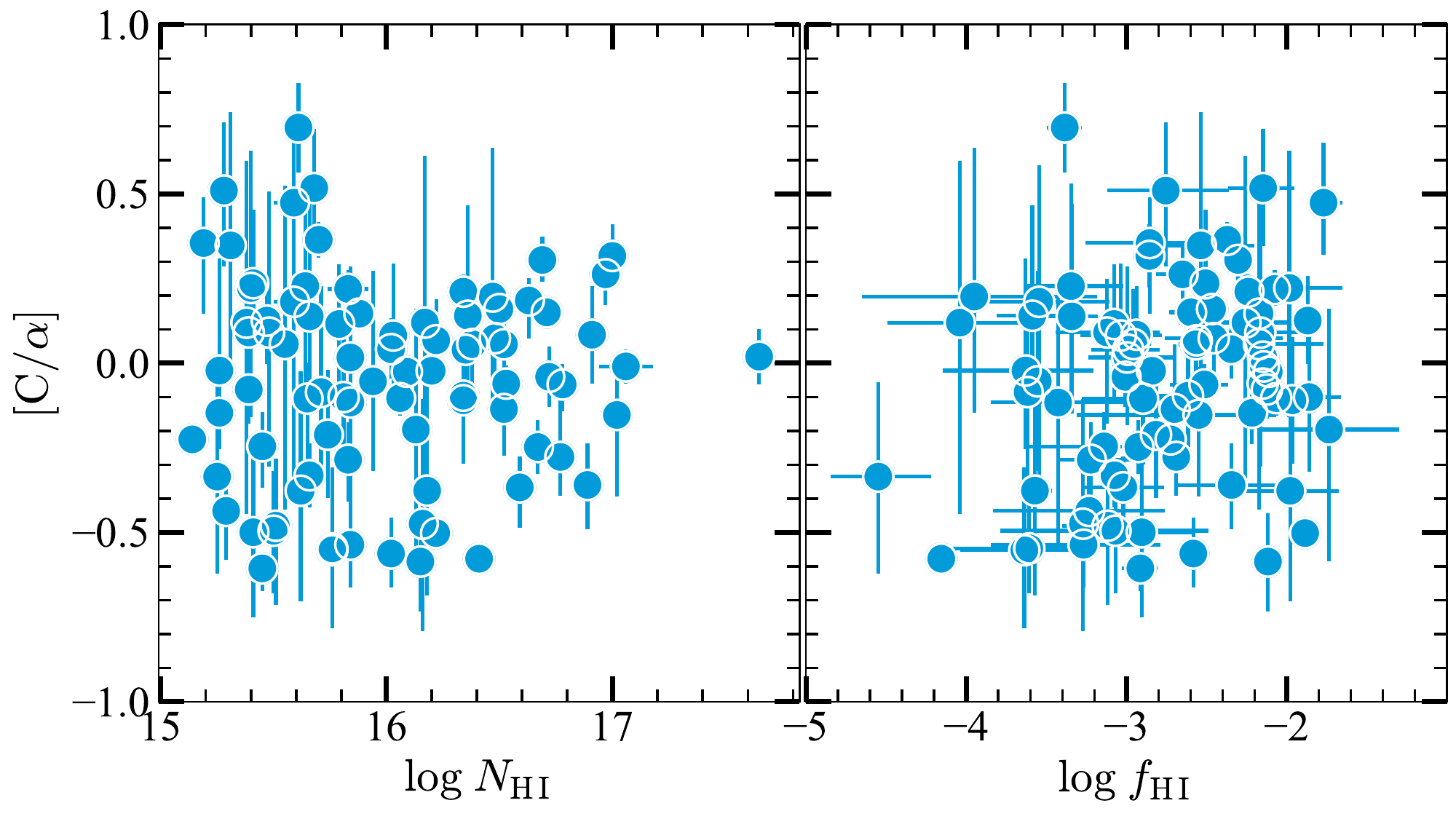}
\caption{The ratio of \ca\ against the \hi\ column density ({\it left}) and neutral fraction ({\it right}). The median values of the \ca\ posterior PDFs are adopted as the central values with 68\% CI. 
\label{f-calpha-nhi}}
\end{figure}

In \citetalias{wotta19}, we also show that the \ca\ distribution for the CCC absorbers (i.e., absorbers with $15<\mlnhi <19$) where a flat prior was used on \ca\  is well-fit by a Gaussian distribution with a mean and standard deviation of $\langle \ca\ \rangle = -0.05 \pm 0.35$ (see Fig.~3 in \citetalias{wotta19}). In Fig.~\ref{f-calpha-nhi}, we show \ca\ as a function of \nhi\ (left panel) and the neutral fraction $f_{\rm H\,I} \equiv \mnhi/(\mnhi+\mnhii)$ (right panel). There is some evidence for a slightly larger scatter in \ca\ for the SLFSs than for the pLLSs or LLSs. However, based on the lack of any relationship between  $f_{\rm H\,I}$ and \ca\ and a similar scatter in  \ca\ for any $f_{\rm H\,I}$, the somewhat larger scatter in \ca\ in the SLFS regime does not seem related to the larger ionization correction at lower \nhi. Therefore the overall model of the \ca\ distribution  with single a Gaussian is adequate. (We note also that there are only small differences in the mean and dispersion of \ca\ between the SLFSs and the pLLSs--LLSs.)

\subsubsection{Results}\label{s-met-det-results}

We summarize the results from the MCMC models in Table~\ref{t-met-sum} for the 262 absorbers in CCC. The new results for the pLLSs and LLSs do not change quantitatively the results from \citetalias{wotta19}, but we have updated the results to be consistent with the newly derived \logU\ Gaussian priors in each \nhi\ intervals (see above, Fig.~\ref{f-logU-vs-h1} and Table~\ref{t-logu-constraint}). As discussed in Footnote~\ref{foot-lost}, we have also removed one LLS from our sample owing to a more uncertain \nhi\ than originally estimated. Finally, we have also updated the \mgii\ column density for the absorber toward J135726.26+043541.3 at $z = 0.328637$ with newly acquired Keck HIRES observations (see \citealt{berg19} for more detail). 

For each absorber, we list in Table~\ref{t-met-sum} the sightline name, the redshift ($z_{\rm abs}$), \nhi, the metallicity, ionization parameter ($\log U$), and C/$\alpha$ ratio (when estimated). Each quantity is reported with the 68\% CI and median values, except in the cases where we derive an upper or lower limit where we instead report 80\% CI (the highest value of that interval corresponds to the 90\% CI). When a colon is present after the median value of $\log U$ in Table~\ref{t-met-sum}, this implies that a Gaussian prior was used to determine the properties of the absorber (see \S\ref{s-met-det}). All the comparison between observations and models and corners plots and all the column densities used in the ionization modeling for all the absorbers in CCC are provided as supplement materials (see Appendix for more details). 

\subsection{Proximate and Paired Absorbers}
\label{s-pp-abs}
As noted in \citetalias{lehner18} and \citetalias{wotta19}, several absorbers in the CCC sample are proximate or paired absorbers, which are absorbers near the redshift of the QSOs or two absorbers that are close in redshift, respectively. Prior to showing and discussing the metallicity PDF of the absorbers, we need to determine if we can include those in the sample or whether they have unique properties that could bias the metallicity PDF.

\subsubsection{Proximate Absorbers}\label{s-proximate}
There are only  13 proximate absorbers in CCC, i.e., absorbers with $\Delta v \equiv (z_{\rm em} - z_{\rm abs})/(1+z_{\rm abs})\, c < 3000$ \km, where $c$ is the speed of light. Among these, 8 are SLFSs and the 5 others are pLLSs. None of these absorbers have an ionization parameter or metallicity that substantially differs from the intervening systems in their measurable properties (see also \citetalias{wotta19}). Since they represent a small-size sample and  their metallicities are not different from the rest of the absorbers, we elect to include them in our sample. We identify the proximate absorbers in Table~\ref{t-met-sum}. 

\subsubsection{Paired absorbers}\label{s-paired}
The paired absorbers are absorbers that are closely separated in redshift space, specifically we define them as $\Delta v \equiv |(z^2_{\rm abs} - z^1_{\rm abs})/(1+z^1_{\rm abs})\, c| < 500$ \km\ (see \citetalias{lehner18}). These absorbers offer the unique opportunity to study metallicity variation over small velocity/redshift scale along the same line of sight. With 30 such paired absorbers, they provide the largest sample where metallicity variation in absorbers closely spaced in velocity can be assessed.

In Table~\ref{t-close}, we list the paired absorbers, giving their \nhi, \zabs, $\Delta v$, and $\xh$. There are only 4 paired-absorbers that involve a combination of pLLS-pLLS (3)  and pLLS-LLS (1). The remaining 26 are SLFS-SLFS or SLFS-pLLS paired-absorbers (13 in each category). As noted in \citepalias{lehner18},  the lack of a larger number of pLLS-LLS and LLS-LLS paired absorbers is due to two effects: 1) there is an observational bias in that it is easier at the COS resolution to separate absorbers  robustly for absorbers with lower \nhi\ than with higher \nhi; and 2) the \hi\ column density distribution function provides a larger number of SLFSs than pLLSs or LLSs, making pairs of lower column density absorbers more likely in a \hi-selected sample with column densities in the range $15<\mlnhi <19$.  Some attempts to study separately the metallicities and physical conditions for a few strong \hi\ absorbers have been made at low $z$  (e.g., \citealt{zahedy19,muzahid15}), but often this requires to make critical assumptions on the width of the lines in heavily blended absorbers to attempt to determine the column densities in individual components of the absorbers.  Among the 30 absorbers in CCC, 3 have upper limits on the metallicities where it is unclear if there is a difference in the metallicities between the paired-absorbers (see Table~\ref{t-close}), and we do not discuss these further. 

\begin{figure}[tbp]
\epsscale{1.2}
\plotone{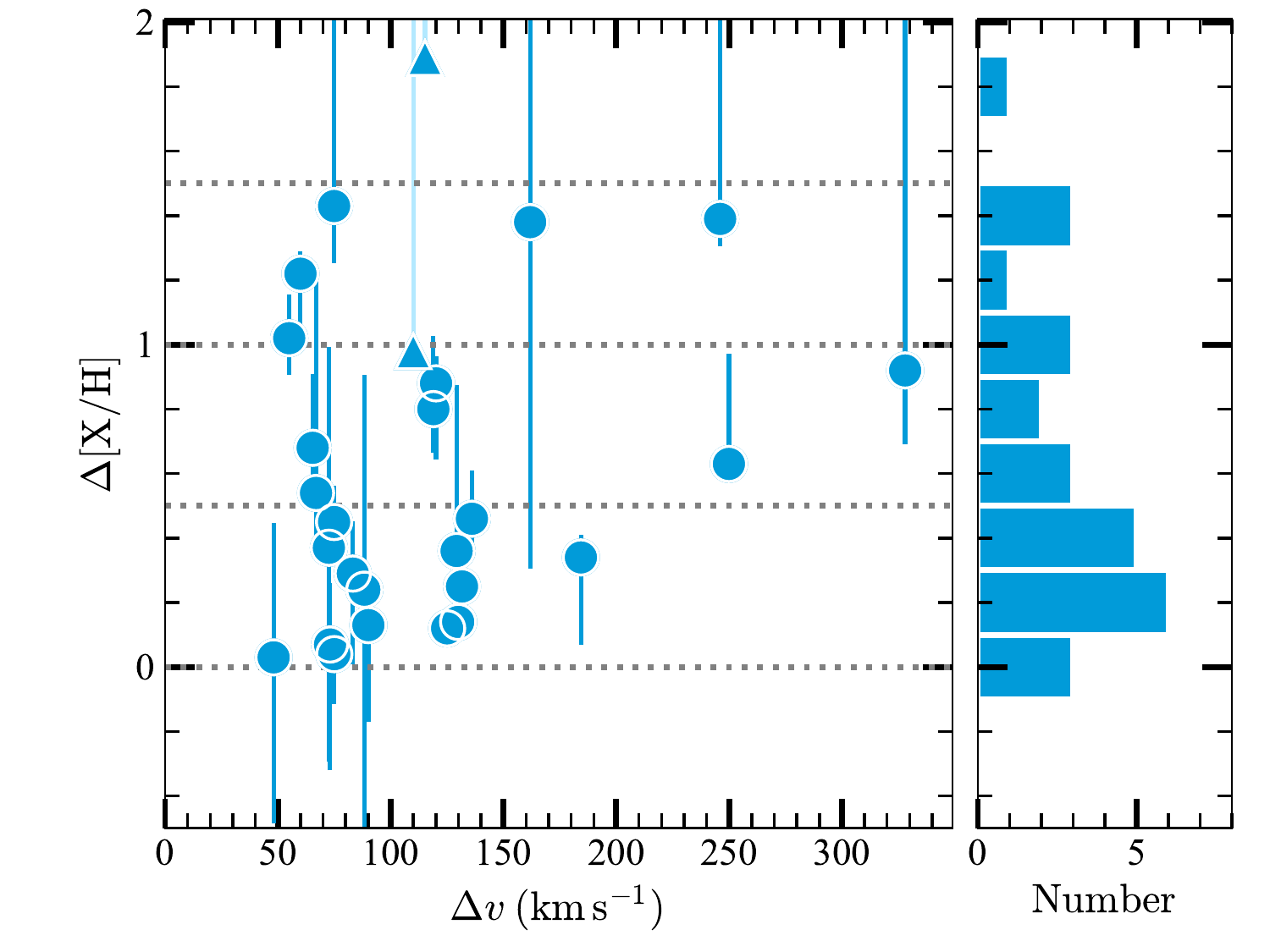}
\caption{{it Left}: scatter plot ({\it left}) of the difference between the higher and lower median metallicities for the closely-redshift separated absorbers (a.k.a. paired-absorbers) as a function of their absolute velocity difference. For the upper limits (blue triangles), we use the lower bound of the 80\% CI to be the most conservative. {\it Right}: distribution of the metallicity differences between paired-absorbers.
\label{f-met-dv}}
\end{figure}

To visualize and to quantify the metallicity variation of the paired absorbers, we show in the left panel of Fig.~\ref{f-met-dv} the absolute metallicity difference ($\Delta \xh$) between the higher and lower median metallicities of the paired absorbers against their absolute velocity separation and in the right panel the distribution of $\Delta \xh$. There is no obvious relation between the velocity separating the paired absorbers and the degree of metallicity variation. At small velocity separation of $\sim$50--150 \km, we find $\Delta \xh$ varies from 0 to $1.7$ dex (i.e., a factor $>50$ variation in the metallicity). Only 3/27 (5\%--25\%, using the Wilson-Score test with a 90\% CI; we apply this test hereafter in this paragraph with the same CI) absorbers show no metallicity variation (6/27; 12\%--38\% if 68\% CI are considered). The majority of the absorbers (14/27; 37\%--68\%) have a metallicity variation in the range $0.2\la \Delta \xh\la 0.7$, i.e., they show significant change in the metallicity from a factor 1.6 to 5. Even more striking is that 10/27 (23\%--54\%) have metallicity variations from a factor 6 to 25 (and at least one with $>50$). Therefore for the majority of the paired absorbers (75\%--96\%), there is evidence for substantial metallicity variations over $\Delta v \la 300$ \km.

The first consequence is  that it is reasonable to treat these absorbers as individual absorbers for the purpose of determining the metallicity PDF, and accordingly we do not treat paired absorbers as single absorbers. The second consequence is, when possible, components in absorbers should be treated individually to derive the true metallicity of the gas rather than a mismatch between very different types of gaseous components.  We discuss in more detail the implication of these findings in \S\ref{s-disc-var}. 

\section{Metallicities of the CGM Absorbers at $\lowercase{z}<1$}\label{s-cgm-prop}
In this section, we focus on the metallicities of the SLFSs. However, to understand their evolution with \nhi\ and $z$, we need to revisit the metallicities of the pLLSs and LLSs as well as the SLLSs and DLAs. We also combine the entire sample to consider the implications of the \ca\ distribution as a function of the metallicities of the absorbers. 

\subsection{Metallicity PDF for the SLFSs and Evolution with \nhi}\label{s-met-pdf}
\begin{figure}[tbp]
\epsscale{1.2}
\plotone{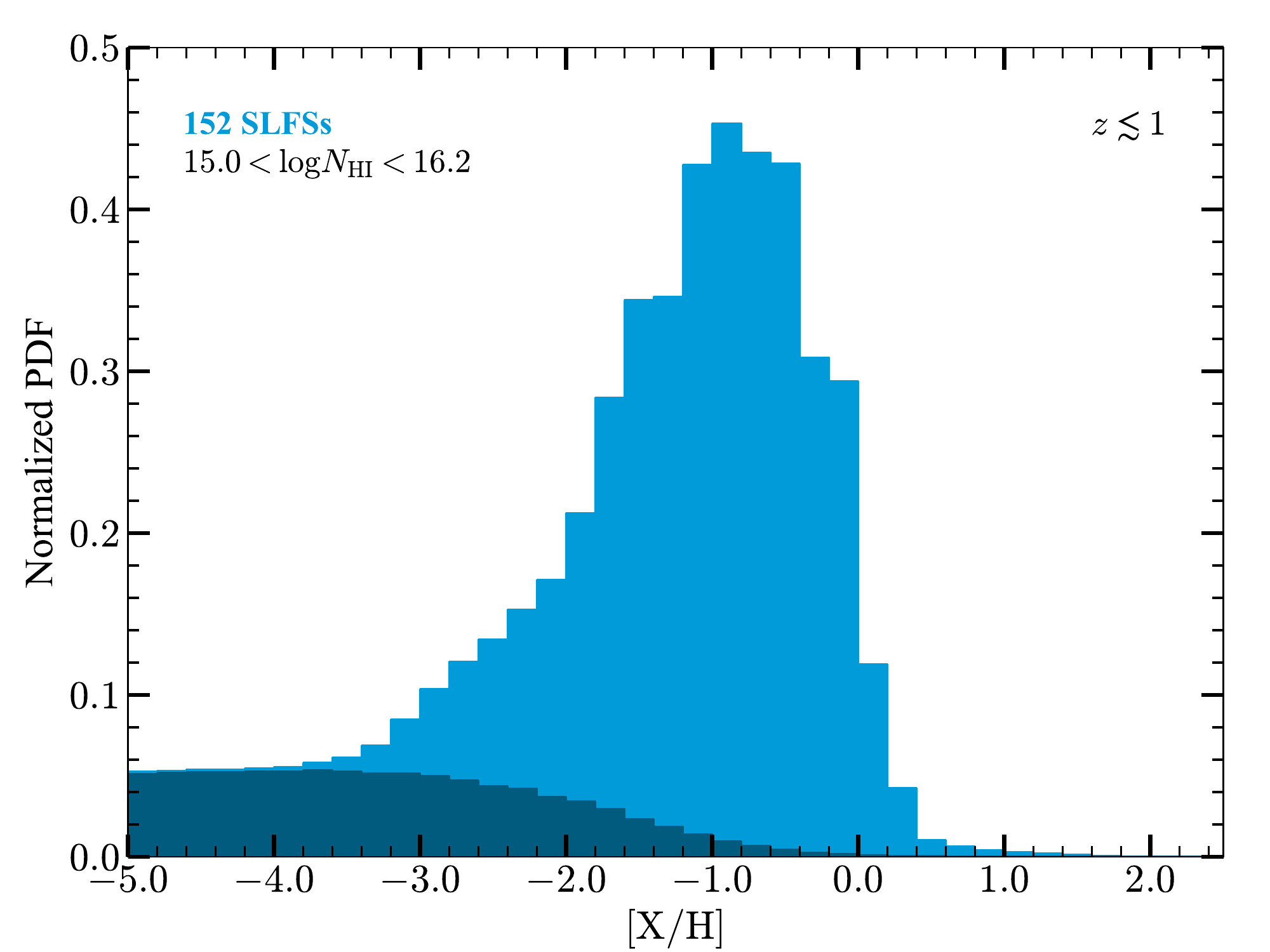}
\caption{Posterior metallicity PDF of the SLFSs at $z \la 1$. The shaded regions indicate the contribution from the upper limits. \label{f-mdf-slfs}}
\end{figure}

In Figure~\ref{f-mdf-slfs}, we show the metallicity PDF of the 152 $z<1$ SLFSs in CCC.  This PDF is constructed by combining the normalized metallicity PDFs of all SLFSs. The metallicity PDF has a strong peak at $\xh\simeq -0.90$ (mode of the distribution) and is negatively skewed to low metallicities with a mean metallicity ($-1.47$ dex) lower than the median metallicity ($-1.18$ dex). As in most of our PDF figures, systems for which the distributions represent limits are shaded differently (e.g., the portion of the PDF tracing upper limits is shaded dark blue). 

\begin{figure}[tbp]
\epsscale{1.2}
\plotone{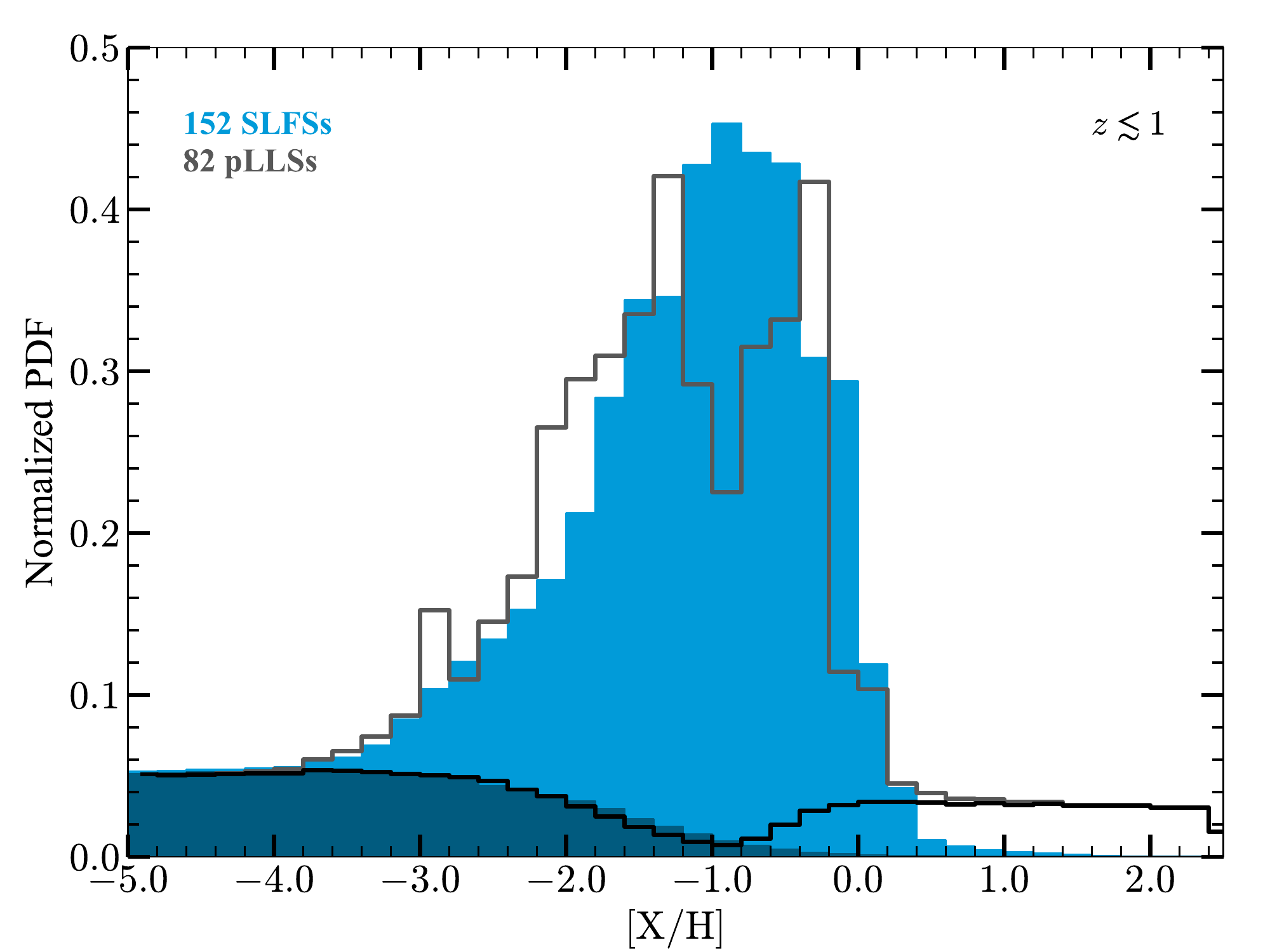}
\caption{Posterior metallicity PDFs of the SLFSs and pLLSs at $z\la 1$. The shaded regions or lower black histograms indicate the contributions from  upper and lower limits. \label{f-comp-slfs-plls}}
\end{figure}

In Table~\ref{t-stat}, we summarize the mean and median metallicities of the SLFSs including or removing the upper limits. We also list in Table~\ref{t-stat} the mean, standard deviation, and median values for the pLLS, LLSs, the combined sample of pLLSs+LLSs, and the entire CCC sample (i.e., the SLFSs+pLLSs+LLSs). Although the mean values are similar for the SLFSs and pLLSs, as we show in Fig.~\ref{f-comp-slfs-plls}, the metallicity PDF of the pLLSs is not consistent with a single unimodal distribution (see also \citetalias{wotta19}). Fig.~\ref{f-comp-slfs-plls} shows in fact how remarkably the metallicity PDF of the SLFSs fills the dip of the metallicity PDF of the pLLSs. There is a slightly larger frequency of very metal-poor absorbers in the pLLS regime compared to the SLFSs, however this is not significant: the fraction of metal-poor absorbers ($\xh < -1$) is 53\%--66\% for the SLFSs compared to 53\%--71\% for the pLLSs; the fraction of very metal-poor absorbers ($\xh < -1.4$, see below) is 34\%--47\% for the SLFSs compared to 39\%--57\% for the pLLSs (all CIs are at the 90\% level). The ranges of metallicities probed by the SLFSs, pLLSs, and LLSs are also quite similar. There is therefore a major change in the shape of the metallicity between the SLFSs and pLLSs, but not in the observed spread of metallicities or the fractions of metal-poor vs. metal-rich absorbers. However, as we show below the difference in the shape of the metallicity PDFs is also redshift dependent.  

\begin{figure}[tbp]
\epsscale{1.2}
\plotone{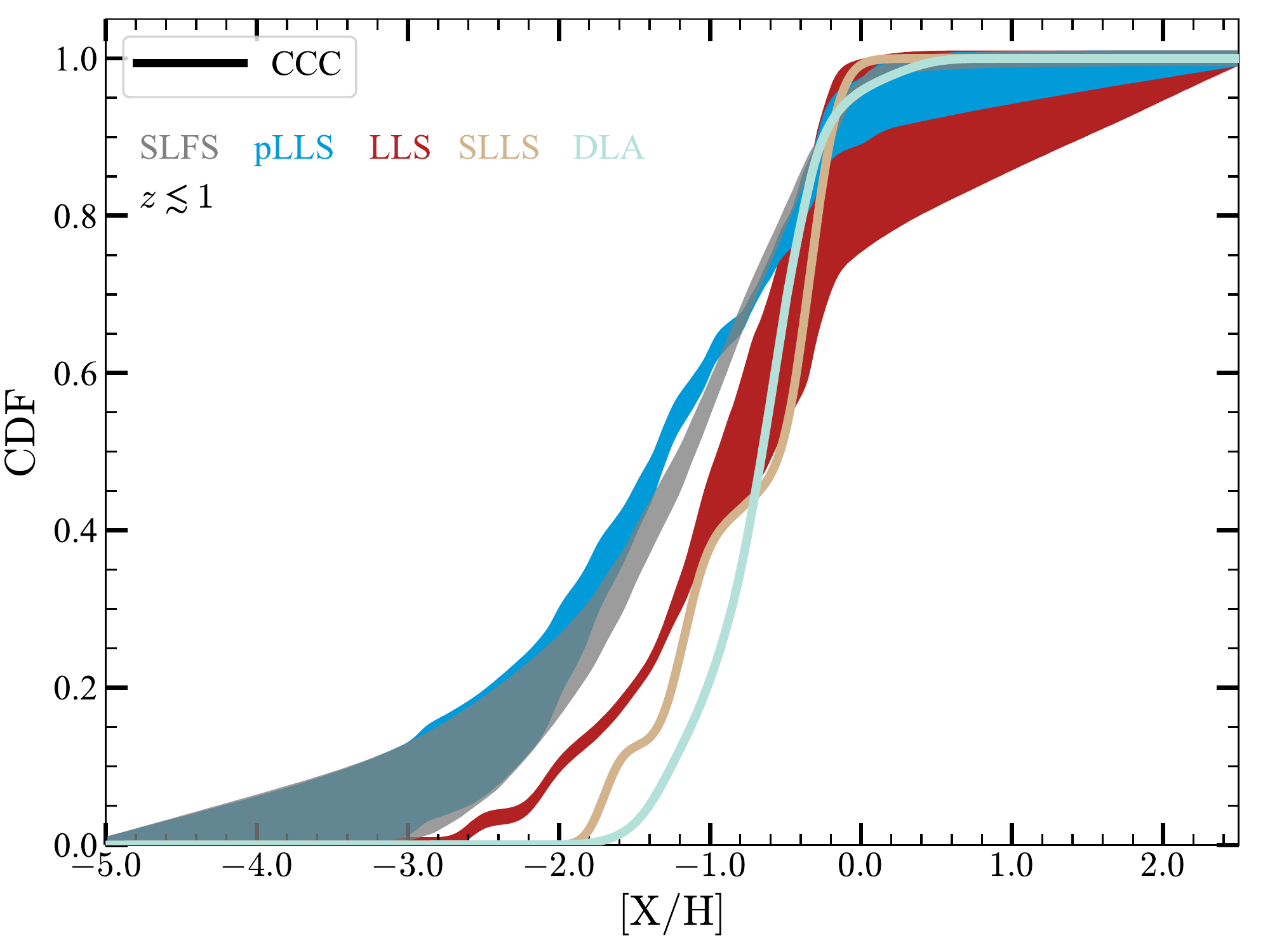}
\caption{The cumulative probabilities of the SLFSs, pLLS, LLS, SLLS, and DLA metallicity PDFs. The wider dispersion in the CDFs at $\xh < -1$ and $\xh >-1$ for the SLFSs, pLLSs, and LLSs indicates the impacts of the upper and lower limits (see \citetalias{wotta19} for more detail). 
\label{f-cdf-abs}}
\end{figure}

The means and medians listed in Table~\ref{t-stat}  further demonstrate a change of the metallicity PDF with \nhi. To observe this across the entire range of absorbers with $\mlnhi > 15$, we show in Fig.~\ref{f-cdf-abs} the cumulative distribution functions (CDFs) of the SLFS, pLLS, LLS, SLLS, and DLA metallicity PDFs (see \citetalias{wotta19} for the description of the estimation of the CDFs including upper and lower limits and the description of the SLLS and DLA samples). The CDF of the pLLSs has  a slightly larger fraction of absorbers with $-2 \la \xh \la -1$ than for the SLFSs and an inflection point is present only in the CDF of the pLLSs corresponding to the dip in the PDF at $\xh\simeq -1$; otherwise both CDFs are comparable. The CDFs  show that the distributions of the metallicity PDFs of the SLFSs and pLLSs have some differences, but the variations of the metallicity CDFs are striking as \nhi\ increases in the LLS, SLLS, and DLA regimes.  

\begin{figure}[tbp]
\epsscale{1.2}
\plotone{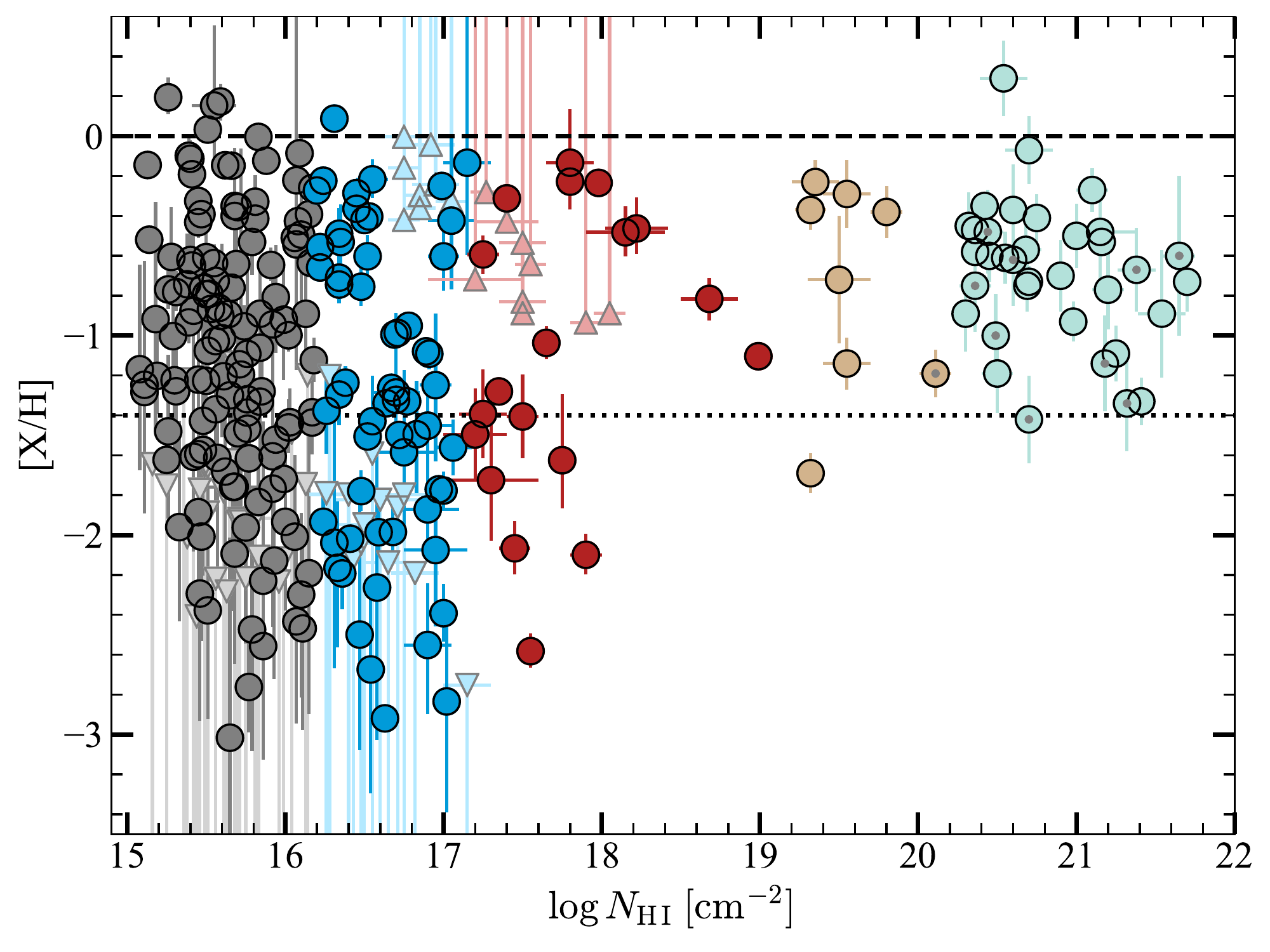}
\caption{Metallicities of the SLFSs ($15<\mlnhi < 16.2$), pLLSs ($16.2\le \mlnhi < 17.2$), LLSs ($17.2\le \mlnhi < 19$), SLLSs ($19\le \mlnhi < 20.3$), and DLAs ($ \mlnhi \ge 20.3$) at $z\lesssim1$ as a function of \nhi. For the SLFSs, pLLSs, and LLSs, the median values (circles) of the metallicity posterior PDFs are adopted with 68\% CI, except for lower (triangles) and upper (down triangles) limits where the lower and upper values represent the $10^{\rm th}$ and $90^{\rm th}$ percentiles, respectively, with the light-colored lines showing the 80\% CI. For SLLSs and DLAs, the best estimated values with their 68\% errors are shown.  The dashed line represents solar metallicity, while the dotted line represent the very low metallicity gas, which is defined as the 2$\sigma$ lower bound of the DLA metallicities (i.e., $\xh <-1.4$). The SLLSs and DLAs with small gray circles have their metallicities derived from \feii\ (and are corrected by $+0.5$ dex, see \citetalias{wotta19}). \label{f-met_vs_nh1}}
\end{figure}

Another way to examine the change of the metallicity with \nhi\ is to simply plot the metallicities of the absorbers as a function of \nhi\ from the SLFS to the DLA regimes, which is shown in Fig.~\ref{f-met_vs_nh1}. This has the advantage that one does not have to make an {\it a priori}\ differentiation between the different absorbers according to a \hi\ column density definition. For the SLFSs, pLLSs, and LLSs with well-constrained metallicities (i.e., not including the lower and upper limits), the central values represent the median of the posterior PDFs, and the error bars represent the 68\% CI. For the upper and lower limits, the down and upward triangles give the $90^{\rm th}$ and $10^{\rm th}$ percentiles while the vertical bar gives the 80\% CI. For the SLLSs and DLAs, the best estimates with their $1\sigma$ error bars are shown (see the tabulated and adopted values in \citetalias{wotta19}). The horizontal dashed line at $[{\rm X/H}]=0$ represents solar metallicity. The horizontal dotted line at $\xh =-1.4$ represents the 2$\sigma$ lower bound of the DLA metallicities at $z\la 1$. Following \citetalias{wotta16} and \citetalias{wotta19}, we define absorbers with $[{\rm X/H}]\le -1.4$ as ``very metal-poor'' absorbers. 

 Figs.~\ref{f-cdf-abs} and \ref{f-met_vs_nh1} reiterate the overall change of the metallicities with \nhi. Very-metal poor absorbers are rarely observed at $\mlnhi > 18$, but are common at  $\mlnhi \la 18$. Absorbers with metallicities in the range $-1.4< \xh \la 0$ are observed for any \nhi\ between $\mlnhi \simeq 15$ and 22. The bulk of the DLAs ($93\%$) and \hi-selected SLLSs (88\%) at $z\la 1$ have their metallicities in the range  $-1< \xh \le -0.2$.\footnote{We note that there is a population of \mgii-selected SLLSs that have solar or super-solar metallicities, see, e.g., \citetalias{lehner13}; \citealt{som15,quiret16,fumagalli16}), which are not represented with the \hi-selection of these absorbers. } In contrast for absorbers with $15<\mlnhi<19$, only 35\% have  metallicities in the range  $-1< \xh \le -0.2$, while 58\% have metallicities $\xh \le -1$.\footnote{We re-emphasize that these results for the absorbers with $15<\mlnhi<19$ are valid for the adopted HM05 EUVB used in the ionization modeling. With HM12, the metallicity would on average increase by $+0.4$ dex for the SLFSs and pLLSs and by $+0.2$ dex for the LLSs \citepalias{wotta19}. The fraction of very metal-poor gas would therefore be smaller. There would also be a larger fraction of super-solar metallicity absorbers especially in the SLFS regime.}

A striking feature from Fig.~\ref{f-met_vs_nh1} is that the lack of data around $-1.2 \la \xh \la -0.8$ is  present in the pLLS and LLS regimes, but not in the SLFS \hi\ column density range. This is remarkable because the separation between SLFSs  and pLLSs is somewhat arbitrary (see \citetalias{lehner18} and \citetalias{wotta19}), and yet for absorbers with $15.1\la \mlnhi < 16.2$, there is no evidence of a metallicity gap in  $-1.2 \la \xh \la -0.8$ range. The paucity of data in the metallicity range $-1.2 \la \xh \la -0.8$ is the only one in Fig.~\ref{f-met_vs_nh1} that is statistically significant and not filled by 1$\sigma$ error bars. 

\subsection{Cosmic Evolution of the Metallicity at $z \la 1$ }
\label{s-met-red}

In the above analysis, we have combined absorbers over all redshifts at $z\la 1$, setting aside any plausible redshift evolution in the metallicity of these absorbers over the redshift range probed by CCC. For the pLLSs and LLSs (see   Fig.~14 in \citetalias{wotta19}), over $0.25 \la z \la 1.1$ where most of the data are, the metallicity-redshift plot does not show significant trends. However, we noted that at $z \ga 0.45$, there is a clear lack of data around $\xh \simeq -1$ (corresponding to the dip in the metallicity PDF of the pLLSs), while at  $z< 0.45$ there is gas present at $\xh \simeq -1$ (see Fig.~15 in  \citetalias{wotta19}), implying that the shape of the metallicity PDFs of these absorbers is redshift dependent. We revisit this in some detail below, comparing the metallicity PDFs of the SLFSs and pLLSs as a function of $z$. 

\subsubsection{Evolution of the Metallicity of the SLFSs at $0.2 < z <0.9$ }
\label{s-met-red-slfs}
Prior to discussing the redshift evolution of the SLFSs, we emphasize that contrary to the pLLSs and LLSs, the SLFSs in CCC are solely from our original survey presented in \citetalias{lehner18}, and hence the redshift interval probed by SLFSs is strictly defined by our survey in the range $0.2 < z <0.9$ (this redshift range is dictated by the need to have both high and low Lyman series transitions to determine accurately \nhi, see \citetalias{lehner18}). For the pLLSs and LLSs, additional absorbers from the literature were used, and therefore the redshift interval is larger---$0.1 \la z \la 1.1$, even though most of the data are also in the range $0.2 < z <0.9$. Therefore for the SLFSs, we do not have any information below $z\la 0.2$ where there could be some evolutionary change in the metallicity distribution as we discuss in \S\ref{s-disc-cos-halos}. 

\begin{figure}[tbp]
\epsscale{1.2}
\plotone{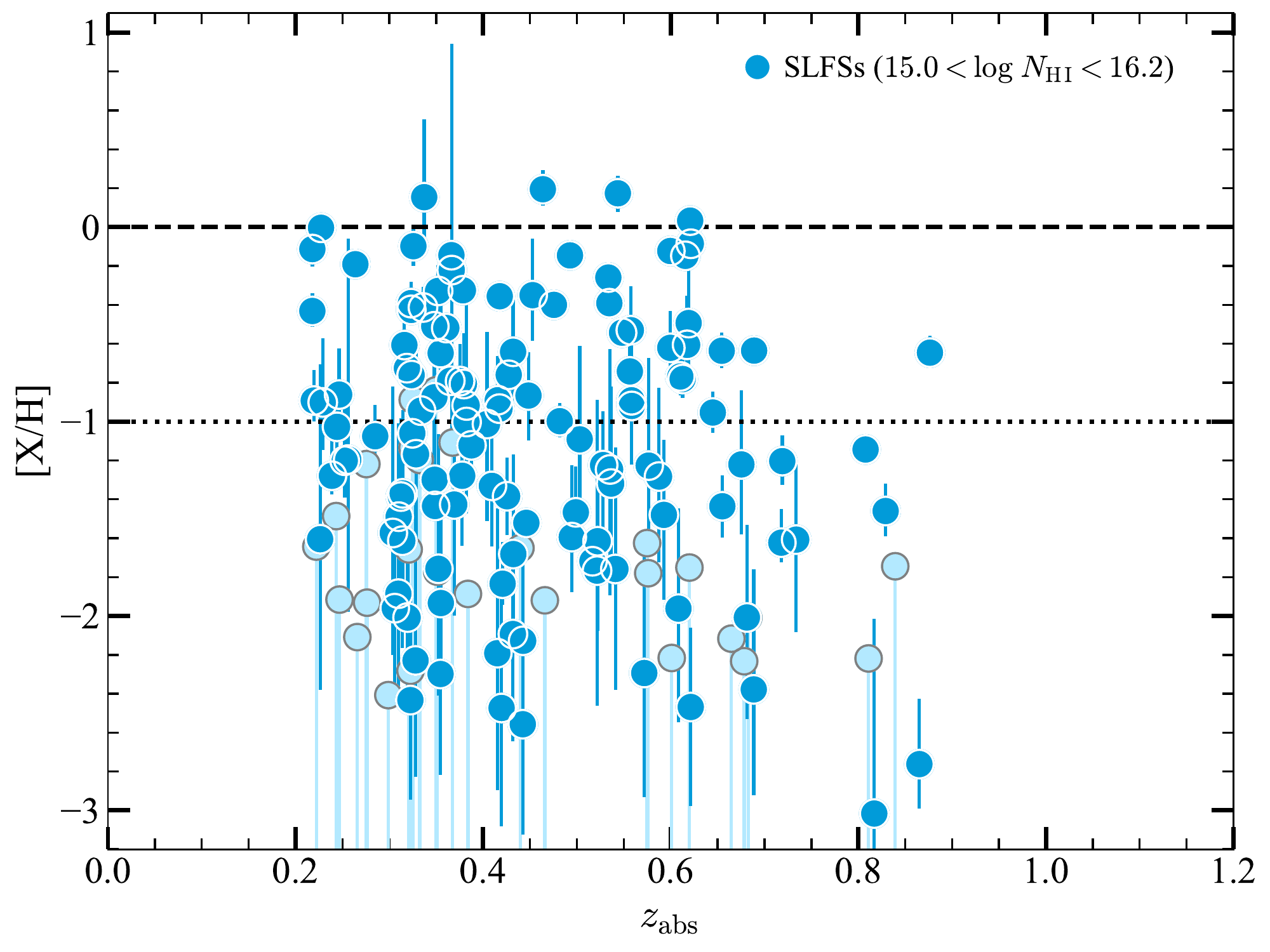}
\caption{Metallicities of the SLFSs as a function of the redshift of the absorbers. The median values of the metallicity posterior PDFs are adopted with 68\% CI, except for lower and upper limits where the lower and upper values represent the $10^{\rm th}$ and $90^{\rm th}$ percentiles, respectively, with the light-colored lines showing the 80\% CI. The dashed line represents solar metallicity. The dotted line represents 10\% solar metallicity, where there is a lack of pLLSs and LLSs at $0.45<z<1.2$ (see \citetalias{wotta19}).  \label{f-met-vs-z}}
\end{figure}

In Fig.~\ref{f-met-vs-z}, we show the metallicity of the SLFSs as a function of their redshift. We follow the same nomenclature for the error bars as in \S\ref{s-met-pdf}. There are two different clusters that are apparent from this figure. In the redshift range $0.2<z\la 0.65$, the metallicity has no clear trend with $z$. Within this redshift range, the metallicity is in the range $-2.6 \la \xh \la 0$. At the higher redshifts $0.65 \la z \la 0.8$, the sample is smaller with 19 absorbers. However, it is notable that there is a lack of  high metallicity absorbers with $\xh \ga -0.6$ with a metallicity range reduced to $-3 \la \xh \la -0.6$.  

A Kolmogorov-Smirnov (KS) test on the data sets at $z<0.65$ and $z\ge 0.65$ implies the metallicity in the two samples is not similarly distributed at better than 95\% CI ($p = 0.023$). Using a survival analysis where the censored data (upper limits) are included \citep{feigelson85,isobe86}, the mean metallicities at $z\ge 0.65$ is $\langle \xh \rangle = -1.86 \pm 0.19$ (error on the mean value from the survival analysis) compared to $\langle \xh\rangle =-1.22 \pm 0.07$ at $z<0.65$. This evolution with $z$ for the SLFSs contrasts with the pLLSs and LLSs where no such trend is observed over the same redshift intervals.

\subsubsection{Evolution of the Metallicity PDFs of the SLFSs and pLLSs/LLSs}
\label{s-met-red-pdf}

\begin{figure*}[tbp]
\epsscale{1}
\plotone{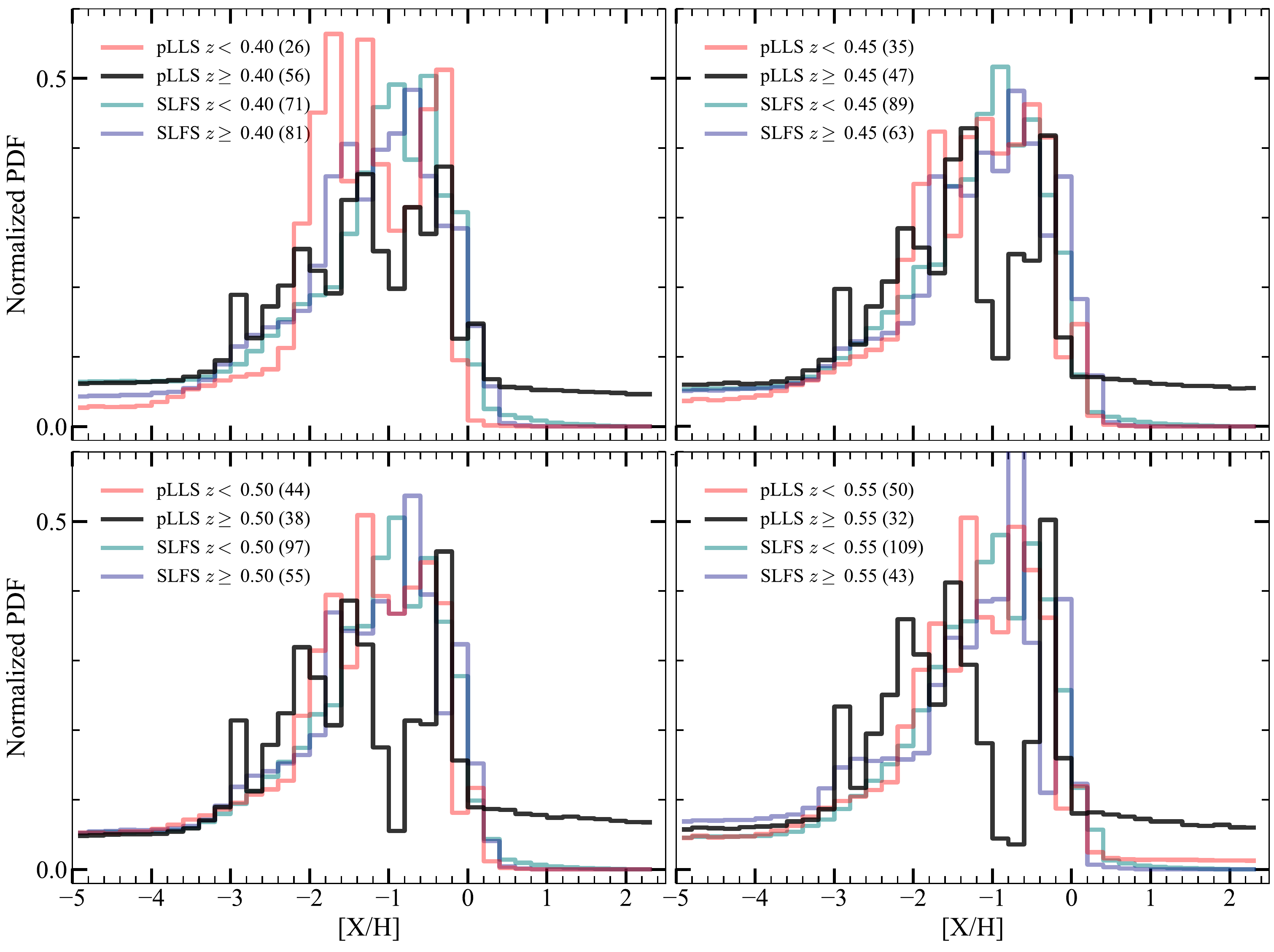}
\caption{Comparison of the metallicity PDFs  of the SLFSs and pLLSs above and below a given redshift threshold, $z_{\rm th} = 0.40$, 0.45, 0.50, 0.55. The numbers between parentheses indicate the number of absorbers in each redshift interval and absorber category. Each sample is about equally affected by upper limits, but only the pLLSs with $z \ge z_{\rm th} $ are affected by lower limits (owing to these absorbers coming from in part from the \citetalias{wotta16} sample where only \mgii\ could be used to estimate the metallicity in this low-resolution survey)---the darker regions showing the upper and lower limits in Fig.~\ref{f-comp-slfs-plls} are not shown here for clarity. 
\label{f-pdfall-vs-z}}
\end{figure*}

In \citetalias{wotta19}, we note that while there is a deficit in the PDFs of the pLLSs and LLSs near $\xh  \simeq -1$ at  $z\ge 0.45$, this is not the case at $0.2< z<0.45$. Of the lower-redshift pLLSs/LLSs, 8 of 44 (about 20\%) have median $\xh$ values in the interval defined by $\xh = -1.0 \pm 0.1$. When plotting the metallicity of the PDFs of the pLLSs, the PDF changes from a bimodal distribution at $0.45 \la z \la 1$ to a unimodal distribution at $z\la 0.45$ (see Fig.~15 in \citetalias{wotta19} and next). In \S\ref{s-met-pdf}, we noted that the metallicity PDFs of the SLFSs and pLLSs are quite different with the metallicity PDF of the SLFSs filling the dip of the metallicity PDF of the pLLS. However, this interpretation does not consider the evolution of metallicity PDF with $z$. 

In Fig.~\ref{f-pdfall-vs-z}, we show the comparison between the metallicity PDFs of the SLFSs and pLLSs above and below several redshift thresholds, $z_{\rm th} = 0.40$, 0.45, 0.50, 0.55. The numbers of absorbers in each redshift bin and absorber category are indicated in the figure (higher and lower values of $z_{\rm th}$ are not shown because the numbers of absorbers would start to be quite small, in the $<10$--20 range---however, we note that the  general trends discussed below are still observed down to $z_{\rm th}=0.30$ and up to $z_{\rm th}=0.65$). Considering first only the pLLSs, this figure reiterates the \citetalias{wotta19} finding that the bimodal nature of the metallicity PDF is redshift dependent and is prominent in the higher redshift bins.  In contrast, the metallicity PDFs of the SLFSs at $z<z_{\rm th}$ and  $z\ge z_{\rm th}$ are remarkably similar for all the threshold redshifts $z_{\rm th}$ shown in this figure; only at $z\ge 0.65$ would one notice an abrupt drop to zero in the metallicity PDF around $\xh \ga -0.6$ not observed for the pLLSs (a finding discussed in \S\ref{s-met-red-slfs} and observed in Fig.~\ref{f-met-vs-z}). Equally remarkable, the metallicity PDF of the pLLSs at $z<z_{\rm th}$ is almost identical to the metallicity PDF of the SLFSs at $z<z_{\rm th}$  or $z\ge z_{\rm th}$ for $ z_{\rm th} \simeq 0.45 \pm 0.10$, implying similar origins for the SLFSs and pLLSs. 

\begin{figure*}[tbp]
\epsscale{1}
\plotone{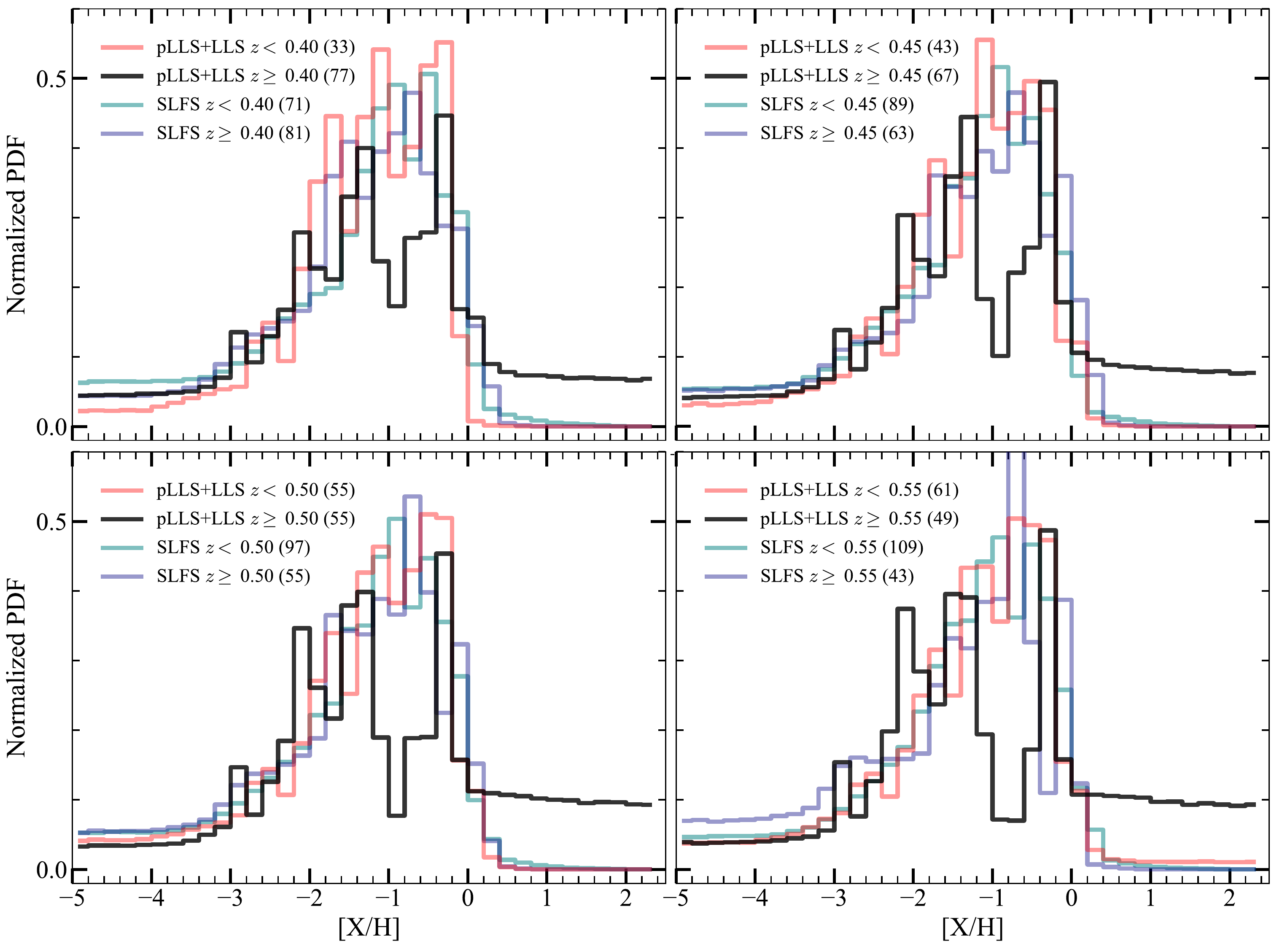}
\caption{Same as Fig.~\ref{f-pdfall-vs-z}, but we now combine the sample of pLLSs and LLSs to compare with the SLFSs. 
\label{f-pdfallv2-vs-z}}
\end{figure*}

In Fig.~\ref{f-pdfallv2-vs-z}, we produce a very similar figure to that Fig.~\ref{f-pdfall-vs-z} but the sample of higher column density absorbers now comprises both the pLLSs and LLSs. LLSs exhibit a similar gap in the metallicity distribution around $\xh \simeq -1$ than the pLLSs at least for LLSs with $17.2\le \mlnhi \la 18$ (see Fig.~\ref{f-met_vs_nh1}). The combined sample of pLLSs and LLSs has a size (110 absorbers) closer to the SLFS sample size (152). Fig.~\ref{f-pdfallv2-vs-z} is quite similar to Fig.~\ref{f-pdfall-vs-z} with better statistics in each redshift bin, implying the same conclusions apply to the combined sample of pLLSs and LLSs. Therefore LLSs+pLLSs, and SLFSs at $0.2 \la z \la 0.55 $ have very similar metallicity PDFs. For some reason(s), the metallicity PDF of the pLLSs or pLLSs+LLSs becomes bimodal at $z\ga 0.45$. With sample sizes in the 50--100 and samples of similar sizes in the low and high redshift intervals, it is very unlikely that this is a random statistical anomaly. Since we employed the same methodology to analyze all the absorbers, it is also unlikely to be a fluke owing to some systematic uncertainties in the ionization correction. Therefore, the redshift evolution in the shape of the metallicity PDFs of the pLLSs+LLSs appears to be real: the bimodal nature of the distribution is confined to the range $0.45 \la z \la 1$. At $z\sim 2.3$--3.5, the metallicity distributions of the pLLSs and LLSs are consistent with a unimodal distribution \citep{lehner16,fumagalli16}, although the sample at these redshifts is still quite small in the range $16.2 \la \mlnhi \la 17.5$. Our KODIAQ Z survey (\citealt{lehner16}) will provide a better characterization of these metallicity PDFs with a much larger sample. The redshift  interval between $z\sim 1$ and $2$ will remain mostly unexplored owing to the Lyman series being in the near-ultraviolet (although a sample large enough may eventually arise from the HST/ACS+WFC3 survey for LLSs that probe the redshift range $1.2\la z \la 2.5$, see \citealt{omeara11,omeara13}).

\begin{figure}[tbp]
\epsscale{1.2}
\plotone{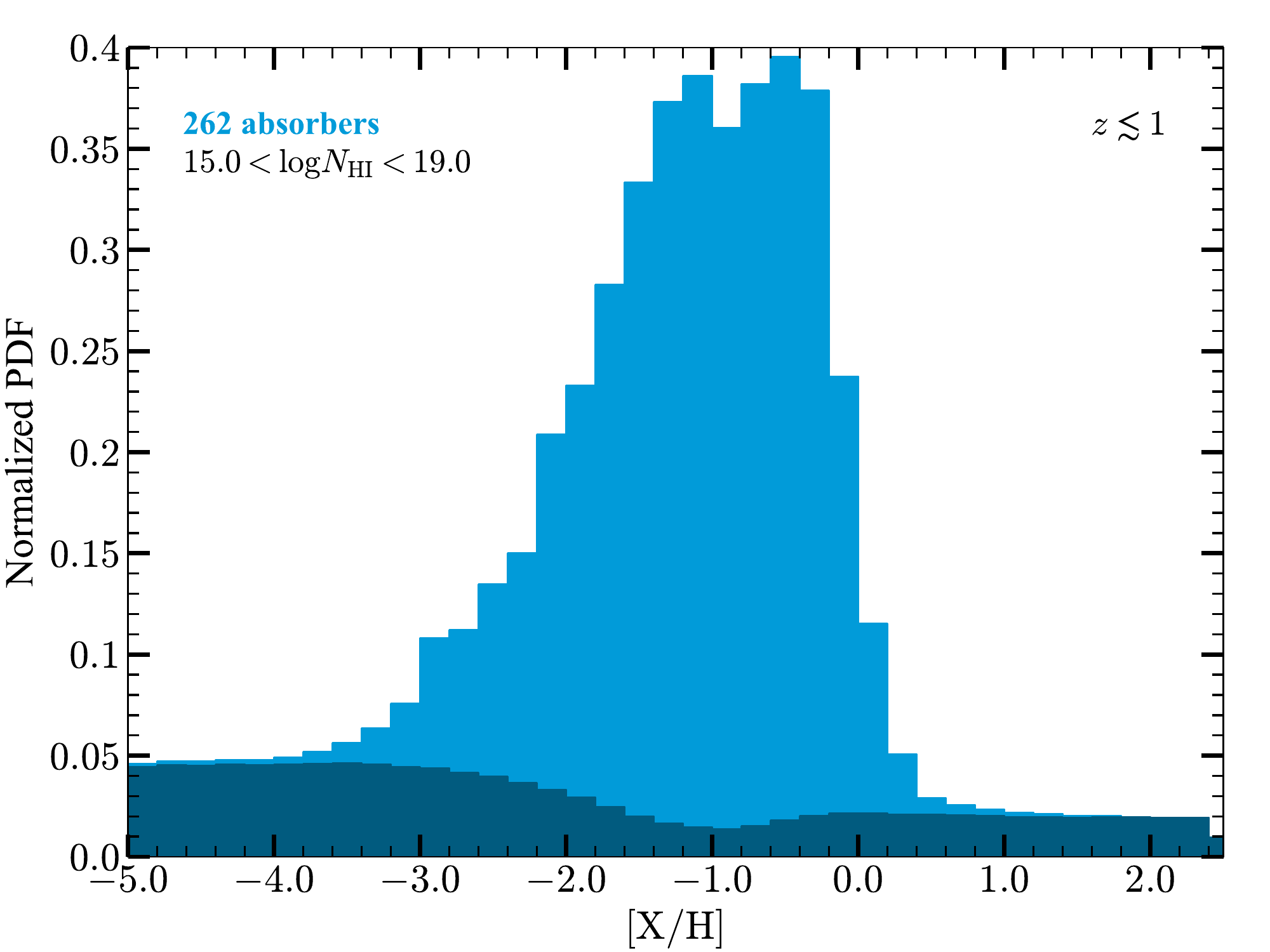}
\caption{Posterior metallicity PDF for the CCC absorbers with $15<\mlnhi <19$. The darker region shows the upper and lower limits. \label{f-mdf-all}}
\end{figure}

\subsection{Metallicity PDF of the CCC absorbers with $15<\mlnhi < 19$ at $z\la 1$}\label{s-met-pdf-all}

If one ignores the \hi\ column density dependence of the metallicity PDFs between the SLFSs, pLLSs, and LLSs, the resulting metallicity PDF of the absorbers with  $15<\mlnhi < 19$ at $z\la 1$ is shown in Fig.~\ref{f-mdf-all}. Not surprisingly, the metallicity PDF of the $15<\mlnhi < 19$ absorbers is not too different to that of the SLFSs since that population of absorbers dominates the sample (the mean and medians of the SLFSs and absorbers with $15<\mlnhi < 19$ are indeed quite consistent, see Table~\ref{t-stat}). The distribution is similarly skewed toward low metallicities, but in contrast to the SLFS metallicity PDF, it has a prominent flat-top in the entire range $-1.4\la \xh \  -0.2$. 

\begin{figure}[tbp]
\epsscale{1.2}
\plotone{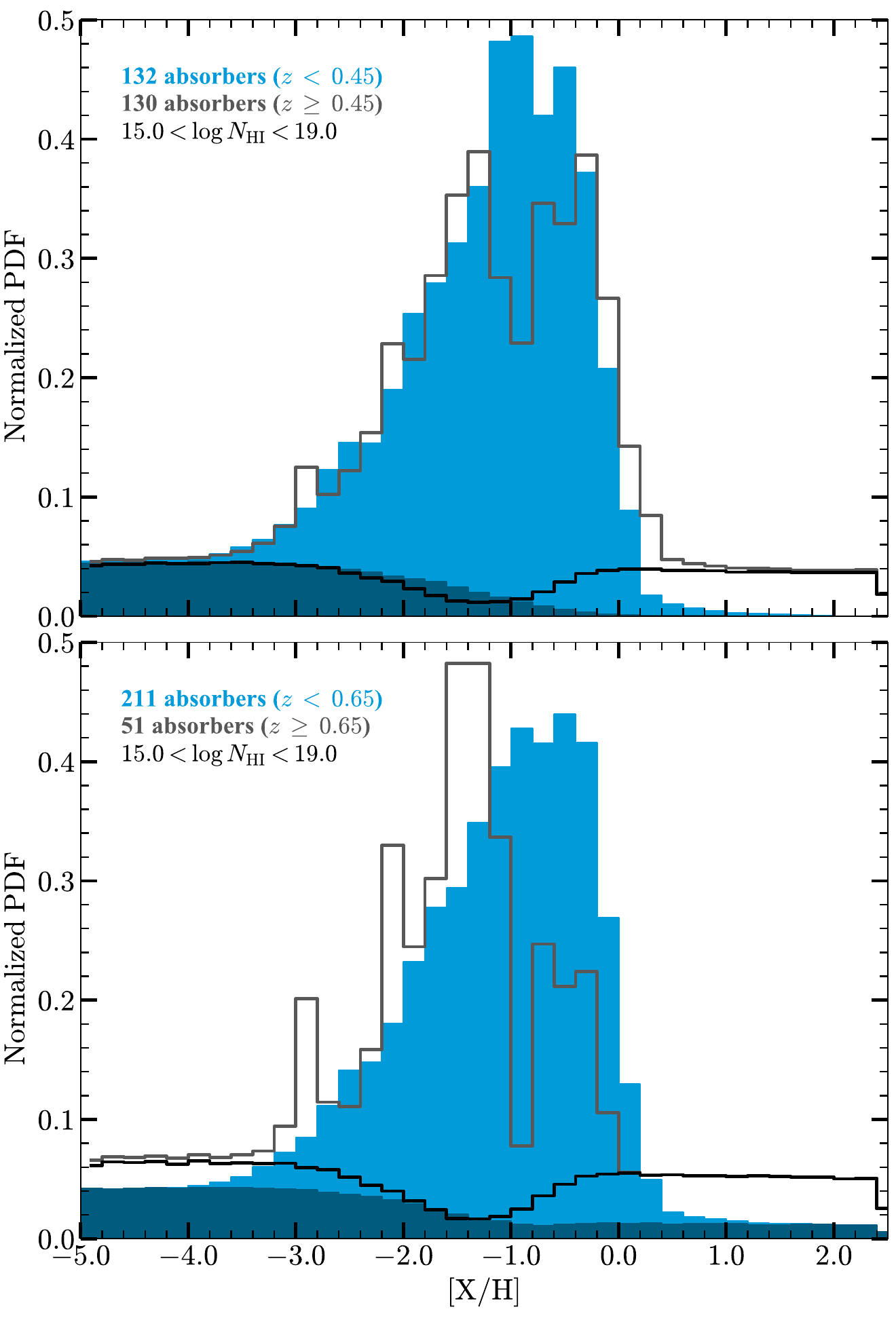}
\caption{Comparison of the posterior metallicity PDFs for the CCC absorbers with $15<\mlnhi <19$ in different redshift intervals. The shaded regions or lower black histograms indicate the contributions from  upper and lower limits. 
\label{f-pdf-vs-z}}
\end{figure}

In Fig.~\ref{f-pdf-vs-z}, we show the metallicity PDFs of the entire sample at $z<z_{\rm th}$  or $z\ge z_{\rm th}$ with $z_{\rm th} =0.45$ or 0.65. We select these two values for $z_{\rm th}$ in view of the redshift evolution seen in the pLLSs and SLFSs, respectively, that is discussed above. For $z_{\rm th} =0.45$, there are about the same numbers of absorbers above and below $z_{\rm th}$. The two distributions are not too dissimilar, but at $z_{\rm th} \ge 0.45$, the imprint of the bimodal nature of the metallicity PDFs of the pLLSs can be observed with two peaks at $\xh \simeq -1.4$ and $-0.4$ and a dip at $\xh =-1$.  For $z_{\rm th} =0.65$, the number of absorbers is a factor 4 times larger in the low redshift bin than in the high redshift interval. Nevertheless with 51 absorbers at $z\ge z_{\rm th}$, the sample is still statistically large enough to compare the two samples. In that case, the two metallicity PDFs are quite different. At $z<0.65$, the metallicity PDF is similar to the one at all $z$ (albeit with a smaller flat-top metallicity range with $-1\la \xh \la -0.2 $). However, at $z\ge 0.65$, the redshift evolution from both the SLFSs and pLLSs/LLSs is clearly imprinted with a strong peak at $\xh \simeq -1.5$ and a strong deficit of absorbers at $\xh >-1$. 

\subsection{The \ca\ ratio as a Function of the Metallicity}\label{s-ca-met}
To this point we have only considered the absolute metallicity distributions. From our photoionization modeling of the absorbers, we have allowed the \ca\ ratio to vary. With our systematic methodology to estimate \ca, we can therefore revisit earlier studies of the \ca\ evolution with the metallicity (\citetalias{lehner13}; \citealt{lehner16}). As discussed in \citetalias{lehner13}, the \ca\ ratio is sensitive to nucleosynthesis effects since there is a time-lag between the production of $\alpha$-elements and carbon (see, e.g., \citealt{cescutti09,mattsson10} for more detail). Fig.~10 in \citet{lehner16} shows that about half of the pLLSs and LLSs at both high ($2<z\la 3.5$) and low $z$ ($z\la 1$) follow the general trend between \ca\ and \ah\ seen in stars \citep[e.g.,][]{akerman04,fabbian09} and SLLSs/DLAs \citep[e.g.,][]{pettini08,penprase10,cooke11a}. That trend is a decrease of \ca\ from about a solar value (up to $+0.2$ dex) near solar or super-solar metallicity to $\ca \simeq -0.6$ at $\xh \simeq -0.5$ where \ca\ plateaus around that metallicity value up to a metallicity of $\xh \simeq -2$ where an upturn is observed in \ca\ with an increase of \ca\ to solar and super-solar value ($\ca \ga +0.2$) at $\xh \simeq -2.8$ (see Fig.~10 in \citealt{lehner16} and references therein for a full discussion of the possible origin(s) of this overall trend). However, as noted by \citetalias{lehner13} and \citet{lehner16}, the other half of the pLLSs and LLSs have \ca\ values clustered around a solar value at any metallicity, i.e.,  \ca\ is 2--5 times larger than observed in Galactic metal-poor stars or high-redshift DLAs at similar metallicities. In \citet{lehner16}, we argue this could be caused by preferential ejection of carbon from metal-poor galaxies into their surroundings.

\begin{figure}[tbp]
\epsscale{1.2}
\plotone{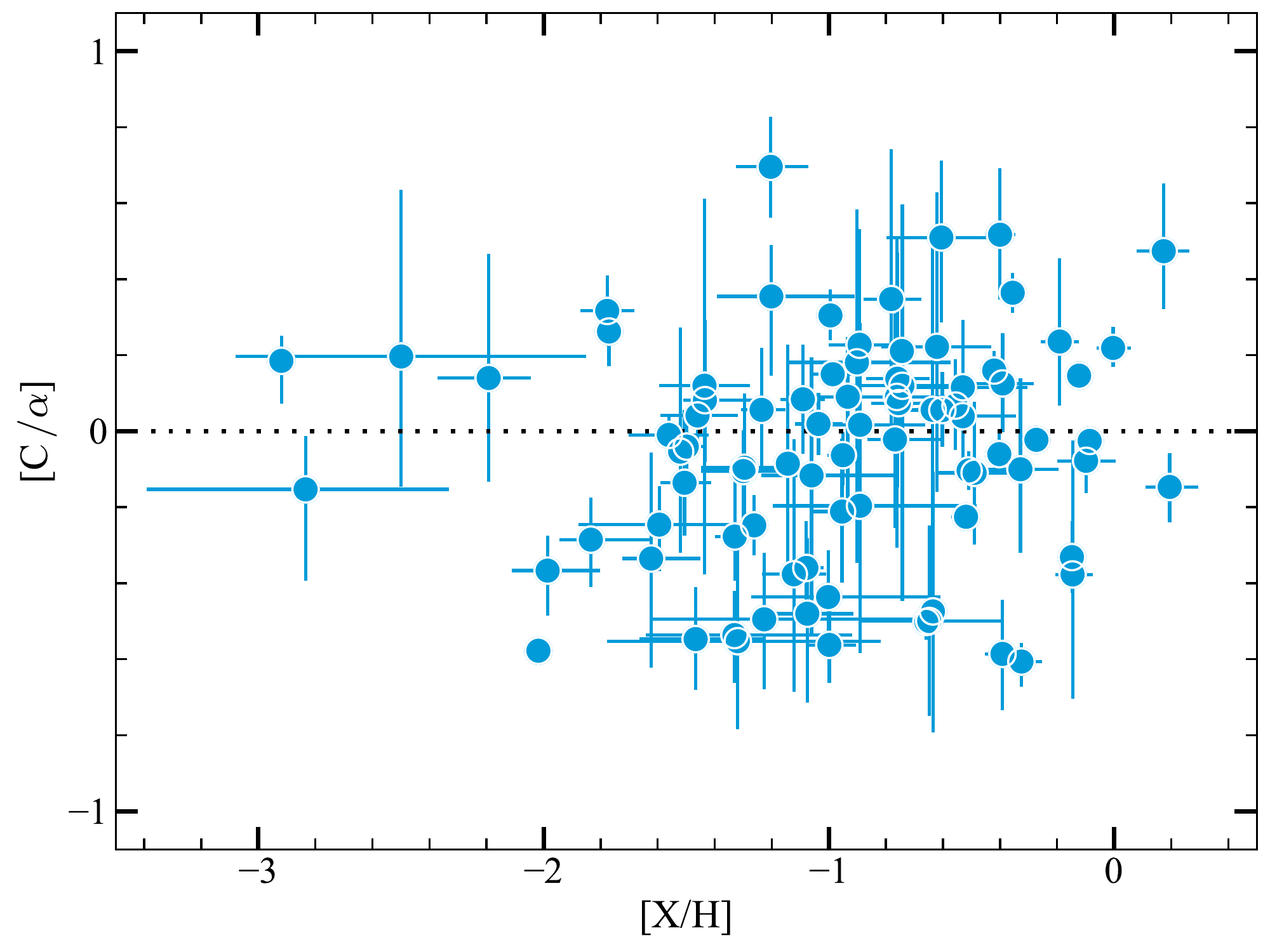}
\caption{The  \ca\ ratio against the metallicity.  The median values of the \ca\ and \xh\ posterior PDFs are adopted with 68\% CI. 
\label{f-calpha-met}}
\end{figure}

 In Fig.~\ref{f-calpha-met}, we show \ca\ as a function of \xh\ for all the absorbers in CCC where the relative abundance and metallicity can simultaneously be constrained. Not too unexpectedly, based on the results discussed in \S\ref{s-met-det} and \citetalias{wotta19} (where we show that the \ca\ distribution  is well fit by a Gaussian distribution with a mean and standard deviation of $\langle \ca\ \rangle = -0.05\pm0.35$), there is no obvious trend observed between \ca\ and the metallicity of the gas at least for metallicities in the range $-1.8 \la \xh \la 0$ where 91\% of the absorbers reside. Since the \ca\ ratio depends on the ions \cii\ and \ciii\ compared to other low and intermediate ions,  the ionization correction could plausibly muddle any trend between \ca\ and \xh. However, while we noted that $U$ moderately increases as \nhi\ decreases (and hence the ionization corrections are larger in the SLFS regime than in the pLLS/LLS regime), there is no significant trend in \ca\ with the neutral/ionization fraction of the absorbers (see Fig.~\ref{f-calpha-nhi}). We also tested the results by using only absorbers with good measurements of \cii\ and \ciii\ (i.e., excluding lower limits or upper limits on the column densities of \ciii\ or \cii), and  we still find the same scatter of \ca\ relative to \nhi, $f_{\rm H\,I}$, or \xh. Therefore the lack of trend of \ca\ with \xh\ appears not to be affected by the large ionization corrections, and hence differs from the observed trends in stars \citep[e.g.,][]{akerman04,fabbian09} or stronger \hi\ absorbers  \citep[SLLSs, DLAs; e.g.,][]{pettini08,penprase10,cooke11a}. 

Nevertheless, there are three worthy features to highlight from Fig.~\ref{f-calpha-met}: 1) the observed lower floor level of the distribution is  $\ca \simeq -0.6$,  similar to that observed for the stars, \hii\ regions, and SLLSs/DLAs (see Fig.~10 in \citealt{lehner16}); 2) the upper ceiling of the distribution is  $\ca \simeq +0.5$, but with a concentration of data around $0< \ca \sim +0.3$, not too dissimilar from the highest values observed in  stars, \hii\ regions, and SLLSs/DLAs; and 3)  there is {\it tentatively}\ an upturn in \ca\ at $\xh \la -2$, which is similar to the one observed in stars and SLLSs/DLAs (see Fig.~10 in \citealt{lehner16}). While the latter is a tentative result for now, our ongoing KODIAQ survey will help better determine (or rule out) this trend for the absorbers  at $z>2$--3 since the metallicity PDFs of the pLLSs and LLSs peak at $\xh \simeq -2$ \citep{lehner16,fumagalli16}. The first two points are the direct result from the fact that the \ca\ distribution is consistent with a Gaussian distribution. Nevertheless, it is striking that despite \ca\ having a flat prior $-1\le \ca \la 1$, the observed range is consistent with that observed in other objects in the universe. This suggests that at least some of the CGM gas probed by these  absorbers follows the same nucleosynthesis processes as those observed in denser environments and stars.

\section{Physical Properties And Cosmic Budgets of the Cool CGM Absorbers at $\lowercase{z}\la 1$ } \label{s-phys-prop}

\begin{figure*}[t]
\epsscale{1.}
\plotone{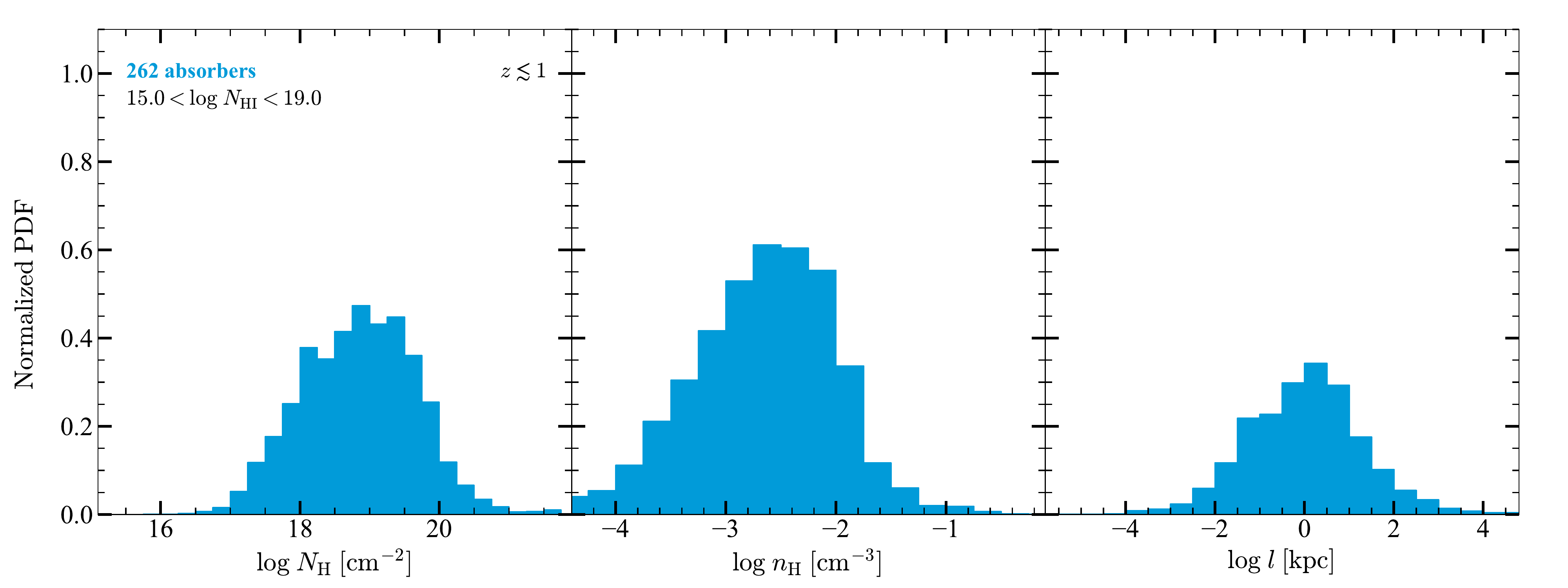}
\caption{Posterior  PDFs of the total H column density ({\it left}), H density ({\it middle}), and path-length ({\it right}) for all the combined absorbers (SLFS+pLLS+LLS). \label{f-pdf-phys-all}}
\end{figure*}
Our ionization models predict physical properties of the absorbers in addition to providing the information needed to derive the metallicities. These include the neutral fraction of the gas ($f_{\rm HI}$), the density of the gas (\nnh), the total  H column density ($\mnh  \equiv \mnhi + \mnhii$), the length-scale of the cloud ($l \equiv \mnnh/\mnh$), and the temperature of the gas ($T$). By considering the behavior of the neutral fraction with column density, we can also deduce the cosmological baryon and metal budgets for these absorbers. As the metallicities,  these physical properties depend directly on the EUVB radiation field. However, as we show in \S\ref{s-met-det-add-const}, \nh\ and \nnh\ (and therefore $l$ and $f_{\rm HI}$) are far more sensitive to the assumptions behind ionization corrections than the metallicity (see \S\ref{s-met-det-add-const}), and are therefore less robust (see also \citealt{fumagalli16}). Comparing 113 CCC absorbers analyzed both with the HM12 and HM05 EUVBs, we find that the mean differences between HM12 and HM05 for \nh, \nnh, and $l$ lead to an average shift of $-0.3$, $-0.2$, and $-0.1$ dex in these values, respectively, i.e., the systematic uncertainty resulting from a change of the EUVB is somewhat smaller than for the metallicity. With these caveats in mind, we now proceed to describe the results for these physical parameters. 

\subsection{Physical Properties of the SLFSs, pLLSs, and LLSs}\label{s-phys}
For each absorber in CCC, we summarize the physical quantities (\nnh, \nh, $l$, and $T$) derived from our modeling in Table~\ref{t-physical_properties} (in the $\mlnnh$ column, a colon next to the median value indicates a Gaussian prior on $\log U$ used for that absorber; see \S\ref{s-met-det}). In Table~\ref{t-phys-stat}, we summarize the statistical properties of the physical properties derived for the entire sample (median, mean, and dispersion of \nhi, \nnh, \nh, $l$, and $T$) and a restricted sample where only a flat prior on $\log U$ was used. Comparing the results from Table~\ref{t-phys-stat} between the entire and restricted samples, it is apparent that the results are not statistically different. The only exception is for the LLS category, but the reason for that is the average \hi\ column densities in the entire and restricted samples change by 0.3 dex. Therefore, below we consider the entire CCC sample, i.e., absorbers that were modeled using the flat and Gaussian priors on $\log U$. 

\begin{figure*}[tbp]
\epsscale{0.9}
\plotone{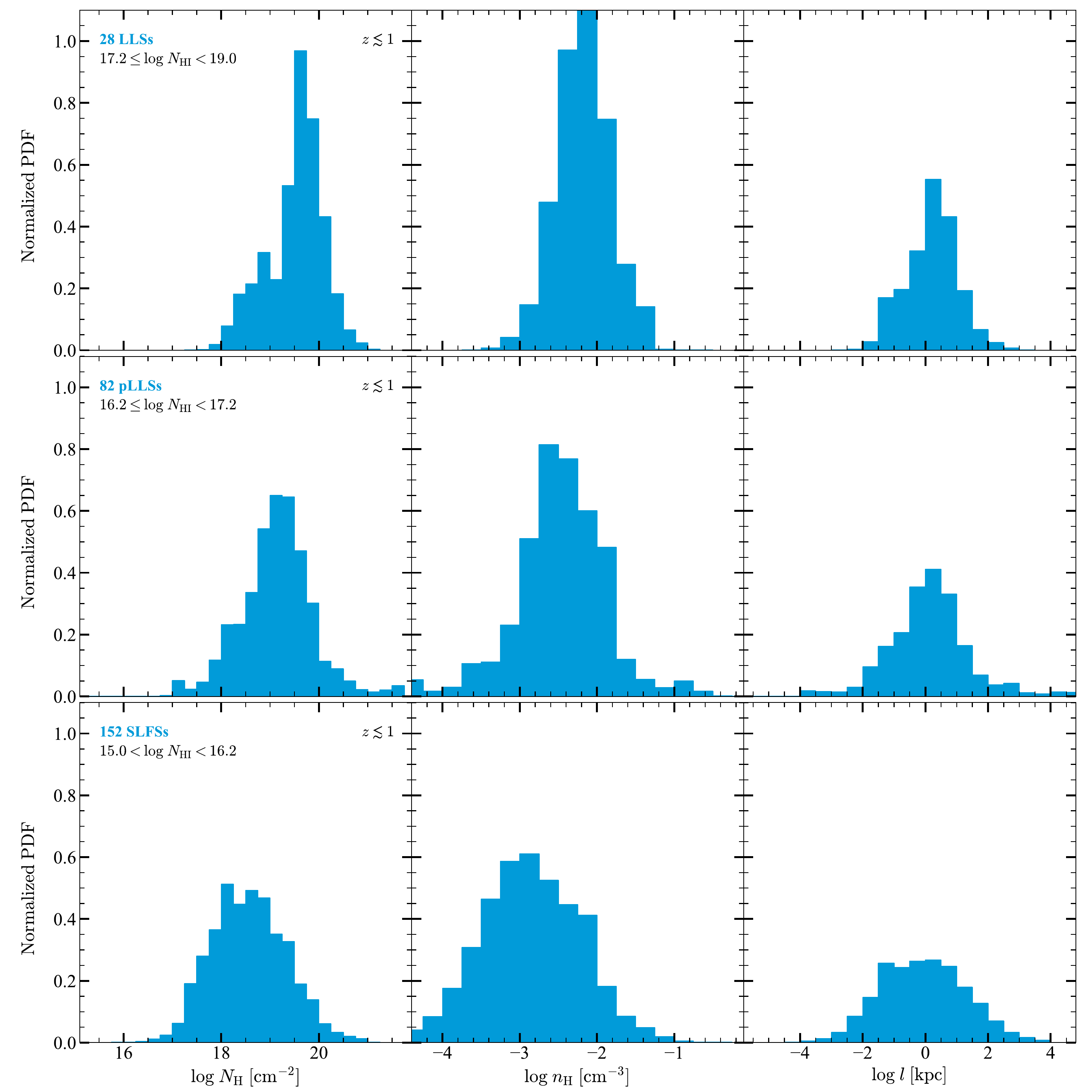}
\caption{Posterior  PDFs of the total H column density ({\it left}), H density ({\it middle}), and path-length ({\it right}) for LLSs  ({\it top}), pLLS  ({\it middle}), and SLFSs  ({\it bottom}). \label{f-pdf-phys}}
\end{figure*}

In Figs.~\ref{f-pdf-phys-all} and \ref{f-pdf-phys}, we show the posterior PDFs of \nh, and \nnh, and $l$ for all the absorbers and the individual categories of absorbers in the CCC sample, respectively. These PDFs contrast remarkably from the metallicity PDFs as there is no evidence of prominent multiple-peak or strongly skewed distribution for \nh, \nnh, or $l$ (as well as the temperature, which we do not show in these figures, but see Table~\ref{t-phys-stat}). For all the absorbers, the temperatures are scattered around $10^4$ K, which is expected for photoionized gas; it is worth noting that this is consistent with the upper limits on the temperature derived from the $b$-values  of the individual components of \hi\ (see \citetalias{lehner18} and also appendix in \citetalias{lehner13}). 

The total hydrogen densities, column densities, and path-lengths decrease with decreasing \nhi, but their distributions largely overlap. This is expected from the neutral fraction relationship with \nhi. The differences between the mean \lnhi\ for SLFSs compared with the pLLSs and LLSs are 1.0 and 1.9 dex (a factor $\sim$10 and 80), respectively. However, the differences in total H are significantly smaller, with differences $\Delta \mlnh \approx 0.7$ and 1.0 dex (a factor $\sim$5 and 10). Thus, despite their lower \nhi\ values, the SLFSs still (moderately) contribute  to the total mass and baryon budgets compared with the pLLSs and LLSs especially since they are more numerous by a factor 5 to 10 relative to the pLLSs and LLSs (e.g., \citealt{lehner07,danforth16,shull17}, and see \S\ref{s-budget}).

Using the entire sample and according to the MCMC Cloudy modeling, the interquartile range of length-scale of the absorbers is $0.3 \la l \la 3.2$ kpc.  Therefore the linear size of these absorbers is consistent with occupying the CGM of galaxies at $z\la 1$. The relatively small scales of these absorbers also signal the need for very high-resolution simulations to resolve these absorbers in cosmological/zoom-in simulations. 

\begin{figure}[tbp]
\epsscale{1.2}
\plotone{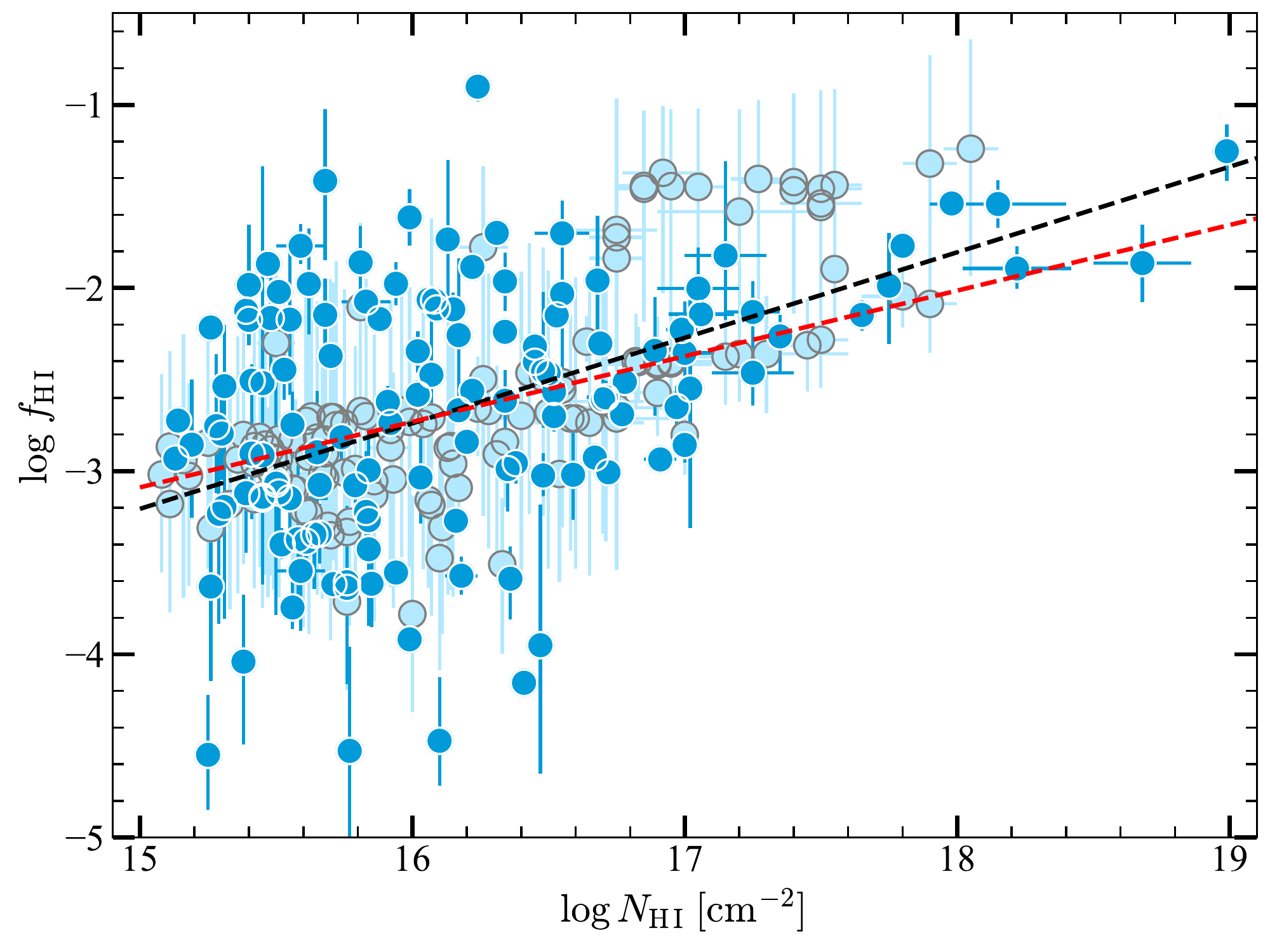}
\caption{The neutral fraction of \hi\ versus \nhi. The median values of the posterior PDFs are adopted with 68\% CI. Light blue data were estimated using a Gaussian prior on $\log U$, while the other data have a flat prior on $\log U$. The black dash-line is a linear fit to the entire set of data, while the red dash-line is a fit to the data with only a flat prior on $\log U$. 
\label{f-fnhi-nhi}}
\end{figure}

Finally, in Fig.~\ref{f-fnhi-nhi}, we show the neutral fraction of the absorbers as a function of \nhi. With  $-3.5 \la \log f_{\rm H\,I} \la -1$, the gas is always at least 90\% ionized, and most of the sample is at least 99\% ionized. Fig.~\ref{f-fnhi-nhi} shows a positive correlation between $f_{\rm H\,I}$ and \nhi, which is confirmed by the Spearman rank-order that shows a moderate positive monotonic correlation between $f_{\rm H\,I}$ and \nhi\ with a correlation coefficient $r_{\rm S} = 0.4$--0.5 and a $p$-value\,$ \ll 0.05\%$. We produce a linear fit between  $\log f_{\rm H\,I}$ and $\log \mnhi$, which is shown by the black dash-line as a fit to the entire set of data ($\log f_{\rm H\,I} = (0.47 \pm 0.05) \mlnhi -10.21 $) and the red dash-line, a fit to the data with only a flat prior on $U$ ($\log f_{\rm H\,I} = (0.36 \pm 0.08) \mlnhi -8.46$). Within $1\sigma$, these two fits are similar and we adopt hereafter an average between these two fits: $\log f_{\rm H\,I} = 0.41 \mlnhi -9.3$. There is, however, a large scatter in $f_{\rm H\,I}$ especially at $\mlnhi \la 17.2$ (i.e., for absorbers with an optical depth at the Lyman limit $\tau_{\rm LL} <1$): at $\mlnhi < 17.2$, the logarithmic value of the neutral fraction varies from $-1.5$ to $<-4$ dex, while for the LLSs ($\mlnhi \ge 17.2$), it is in the range $-2.2 \la \log f_{\rm H\,I} \la -1.2$. This fit is significantly shallower than the result reported at high-redshift ($2.5 \la z \la 3.5$) by \citet{fumagalli16}, who find $\log f_{\rm H\,I} \propto 0.99 \mlnhi$, although this may in part be due to a significant contribution from higher column density absorbers with $\mlnhi \ge 19$.

\subsection{Cosmic Budgets}\label{s-budget}

Although the baryon budget in the CGM of low redshift is not thought to be a major source of baryons of the universe ($5\% \pm 3\%$ according to the accounting by \citealt{shull12}), using the results from CCC, we can re-assess the impact of the SLFSs, pLLSs, and LLSs on the cosmological baryon and metal budgets.

\subsubsection{Cosmological Baryon Budget}\label{s-budget-baryon}
 Following the standard methodology to estimate the  mean gas density relative to the critical density, $\Omega_{\rm g}$ \citep[e.g.,][]{tytler87,lehner07,omeara07,prochaska10}, we can write:
\begin{equation}
\Omega_{\rm g} =  \frac{\mu_{\rm H} m_{\rm H} H_0}{\rho_c\, c } \int \frac{\mnhi}{f_{\rm H\,I}} f(\mnhi) d \mnhi\,,  \\ 
\end{equation}
where $m_{\rm H} = 1.673\times 10^{-24}$ g is the atomic mass of hydrogen, $\mu_{\rm H} = 1.3 $ corrects for the presence of helium, $c = 2.9979\times 10^{10}$ cm\,s$^{-1}$ is the speed of light, $\rho_c = 3 H^2_0/(8\pi G) = 8.62 \times 10^{-30}$ g\,cm$^{-3}$ is  the current critical density, $H_0 = 2.20\times 10^{-18}$ s$^{-1}$ is the Hubble constant,\footnote{We adopt here the Planck 2015 results (\citealt{planck16}).} and $f(N_{\rm H I})$ the column density distribution function \citep[e.g.,][]{tytler87,lehner07,omeara07,prochaska10,ribaudo11,shull17}. The column density distribution function can be approximated by a power-law as $f(N_{\rm H I}) = C_0 \mnhi^{-\beta}$ with $C_0 \simeq 4\times 10^7$ and $\beta =1.48$, adopting the results from \citet{shull17} who have recently determined the  column density distribution function for the absorbers that include the SLFSs, pLLSs, and LLSs at similar redshifts probed by CCC. The neutral fraction is modeled with a linear fit $\log f_{\rm H\,I} = 0.41 \mlnhi -9.3 $ (see \S\ref{s-phys}). Integrating the above equation over the different \nhi\ range of the SLFSs, pLLSs, and LLSs, we find that $\Omega_{\rm g} \simeq (1.8, 2.3,6.0)\times 10^{-4}$ for the SLFSs, pLLSs, and LLSs, respectively. For the entire range $15.1\le \mlnhi \le 19$,  $\Omega_{\rm g} = 1.0\times 10^{-3}$ or $\Omega_{\rm g}/\Omega_{\rm b} = 0.021$ where $\Omega_{\rm b} = 0.0486$ is  the ratio of the total baryon density to the critical density, i.e., the gas of the SLFSs, pLLSs, and LLSs contributes to only about 2.1\% of the cosmic baryons at $z\la 1$. 

While these absorbers are a very small part of the overall baryon budget of the universe, only $\sim 5\%$ of the baryons are thought to be in the CGM of galaxies \citep{shull12}. Thus, the combined SLFSs, pLLSs, and LLSs account for about 40\% of the circumgalactic baryons. In the COS-Dwarfs and COS-Halos surveys, a baryon accounting was undertaken, but in this case the absorbers were selected to be within the CGM of sub-$L^*$ and $L^*$ galaxies at $z\la 0.2$. Using the minimal and more conservative  values from these surveys, they find  a similar fraction  (20\%--50\%) for the low-ionization gas of the CGM (see Fig.~8 in \citealt{tumlinson17}; and see \citealt{werk14,bordoloi14}). The remaining baryons in the CGM are thought to be predominantly in hotter gas \citep[e.g.,][]{shull12,werk14,prochaska17,tumlinson17,bregman18}.

\subsubsection{Cosmological Metal Budget}\label{s-budget-metal}

The metal-mass density of the absorbers can be simply estimated using the baryon density via $\Omega_{\rm m} = \Omega_{\rm g} Z$ where $ Z =10^{\xh} Z_\sun $ is the metallicity of the gas in mass units and $Z_\sun = 0.0142$ is  the solar metallicity in mass units from \citet{asplund09}. Using the mean metallicity from Table~\ref{t-stat} (using the entire sample), we find $\Omega_{\rm m} \simeq (0.09, 0.13,2.6)\times 10^{-6}$ for the SLFSs, pLLSs, and LLSs, respectively. These values can increase or decrease by a factor 1.4, 1.8, and 2.3, respectively, based on a 90\% CI of the population mean of the metallicity. The LLS metal-mass density dominates the metal budget because the mean metallicities and mean \nh\  are substantially larger  than those of the SLFSs and pLLSs. At  at $z\simeq 2.5$--3.5, the LLSs with $\mlnhi <19$ yield $\Omega_{\rm m} \simeq 5.1 \times 10^{-7}$  \citep{fumagalli16}. Therefore, $\Omega_{\rm m}$ has increased by a factor 5 from $z\simeq 2.5$--3.5 to $z\la 1$.

\citet{peeples14} estimated that star-forming galaxies of stellar mass $10^{8.5}$--$10^{11.5}$ M$_\sun$ have produced metals at a cosmic density $\Omega_{\rm m} = 5.9 \times 10^{-5}$, with about 80\% of these metals no longer in galaxies, i.e, $\Omega^{\rm lost}_{\rm m} = 4.6 \times 10^{-5}$. From the entire sample, the cool CGM gas probed by SLFSs, pLLSs, and LLSs only contributes to approximately 4\% of $\Omega^{\rm lost}_{\rm m}$. About 40\% of the SLFSs+pLLSs+LLSs have $\xh >-1$ with an average metallicity around $-0.4$ dex above this threshold (see Fig.~\ref{f-met_vs_nh1}). These high metallicity absorbers alone contribute to about 83\% of the  4\% of $\Omega^{\rm lost}_{\rm m}$. As for the baryonic budget, the minimal values of the cool CGM found in the COS-Dwarfs and COS-Halos surveys is similarly low (see Fig.~9 in \citealt{tumlinson17}; and see \citealt{peeples14}). Our summary of the metal mass density of SLFSs, pLLSs, and LLSs accounts only for the cool photoionized phase of these absorbers. A full description of the metal content would require an accounting of the higher ionization/hotter material, higher \hi\ column density gas, and dust \citep{menard10,menard12,peeples14,tumlinson17}.

\section{Implications and Discussion}\label{s-disc}

We have studied a sample of 262 \hi-selected absorbers with $15.1\la \mlnhi <19$ at $z\la 1$ (most of the CCC absorbers being at $0.2 \le z \le 0.9$) and in this paper we have presented, for the first time, a sample of 152 SLFSs with $15<\mlnhi<16.2$ at  $0.2 \la z \la 0.9$. Each absorber in our sample has been uniformly analyzed to determine the column densities of \hi\ and metal ions \citepalias{lehner18} and  using the same EUVB (HM05) and the same Bayesian MCMC analysis technique to determine ionization correction in order to derive the metallicity, which reduce significantly any possible systematic uncertainty when comparing the properties of these absorbers across several orders of magnitude in \nhi. We also show that the metallicity estimates are less sensitive to the assumptions behind ionization corrections than other physical parameters such as the hydrogen density, length-scale, or total column density of hydrogen, making the metallicity a unique parameter to probe the origins and chemical enrichment of the gas associated with $15.1\la \mlnhi <19$ absorbers. Below we discuss some of the implications of our findings and compare them with recent cosmological simulations. 

\subsection{Plenty of Primitive Gas  at $z\la 1$, but No Evidence for Pristine Gas}\label{s-disc-pristine}

A striking result from CCC is that very low metallicity gas is common in the CGM of $0.2 \la z \la 1$ galaxies: 40\%--50\% of the SLFSs, pLLSs, and LLSs with $\mlnhi \la 18 $ have metallicities $\xh < -1.4$, metallicities rarely observed in higher \hi\ column density absorbers over the same redshift interval. About 13\%--21\% of the CCC absorbers with $15 < \mlnhi <19$ at $0.1 \la z\la 1$ have even metallicities $\xh \la -2$. A metallicity $\xh \simeq -2$ is the peak of the metallicity distribution of the pLLSs and LLSs at $2.3\la z\la 3.5$ \citep{lehner16,fumagalli16}, implying that gas in the overdense regions around galaxies has had little or no additional chemical enrichment over $\sim$6 Gyr (see also discussion in \citetalias{wotta19}). There is even  0.5\%--$2.8\%$ of extremely low metallicity absorbers with $\xh \simeq -3$ in CCC, which even at  $2.3\la z\la 3.5$ would be in the lower tail of the metallicity PDF. Since the most likely origin of this very metal-poor gas is the IGM, it also implies that a large fraction  of the IGM at $0.2\la z \la 1$ has remained largely unpolluted by the successive star-formation episodes in galaxies over several billions of years. 

While very-metal poor gas probed by absorbers with $15<\mlnhi<19$ is common at $z<1$, the evidence for pristine absorber (absorbers with no detectable metal absorption in high-quality data) is lacking. \citet{fumagalli11b} reported the detection of two pristine LLSs at $z\sim 3$ deriving metallicities $\xh < -4.2 $ and $<-3.8$. Recently, \citet{robert19} reported another pristine LLS at $z\simeq 4.4$ with $\xh <-4.1$ in a dedicated search of extremely low metallicity gas at high redshift. The KODIAQ Z survey of the pLLSs and LLSs at $2.3 \la z \la 3.3$ \citep{lehner16} sets a strong upper limit on the frequency of such pristine absorbers: the fraction of pLLSs/LLSs with $\xh \la -4$ is only 3\%. This implies that pristine gas at $2.3 < z < 3.3$ is rare, but exists. 

In CCC, there is not a single absorber that can be defined as pristine since all the absorbers have some detection of  metal absorption when strong transitions are covered by the observations. This includes the ``pristine absorber'' reported in \citet{zahedy19} and \citet{chen19} (J135726.26+043541.3 at $z = 0.328637$ in CCC) since we detect unambiguous \mgii\ and \feii\ absorption features (see \citetalias{lehner18} and also new high-resolution Keck observations presented in \citealt{berg19}). We derive a metallicity $\xh = -2.58 \pm 0.09$ for this absorber (but see \citealt{berg19} for a detailed analysis and discussion), a value consistent with the upper limits reported in \citet{zahedy19} and \citet{chen19}. Even though this is low, this is nevertheless a factor $>20$--60 higher than the metallicities of the pristine LLSs at $z\sim 3$. It is also no the lowest metallicities in CCC. The three lowest metallicity absorbers in CCC are 1 SLFS with $\xh = -2.76 \,^{+0.34}_{-0.23}$ (J044011.90−524818.0 at $z=0.865140$) and 2 pLLSs with $\xh =-2.92 \pm 0.05$ (J152424.58+095829.7 at $z=0.728885$) and $-2.83 \pm 0.50$ (J055224.49−640210.7 at $z=0.345149$) (see also Fig.~\ref{f-met_vs_nh1}).\footnote{Another SLFS at $z=0.817046$ toward -J140923.90+261820.9 has $\xh \sim -3$, but it has a large 1 dex uncertainty.} These 3 extremely low metallicity absorbers have all detected absorption in at least one metal ion (e.g., \oiv, \ciii, and/or \mgii). 

With the entire CCC sample of 262 absorbers, a 90\% CI on the fraction of absorbers with no metal is $<1\%$, implying that truly-pristine gas at $z\la 1$ is scant. We, however, emphasize that with the typical SNRs of the COS and ground-based QSO spectra, it would be difficult to derive metallicities at $z<1$ as low as those limits reported at high $z$, especially for absorbers with $\mlnhi <17$. Nevertheless the presence of metal-line absorption in CCC absorbers precludes such low metallicities anyway.

\subsection{Comparison with COS-Halos: a Redshift Evolution at $z\la 0.2$--$0.3$?}\label{s-disc-cos-halos}
The selection of the absorbers in the COS-Halos survey is galaxy-centric. The absorbers are selected in a homogeneous sample of sightlines piercing the CGM of field, $\sim L^*$ (the full range being $0.3<L/L^*<2$) galaxies at $z \sim 0.2$ with projected distances $<160$ kpc \citep{tumlinson11,tumlinson13,werk13,werk14}. This selection contrasts from the absorber selection in CCC where the absorbers are selected to have strong enough \nhi\ (specifically $15.1 \le \mlnhi \le 19$) to most likely probe the CGM/environment of galaxies at $z<1$. This conclusion is based on the following findings from surveys of galaxies-absorbers. First, when surveys on the galaxy environments of these absorbers have been undertaken, nearly all the time at least one galaxy has been found within about 30--200 kpc at the redshift of the absorbers  \citep[e.g.,][]{lehner17,lehner13,lehner09,kacprzak12,cooksey08}. Second, studies of the  galaxy--absorber two-point cross-correlation function have shown significant clustering between galaxies and absorbers with $\mlnhi \ga 15$ while a weak or absent clustering signal for weaker \nhi\ absorbers \citep[e.g.,][]{lanzetta95,chen00,chen01a,bowen02,prochaska11c,tejos14}. With different selection criteria between CCC and COS-Halos, a comparison between these two surveys may yield further insight on the origins of the absorbers in CCC.

As noted in \citetalias{lehner18}, although several absorbers in CCC are also found in spectra that were originally obtained for the COS-Halos survey, the vast majority of these are not the absorbers that were initially targeted to probe  the CGM of the COS-Halos galaxies. There are in fact only 9 absorbers in common between CCC and COS-Halos surveys (see \citealt{lehner18} for the underlying reasons).\footnote{Among the reasons are  both the redshift cutoff in CCC---typically $z>0.2$, i.e., higher redshifts than most of the COS-Halos absorbers--- and the use of only G130M and G160M observations to derive \nhi---subsequent COS G140L spectra were obtained by the COS-Halos team to add some constraints on \nhi\ from the break at the Lyman limit for several absorbers (see \citealt{prochaska17}).} Our identification, analysis of the absorbers, including velocity component definition and estimation of the metallicities, and data coaddition are fully independent from the COS-Halos analysis. As we show in \citetalias{wotta19}, the choice of the HM12 over the HM05 EUVB \citep{haardt96,haardt12} can lead to a difference in the metallicity on average of 0.4 dex for the SLFSs and pLLSs and 0.2 dex for the LLSs. Thus, it is critical to adopt the same EUVB to model the ionization in order to derive metallicities when comparing different samples.

Here we adopt the \citet{prochaska17} results that recently re-estimated \nhi\ of several absorbers for COS-Halos absorbers and estimated the metallicities using the same MCMC Bayesian approach than we did in CCC. We use these new \nhi\ results, but we re-derive the COS-Halos metallicities using the same the HM05 radiation field and similar ions. In Table~\ref{t-compcos}, we show the comparison of \nhi\ and metallicities between CCC and COS-Halos \citep{prochaska17} for the 9 absorbers in common. Overall, there is a reasonable agreement between the two independent analyses; for  3 absorbers with quite different \nhi, we explain these differences in the footnote of this table.

This overall agreement differs from the results of \citet{prochaska17}, who found that find a difference in the metallicity PDFs compared to that of the pLLSs/LLSs in \citepalias{wotta16} at the 95\% level. This is largely explained by the use of the HM12 EUVB in \citet{prochaska17} and the use of the HM05 EUVB by \citetalias{wotta16}. However, there may also be a difference caused by a  redshift evolution since the COS-Halos absorbers are mostly at $z\la 0.2$, while the absorbers in the \citetalias{wotta16} survey are mostly at higher redshift ($\langle z \rangle \simeq 0.6$). Although in \citetalias{wotta19} we argue that for the pLLSs and LLSs there is no strong evolution in the mean metallicity with $z$ at $z\la 1$, we show in \S\ref{s-met-red} there is some redshift dependence for the SLFSs with a higher fraction of metal-poor absorbers at $z \ga 0.65$ than at lower $z$. Therefore, to alleviate any potential redshift evolution, we also restrict the CCC absorbers to be at $z \la 0.28$ (a redshift range chosen to best match the COS-Halos redshift range with the exact same number of absorbers, see below). 

\begin{figure}[tbp]
\epsscale{1.2}
\plotone{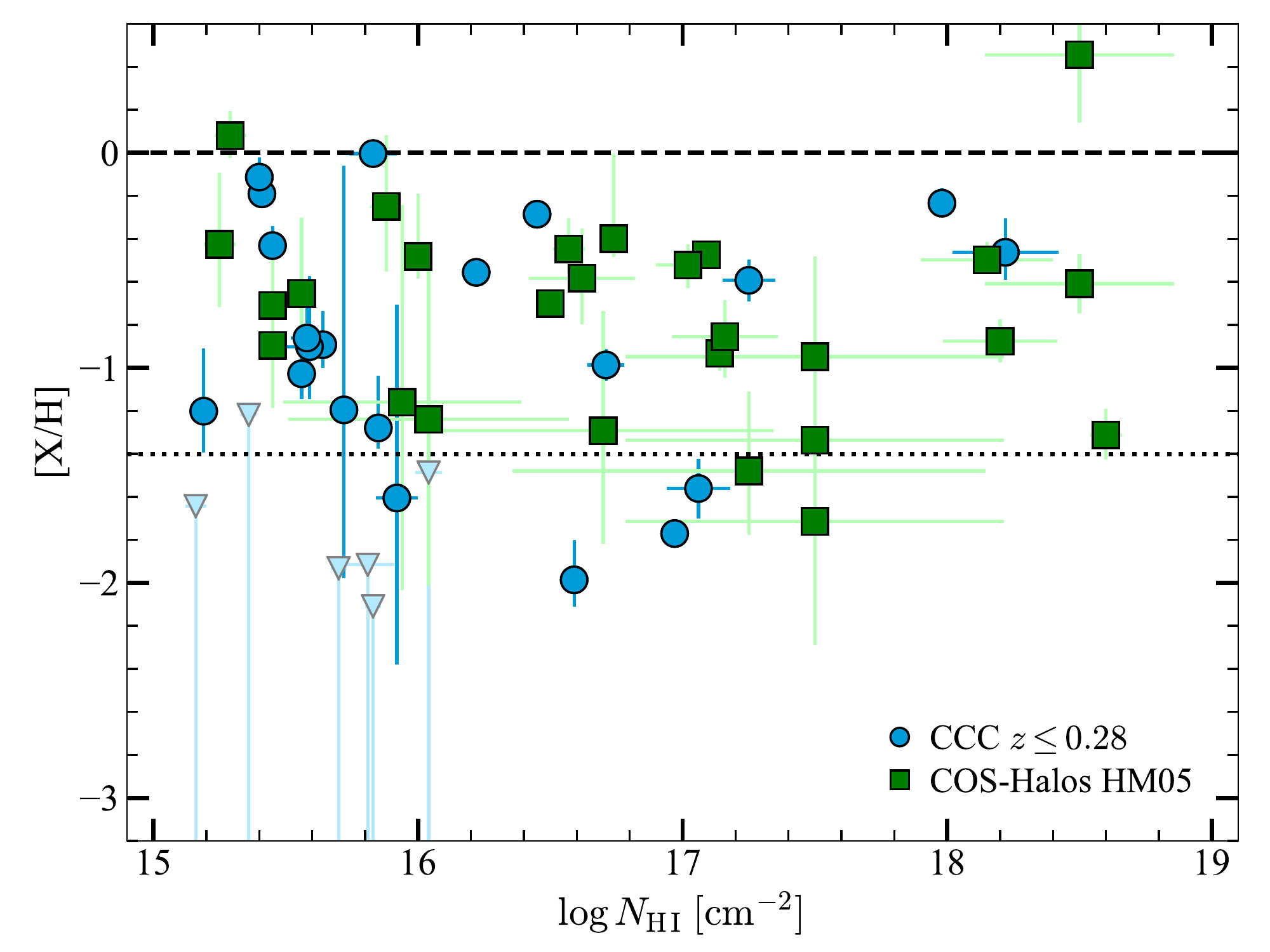}
\caption{Metallicities of the absorbers as a function of the \nhi\ for the CCC and COS-halos absorbers. For CCC, we only show absorbers at $z\le 0.28$, which match better the redshift distribution of  COS-Halos absorbers. For COS-Halos, we used the column densities from \citet{prochaska17}, but we re-estimated all the metallicities using the HM05 EUVB (see text for more detail). \label{f-cos-halos}}
\end{figure}

In Fig.~\ref{f-cos-halos}, we show the comparison of the metallicities of the CCC absorbers at $z\le 0.28$ and COS-Halos against \nhi. Using the $z$ cutoff on CCC reduces the sample to 27 absorbers, which is the same size than the COS-Halos sample where the metallicity has been estimated.\footnote{The COS-Halos size  is different from the sample of 32 absorbers in \citet{prochaska17} on account of 2 and 3 absorbers having $\mlnhi <15$ and $>19$, respectively.} The first  result from Fig.~\ref{f-cos-halos} is that the two surveys are not quite as different as originally shown in \citet{prochaska17}, owing in part to the use of the same HM05 EUVB in our comparison. There is still an apparent lack of very low metallicity absorbers ($\xh \le -1.4$) in the COS-Halos sample, especially for absorbers in the \hi\ column density range $15\la \mlnhi \la 17$: only $<16$\% (90 \% CI) of the COS-Halos absorbers in this range have  $\xh \le -1.4$ versus 22\%--51\% of the CCC absorbers. Although the mean metallicity of this CCC sample is $-1.1$ dex compared to $-0.7$ for COS-Halos, a two-sided KS-test implies that the null hypothesis that the samples are drawn from the same distribution cannot be rejected at a strong level of statistical significance  ($p=0.224$). The second result is that the metallicity--\nhi\ plot for the COS-Halos (and CCC) absorbers is a scatter-plot, which differs from \citet{prochaska17} where they find an anti-correlation between the metallicity and \nhi\ for absorbers with $15\la \mlnhi \la 18$ when the ionization models use the HM12 EUVB (see their Fig.~8): this anti-correlation appears to be purely driven by the use of the HM12 EUVB.  

If instead we compare the entire CCC sample and COS-Halos, a two-sided KS-test implies that the null hypothesis that the samples are drawn from the same distribution can be rejected at a clear level of statistical significance with $p=0.005$. This suggests that both the redshift evolution and use of a different EUVB can explain the differences noted by \citet{prochaska17}. When choosing the same EUVB and a comparable redshift interval, the metallicities in an \hi-selection and galaxy-centric selection of the absorbers with similar \nhi\ more closely overlap. While only marginally different, we still note a lack of very metal-poor absorbers in COS-Halos relative to CCC, possibly pointing to a different origin for these absorbers (maybe gas associated with less massive galaxies than probed COS-Halos galaxies) or a strong redshift evolution that occurs at $z\la 0.2$ since 85\% (23/27) of the CCC absorbers in this limited sample are at $0.2<z\le 0.28$ (see also below). For the absorbers with overlapping metallicities, the CCC absorbers may have the same origin as the COS-Halos absorbers. 

As just alluded to, this comparison reveals that there might be an evolutionary  change around $z\simeq 0.2$ with a smaller frequency of metal-poor absorbers. In \citetalias{wotta19}, we already noted that only 1 out of 4 pLLSs/LLSs at $z<0.2$ has $\xh <-1$; if we consider the $z\la 0.25$ range, the sample of absorbers increases to 7 with still only 1 absorber with $\xh <-1$. Obviously, this could be the result of small number statistics (a 90\% CI allows for a fraction ranging from 3\%--45\% for a sample size of 7 and a hit rate of 1/7). However, combined with the COS-Halos results, the implication is that while the metallicity distribution of the pLLSs/LLSs did not change much over the redshift range $\sim$0.9 to $\sim$0.2 (i.e., over about 5 Gyr), it seems to evolve at $z \la 0.2$ (over the last $\sim3.5$ Gyr), with metal-poor gas much rarer in the CGM of galaxies at $z\la 0.2$.

We finally note that two independent surveys of the CGM about luminous red galaxies (LRGs) at $z\sim 0.4$ also reveal a mixture of low- and high-metallicity CGM absorbers \citep{berg19,zahedy19}. Thus, even the most massive galaxies have low and high metallicities in their CGM. These works do not yet find a strong difference in the prevalence of high and low metallicity in these galaxies compared with lower-mass galaxies, but the small number of systems in the present sample precludes to make any strong conclusion.

\subsection{Inhomogeneous Metal Mixing around $0.2 \la z\la 1$ Galaxies}\label{s-disc-var}
For the absorbers with $16.2<\mlnhi < 19$ in CCC, metal-enriched and metal-poor absorbers are observed at $0.2 \la z \la 1$. As noted in the previous section, there is some evidence of a turnover around $z\la 0.2$ with a smaller frequency of metal-poor pLLSs and LLSs, implying that the metals in the halos of $z\la 0.2$ galaxies may be more well mixed and overall be more metal-enriched than at $0.2 \la z \la 1$. The metallicity range of the pLLSs and LLSs at $0.2 \la z \la 1$ is $-3 \la \xh \le 0.0$, a factor of $1,000$ variation between the highest and lowest metallicities. A similar metallicity range is found for the SLFS, but only at $0.2<z\la 0.65$ (see \S\ref{s-met-red-slfs}). At $z\ga 0.65$, the metallicity range of the SLFSs is reduced to $-3 \la \xh \le -0.6$, still a factor 250 variation, but there is a lack of more metal-enriched gas with $\xh >-0.6$. Although there is some redshift evolution in the metallicity within $z\la 1$, the large metallicity variation  at any $z$ between $z\simeq 0.2 $ and $z\simeq 1$ is remarkable.

As discussed above  (see also  \citetalias{lehner13,wotta19}; \citealt{lehner17}), several pieces of evidence indicate that SLFSs, pLLSs, and LLSs most likely probe the CGM of galaxies or small group of galaxies at $z\la 1$.\footnote{Preliminary analysis of integral field unit observations indicate that in several cases---but not always---$\ga 2$--3 galaxies can be found at the redshift of the pLLSs or LLSs, see \citealt{lehner17} and references therein, and M.~Berg et al.~(in preparation).} With the preliminary analysis of the galaxies in the field of these absorbers, there is no evidence yet that different mass galaxies probe predominantly low or high metallicity absorbers. In fact, both metal-enriched and very metal-poor absorbers have now been discovered in the CGM of the most massive galaxies at $z\sim 0.4$ \citep{zahedy19,berg19}. Therefore, the large variation of metallicities of the SLFSs, pLLSs, and LLSs implies seems so far unrelated to the types of galaxies that the absorbers are associated with. Instead the large scatter in the metallicities most likely arise from the wide range of physical processes that these absorbers probe, including metal-poor accretion, metal-enriched recycled inflows, or metal-rich outflows. The large metallicity scatter also implies that the  metals in the CGM or intra-cluster medium of small group of galaxies at  $0.2 \la z \la 1$ are poorly mixed. 

The large variation in chemical abundance in the surrounding of galaxies at $z\la 1$ is not only apparent from the large difference of metallicities in the SLFSs, pLLSs, and LLSs along different sightlines, but is also commonly observed in closely redshift-separated absorbers along the same sightlines. As noted in \S\ref{s-paired} (see also Fig.~\ref{f-met-dv}), there are 27 paired-absorbers in CCC for which we were able to derive the metallicity for absorbers that are separated in velocity by $<300$ \km. For the majority of these paired-absorbers (75\%--96\%), the metallicity varies  significantly between a factor 2 to 25 over a velocity difference $\Delta v \la 300$ \km\ (for the majority of these, the velocity separation of the paired absorbers is even smaller with $\Delta v \la 150$ \km, see Fig.~\ref{f-met-dv} and Table~\ref{t-close}) or  a redshift difference of $\Delta z \la 10^{-3}$ ($\Delta z \la 5\times 10^{-4}$ for the majority of the paired absorbers). Although we use a search window of $\Delta v = 500$ \km\ to identify paired absorbers, most of the paired absorbers are separated by 50--150 \km\ rather being randomly distributed between 50 and 500 \km, strongly suggesting that these absorbers are not randomly distributed and may instead probe the same galaxy environment along each sightline (which may be within a single halo or the halo of a galaxy group). For $50 \la \Delta v \la 150$ \km, the gas does not escape galaxy halos with $M_ {*}\ga 10^{9.5}$ M$_{\sun}$. Therefore, these paired absorbers show that in most cases the gas associated with a single galaxy halo or at least a galaxy-scale structure (e.g., small galaxy groups) have large chemical inhomogeneity. This is also observed in the CGM of LRGs at $z\sim 0.4$ where \citet{zahedy19} found metallicity variation by a factor 2--10 in the velocity components of about 7 out of 10 LRGs when this experiment could be done.  Individual studies of single galaxy halos have found similar metallicity variations of higher redshift galaxies at $z\sim 0.9$ \citep{tripp11,rosenwasser18} and $z\sim 2.4$--3.5 \citep{prochter10,crighton13a}. 

Thus, within a single galaxy or group mass-scale halo, the CGM gas is frequently chemically inhomogeneous where both metal-poor and metal-rich gas can be observed at least at $0.2 \la z\la 1$. This is consistent with the overall picture we first advanced in \citet{ribaudo11} and \citetalias{lehner13} that  metal-poor and metal-rich gas probed different physical processes (e.g., metal-poor accreting gas versus metal-rich ejected large-scale galaxy winds or metal-rich recycling gas). The gas is not well-mixed over scales of 30--250 kpc corresponding to the typical projected distances of these absorbers to the associated galaxies \citep{lehner17}. Since such inhomogeneity appears to be common,  high-resolution spectra and detailed ionization modeling of each component are critical to accurately determine the distribution of metals in the CGM of galaxies at any $z$.
 
\subsection{Comparison with Cosmological Simulations}\label{s-disc-comp}
\begin{figure*}[tbp]
\epsscale{1.}
\plotone{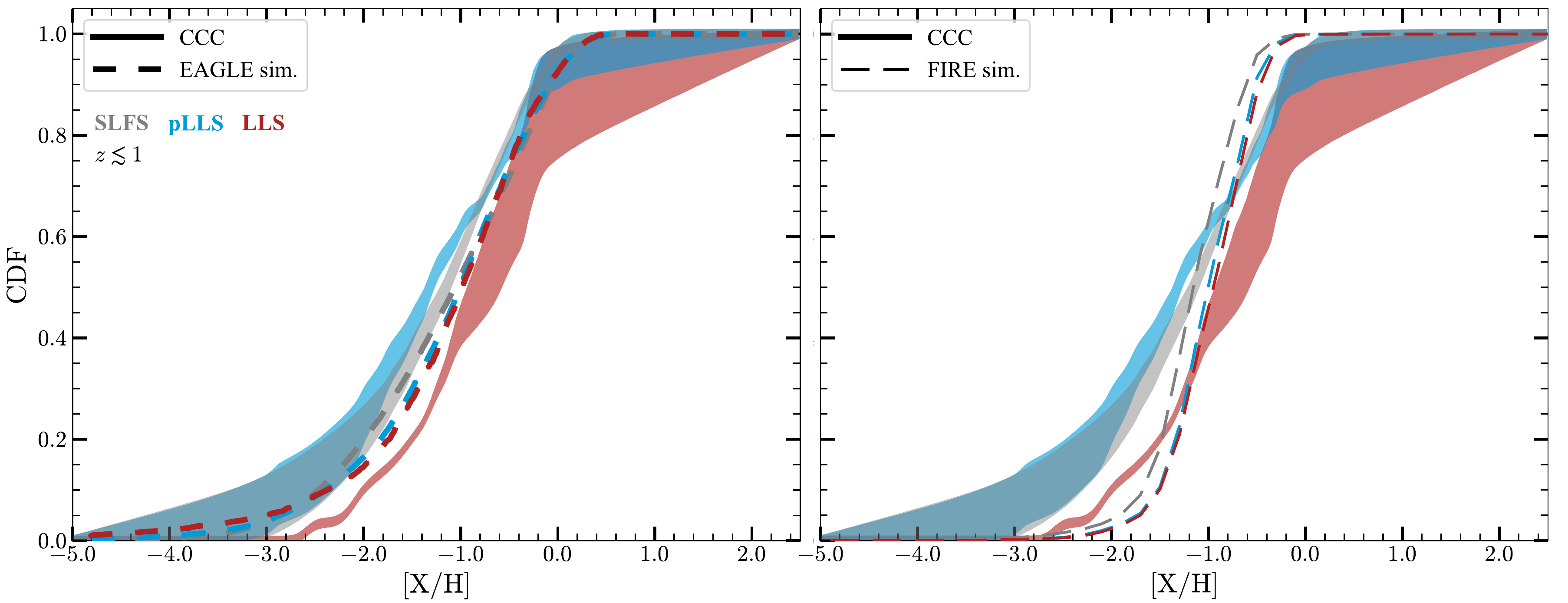}
\caption{Comparison of the  CDFs of the SLFS, pLLS, and  LLS metallicity PDFs as a function of metallicity between CCC and EAGLE ({\it left}---this paper and see \citealt{rahmati18}) and FIRE ({\it right}---from \citealt{hafen17} and Z. Hafen 2018, private communication) simulations. 
\label{f-cdf-comp}}
\end{figure*}

Following \citetalias{wotta19}, we compare our results to cosmological zoom results from the Feedback In Realistic Environments (FIRE) simulations \citep{hafen17} and the EAGLE simulations \citep{rahmati18}. These simulations have reproduced successfully a variety of galaxy observables \citep[e.g.,][]{hopkins18,schaye15} and are the first ones to explore quantitatively the metallicity distributions of the CGM of galaxies at $z<1$. We refer the reader to the original papers and  \citetalias{wotta19} for more details on these simulations. Briefly, the zoom FIRE simulations of \citet{hafen17}  survey the CGM of 14 simulated galaxies at $z \la 1$. The selection of the pLLSs and LLSs in the CGM of these galaxies follows the  column density distribution function.   

The EAGLE simulations \citep{schaye15,crain15} are publicly available in the data release of \citet{mcalpine16}. Here we use the EAGLE \emph{Recal-L025N0752} high-resolution (HiRes) volume, which was initially used by \citet{rahmati18} to determine the metallicity distribution of the pLLSs at and LLSs.  For \citetalias{wotta19} and the present work, we select \hi\ absorbers from this 25-Mpc comoving volume for comparison against the sample of CCC, choosing sightlines directly matching the \hi\ column densities and redshifts for each absorber in CCC.  Both sets of simulations use the same method to calculate \nhi\ and $[{\rm X/H}]$, where both \hi\ density and \hi-weighted metallicity are summed onto a two-dimensional grid.  While the gridding technique is not the same method used on the observed data where spectra are used, this technique has been proven to reproduce reasonably well the column densities derived from mock spectra through simulations \citep{altay11}. It also must be noted that the underlying \hi-weighted metallicity in simulations is not homogeneous, and that a subset of gas particles can preferentially contain much of the very metal-rich neutral gas.  

In Fig.~\ref{f-cdf-comp}, we show the CDF for the EAGLE ({\it left}\ panel) and FIRE ({\it right}\ panel) simulations and the CCC observations for the SLFSs, pLLSs, and LLSs (see \citetalias{wotta19} for the SLLSs and DLAs). As noted in  \citetalias{wotta19}, the comparison of our results for the pLLSs and LLSs with those from the FIRE simulations shows that these simulations severely under-predict the amount of low-metallicity gas probed by these absorbers. This also applies for the SLFSs. However, while the metallicity CDFs of the pLLSs, LLSs, SLLSs, and DLAs are very similar in the FIRE simulations (i.e., there is essentially no evolution of the metallicity with \nhi\, see \citetalias{wotta19}), the SLFS metallicity CDF departs from that of the pLLSs and LLSs, with on average a lower metallicity by about 0.2 dex, showing some mild metallicity evolution between the SLFSs and the stronger \hi\ absorbers.  

The EAGLE simulations produce a much broader range of metallicities in the SLFSs, pLLSs, and LLSs than FIRE. The EAGLE simulations overall compare better with the CCC results, especially at low metallicity. Although we do not show it here, there is, however, a much stronger redshift evolution  of the metallicities of absorbers between $z\sim 0$ and 1 in the EAGLE simulations than observed in CCC (see Fig.~22 in \citetalias{wotta19} and also \citealt{rahmati18}). The observations of the metallicity CDFs of the SLFSs, pLLSs, and LLSs are also  nearly identical, still showing no metallicity change with \nhi\ even the lower \nhi\ end.  

Not surprisingly, similar issues as those described in \citetalias{wotta19} between the models and the observations are found. In both simulations, strong feedback could cause an overabundance of metal-enriched gas. Furthermore, the limited resolution of these zoom simulations may also be an issue, especially since they do not fully resolve the cool CGM probed by the SLFSs, pLLSs, and LLSs. As we discuss and show in \S\ref{s-phys-prop}, the linear scale of the SLFSs, pLLSs, and LLSs from our models at $z\la 1$ ranges from sub-kpc to several kpc scales (see also Figs.~\ref{f-pdf-phys-all} and \ref{f-pdf-phys}).  Considering the entire CCC sample, the mean and median linear scales are about 1 kpc with $1\sigma$ dispersion ranging from about 100 pc to 10 kpc (see Table~\ref{t-physical_properties}). Furthermore as discussed in \S\S\ref{s-disc-cos-halos}, \ref{s-disc-var}, the CGM of galaxies at $z\la 1$ is highly inhomogeneous, which requires high resolution to model it in simulations. 

Several very high-resolution zoom simulations have been recently published \citep{vandevoort18,peeples19,corlies19,suresh19,hummels19}. These new simulations achieve unprecedented mass ($<100$--$2\times 10^3$ M$_\sun$) and spatial (100--1000 pc) resolution in the CGM; in mass resolution, this is a factor 10--100 improvement in the CGM compared to the standard zoom-simulations described above. While these works have not yet provided a detailed analysis of the metallicity of the CGM probed by pLLSs and LLSs (but see \citealt{suresh19}), they demonstrate some quantitative changes, which can impact  the mixing of metals, especially in the cooler and denser regions of the CGM. For example, \citet{vandevoort18} show that  resolution directly affects the radial profile of the \hi\ column density at $z=0$, enhancing it at impact parameters $>40$ kpc and doubling the covering factor of the LLSs. \citet{peeples19} show that most of the \hi\ absorbers at $z\sim 2$ are in structures with masses $\la 10^4$ M$_\sun$, well below the CGM mass resolution of the standard zoom simulations,  in particular causing a more natural metal-mixing in the CGM structures and allowing low-metallicity gas to reach regions much closer to galaxies in the well-resolved resolution simulations (see, e.g., Fig.~4 in \citealt{peeples19}).  \citet{suresh19} present and discuss the metallicity distribution of the cold CGM phase of their single simulated galaxy, finding a strong bimodality distribution in their ``cooling only'' run. Interestingly, while the metallicity distribution is more flattened when feedback is included, there is still some bimodality in the CGM  probed by the pLLSs and LLSs. The bimodality in their simulations is caused by  two distinct populations of CGM gas. One population with metallicities $\xh\ga -1$ has cycled many times through the central galaxy and has been ejected through the galaxy wind or stripped from satellites. The other population with very low metallicity gas ($\xh \la -2$), has never been accreted by the galaxy. Although this result generally confirms our  interpretation of the bimodal metallicity distribution of the pLLSs and more generally the presence of both low and high metallicity gas in the CGM of galaxies at $z\la 1$, there are some major caveats to these findings. The \citeauthor{suresh19} simulation is for a single massive halo at $z=2.25$. The KODIAQ-Z \hi-selected sample of pLLSs and LLSs at $z\sim 2$--3.5  does not have a prominent high metallicity peak at $\xh \sim -0.3$ as observed in these simulations (\citealt{lehner16}, and see also \citealt{fumagalli16,cooper15,glidden16} for strong LLSs and SLLSs). The \hi-selected observations almost certainly draw from a very different galaxy mass distribution, and the sample is still small (31 absorbers). This makes detailed comparisons premature. We are currently increasing the size sample of absorbers at high $z$, which should reach a sample size similar to CCC and cover the same range of \hi\ column densities; so we will be able to determine more robustly the metallicity PDF of these absorbers at $z\sim 2$--3.5. 

Some of the conclusions from these high-resolution simulations are still tentative in view of a single halo studied and having not always a realistic feedback implementation. Yet, it is encouraging to see that the forced refinement simulations may have reached a resolution in the CGM that is sufficient to more realistically characterize in the future the nature and origin of the metal distributions in the cool CGM gas of simulated galaxies. 

\subsection{Location of the Strong \hi\ Absorbers in the EAGLE Simulations}\label{s-disc-orig}
\begin{figure*}[tbp]
\epsscale{1}
\plotone{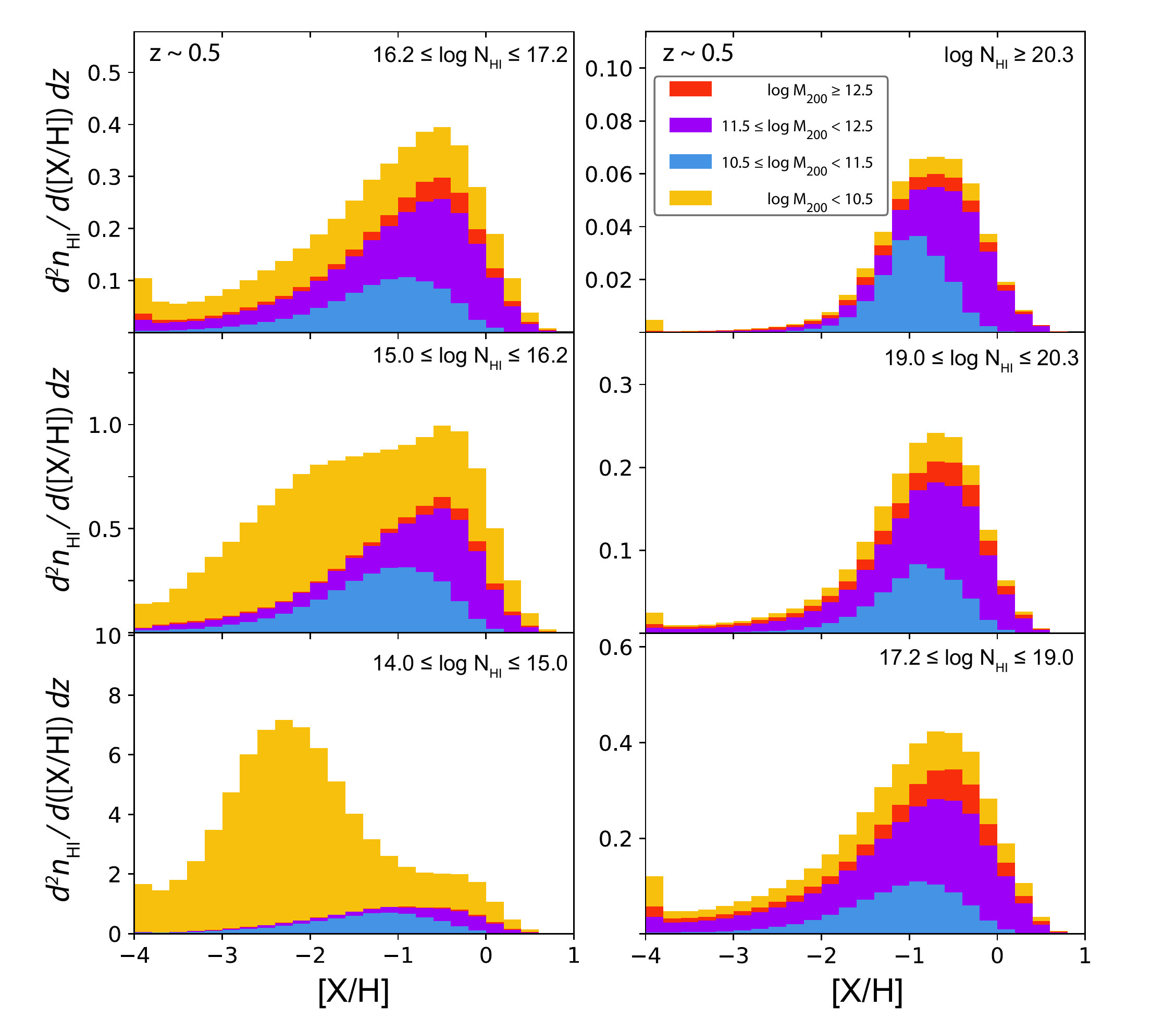} 
\caption{Origins of \hi\ absorbers in the EAGLE simulations. Metallicity distribution histograms are shown  at each column density range into halo mass categories of $1$ dex difference (the $M_{200}\le 10^{10.5}$\,M$_\sun$ interval includes dwarf galaxies and IGM).  Note that the small peak at $\xh =-4$ is an artificial grouping of all absorbers with this metallicity and below (we also caution that EAGLE does not include metal diffusion nor can adequately track the pre-reionization enrichment from the first stars and galaxies, which makes predictions for these very low metallicities unrealistic and likely under-estimated).
\label{f-HI-abs-eagle}}
\end{figure*}

Although we identify some issues regarding the metallicities of the absorbers between the EAGLE HiRes cosmological simulations and the CCC observations (see \S\ref{s-disc-comp}), it is still useful to use the cosmological volume probed by these simulations to learn about the location and origin of the strong \hi\ absorbers, especially since these simulations have otherwise shown a reasonably good level of agreement for a wide variety of observations, including the \hi\ distribution of absorbers above $10^{16}$ \cmm\ \citep{rahmati15} and metal absorber column density distributions \citep{rahmati16}. This model is calibrated to reproduce galaxy properties, and in this context, we explore below the location of the strong \hi\ absorbers within these simulations. 

The mean (standard deviation) and median redshift of the CCC absorbers are $\langle z_{\rm abs}\rangle = 0.49 \pm 0.19$ and 0.45. We therefore calculate absorber metallicities as described in \citet{rahmati18} (and see \S\ref{s-disc-comp}) using the closest snapshot to the mean redshift in CCC,  $z=0.503$.  Briefly, the column density maps of $10,000\times 10,000$ pixels are calculated for each of 8 slices corresponding to $1/8$th the simulation volume ($\sim 2$ Mpc in depth), and \hi-weighted metallicities are calculated.  The \citet{haardt01} ionization background is applied with the \citet{rahmati13} self-shielding correction.\footnote{It is crucial for the model to reproduce the number density of Ly$\alpha$ forest \citep{danforth16}, the average Ly$\alpha$ flux decrement \citep{kirkman07}, as well as stronger absorbers where measurements are available in the pLLS range \citep{shull17}.  For the EAGLE HiRes volume, we find the \citet{haardt01} background does a very good job of reproducing these three observations of \hi\ in the low-redshift ($z<0.5$) Universe, motivating the use of this field intensity.  Although the ionization modeling in this work applies the HM05 field, only the intensity matters for the simulations since \hi-weighted metallicities are plotted.} We also determine the environment of an absorber by determining whether it arises from inside a halo of mass greater than $M_{200} = 10^{10.5}$\,M$_\sun$, where $M_{200}$ is the halo mass calculated summing the mass within a region with $200\times$ the critical density of the Universe.  We sub-divide our metallicity distribution histograms (MDHs) at each column density range into halo mass categories spanning $1$ dex.  Any absorber that is outside a halo or associated with a halo less massive than $M_{200}=10^{10.5}$\,M$_\sun$ is grouped into the ``IGM" category (which is the reason we use systematically IGM between quotation mark in this section).  

Six MDHs corresponding to different column density ranges are displayed in Fig.~\ref{f-HI-abs-eagle} from  Ly$\alpha$ absorbers (absorbers with $\mlnhi < 15$) in the lower left to the DLAs in the upper right. Only absorbers with $14 \le \mlnhi \le 15$ have a positive skewed distribution with a mode at $\xh \sim -2.3$ compared to a mode of $-0.6$ dex for the stronger \hi\ absorbers. The absorbers with $15<\mlnhi <19$ have a skewed metallicity distribution, but with a tail to low metallicity that becomes more prominent as \nhi\ decreases (which is reminiscent of the metallicity PDF that combines all the CCC absorbers shown in Fig.~\ref{f-mdf-all}).  However, in the EAGLE simulations, there is  a larger preponderance of metallicities $\xh<-2$ for the SLFSs than the pLLSs that is not observed in CCC (we also note that for the SLFSs, we are less sensitive to gas with $\xh <-2.5$, so it is unknown if we would observe such a large preponderance of $\xh<-2.5$ metallicities  in this regime, especially if it is dominated by $15<\mlnhi <15.3$ absorbers). The DLAs have a normal distribution (also consistent with the observations).

When sub-divided by environment, the MDHs becomes dominated by ``IGM" absorbers (IGM or galaxy halos with  $M_{200} < 10^{10.5}$\,M$_\sun$) for the typical Ly$\alpha$ absorbers and the SLFSs, especially those with low metallicities ($\xh <-1$).  In fact a prediction of the simulations is that MDHs inside halos (the CGM) have relatively similar metallicity distributions for $\mnhi \approx 10^{19}$ \cmm\ and below, but ``IGM" absorbers with lower metallicities than $\xh \la -1.5$ become more significant for SLFSs and dominant in the Ly$\alpha$ forest.  Sub-dividing within the CGM, we see metallicities peak at $\xh \approx -1$ for low-mass halos hosting sub-$L^*$ galaxies ($M_{200} = 10^{10.5}$--$10^{11.5}$\,M$_\sun$), while for more massive halos ($M_{200}>10^{11.5}$\,M$_\sun$), the metallicities peak  at $\xh \approx -0.6$.   It is also interesting that sub-$L^*$ halos have more DLAs than pLLSs/LLSs relative to higher mass halos. 

The EAGLE HiRes simulations therefore further demonstrate  that a majority of the pLLSs and LLSs arise in the halos of $M_{200} > 10^{10.5}$\,M$_\sun$  galaxies at $z\la 1$. For the SLFSs, a large fraction also arises in the  $M_{200} > 10^{10.5}$\,M$_\sun$  galaxies, but an important portion (especially for the SLFSs with $\xh <-1$) originates in lower mass halo galaxies or the IGM. We look forward to the impending observational surveys  that will begin associating \hi\ CCC absorbers with the properties of galaxies at $z\la 1$ to further learn about the origins of these absorbers and their origins compare with cosmological simulations.  

\section{Summary and Concluding Remarks}\label{s-sum}
Using the archive of \hst/COS G130M and G160M data  as well as additional data from \hst/STIS, \fuse, and \hst/COS G140L, we have built the largest database to date of \hi-selected absorbers with $15<\mlnhi < 19$ at $z\la 1$ (with a mean and standard deviation redshift of  $\langle z_{\rm abs}\rangle = 0.49 \pm 0.19$ and most of the absorbers being in the redshift range $0.2\le z \le 0.9$), which constitutes CCC. About 88\% of the CCC data (231 absorbers) come from  COS G130M and/or G160M, showing how the medium resolution grating of COS has revolutionized this field thanks to the COS sensitivity in the FUV mode. For about half of our sample observed with COS G130M and/or G160M, we have also collected or used archival high-resolution Keck/HIRES and VLT/UVES observations that cover the strong NUV transition of \mgii\ and \feii. 

The \hi\ selection and the \hi\ column density range both ensure that no bias is introduced in the metallicity distribution of these absorbers, i.e., we are sensitive to any absorbers with $\xh \ga -3$. For each absorber of the 152 SLFSs, 82 pLLSs, and 28 LLSs  in CCC, we estimated the column densities of \hi\ and low and intermediate ions that probe the cool photoionized gas probed by the SLFSs, pLLSs, and LLSs (summarized in \citetalias{lehner18}). Importantly, as much as possible, we estimate the column densities in individual components defined by the velocity width of the weak, typically unsaturated \hi\ Lyman series transitions (at the COS resolution) and therefore evaluate the metallicity in individual components. 

In this paper and \citetalias{wotta19}, we employ a Bayesian formalism that uses MCMC sampling of a grid of Cloudy photoionization models (where we assume photons from the HM05 EUVB provide the source of photoionization) to derive the posterior PDFs of the metallicities and other physical quantities. This technique provides robust posterior PDFs and uniformly estimates the confidence intervals of each physical quantity and metallicity.  As part of our ionization correction modeling, we have developed a new method (the ``low-resolution" method) to derive the metallicities that can be used for absorbers for which the constraints from the metal ions are less than ideal (\citetalias{wotta16,wotta19}; this paper). Using absorbers with reliable constraints from the metal ions,  we find that \logU\ can be reasonably well modeled by a Gaussian  for the absorbers in CCC (see \citetalias{wotta19}). Applying this method to a sample with well-constrained absorbers, we demonstrate a good match in the results between the low-resolution and detailed approaches. We caution, however, to not use this method blindly at higher redshifts (where the \logU\ PDFs are different, \citealt{lehner16,fumagalli16}) or other \nhi\ ranges (where the \logU\ PDFs might also be different). We also explore the effects of changing the EUVB to estimate the metallicities, which can be as large as 0.4 dex and 0.2 dex for the SLFSs/pLLSs and LLSs on average, respectively (when comparing the results derived using the HM12 and HM05 EUVBs; the magnitude of this difference is likely to change with other EUVBs). We also show that the metallicity is much less sensitive to the ionization parameter than the densities, total hydrogen column densities, or length-scales of the absorbers (see \S\ref{s-met-det-add-const}). Finally,  while we make the assumption that the modeled ions are in a single gas-phase, we show that the situation can be more complicated when several ionization stages are observed (e.g., \oii, \oiii, \oiv) with different ionized gas-phases (see Fig.~\ref{f-multi}). However, in this case, based on our study of the \oiv\ and modeling of absorbers using only low ions, we estimate that the effect of the multiple gas-phases on the metallicity is on average not larger than 0.1--0.2 dex, making the metallicity a robust probe of the chemical enrichment and origins of the absorbers. 

In this paper, we focus on the gas probed by 152 SLFSs at $z<1$. Beyond their redshift evolution, this class of absorbers has never been studied in any detail before. With \hi\ columns intermediate between the \lya\ forest absorbers (IGM) and the stronger \hi\ absorbers probed the pLLSs and LLSs, the SLFSs allow us to characterize and understand the transition from the inner, denser, and often closer regions of galaxies to the more extended, more diffuse regions around galaxies. As part of this analysis, we revisit the metallicity PDFs of the stronger absorbers (pLLSs, LLSs in particular). We also compare our results with the COS-Halos survey and with results from state-of-the-art cosmological simulations. Our main findings can be summarized as follows.

\begin{enumerate}[wide, labelwidth=!, labelindent=0pt]
\item We derive the metallicity of each SLFS at $z\la 1$ and combine them,  resulting into a unimodal skewed distribution to low metallicities with a mode $\xh \simeq -0.90$. The mean and median metallicities are $-1.47$ and $-1.18$ dex, respectively.  Over the entire redshift range at $z\la 1$, the functional form of the metallicity PDF for the SLFSs is remarkably different from the bimodal metallicity PDF of the pLLSs with the SLFS metallicity PDF peaking in the dip of the pLLS metallicity PDF. 

\item We find there is some evolution of the metallicities with redshift. For the SLFSs, at $z\ge 0.65$, there is a lack of metal enriched ($\xh >-1$) absorbers with a mean metallicity  $\xh = -1.86 \pm 0.19$ (error on the mean) compared to  $-1.22 \pm 0.07$ at $0.2\la z < 0.65$. This is not observed in the pLLSs or LLSs. However, for the pLLSs and LLSs, the metallicity PDF changes with $z$: at $z\ga 0.45$, the metallicity PDF of the combined sample of pLLSs and LLSs (or the sample that includes only the pLLSs) is bimodal, while at $z\la 0.45$, it is not only unimodal, but it also matches the metallicity PDF of the SLFSs. Therefore the bimodal nature of the metallicity PDF of the pLLSs is redshift dependent and occurs only at $0.45 \la z \la 1$. 

\item The range of metallicities ($-3 \la \xh \le +0.4$) and the fractions of metal-poor (50\%--70\% for $\xh <-1$ gas) and very metal-poor (40\%--50\% for $\xh <-1.4$ gas) absorbers are quite similar in the SLFS, pLLS, and LLS (with $\mlnhi\la 18$) regimes. The photoionized gas around $z\la 1$ galaxies probed by absorbers with $15<\mlnhi \la 18$ is therefore highly inhomogeneous and contains a large reservoir of primitive gas.

\item Our sample contains 30 closely redshift-separated absorbers ($\Delta v \la 100$--300 \km) along the same sightlines. For the majority of these paired-absorbers (75\%--96\% at 90\% CI), the metallicity variation between the paired absorbers is significant, varying by a factor 2 to 25. Both the large metallicity range and metallicity variation between paired absorbers imply that the gas around $z\la 1$ galaxies (i.e., the CGM, intragroup gas between small group of galaxies) is therefore not chemically homogeneous. 

\item While there is plenty of {\em primitive} gas around $z\la 1$ galaxies, the evidence for {\em pristine} gas (i.e., no absorption from strong metal transitions in good quality spectra) is scant. The fraction of pristine absorbers is $<1$\% at the 90\% confidence level over the range $15<\mlnhi <19$. 

\item We estimate the posterior PDFs of \nh, and \nnh, and $l\equiv \mnh/\mnnh$ for all the individual absorbers and categories of absorbers in the CCC sample. These PDFs are consistent with unimodal distributions,  contrasting remarkably from the metallicity PDFs of the same absorbers. The H densities, H column densities, and path-lengths decrease with decreasing \nhi, but greatly overlap. This is expected from the neutral fraction relationship with \nhi. The interquartile range of the path-length of the CCC absorbers is   $0.3 \la l \la 3.2$ kpc, consistent with galaxy CGM-scales.

\item We estimate that the combined photoionized components of SLFSs, pLLSs, and LLSs account for about 40\% of the CGM baryons, while contributing only to about 2.1\% of the cosmological baryon budget. The cool CGM gas probed by SLFSs, pLLSs, and LLSs only contributes to 4\% of the cosmic metal density lost by galaxies, a value similar to those found in the COS-Dwarfs and COS-Halos surveys. 

\item  We compare the CCC results with the COS-Halos survey. Using the same ionizing background and restricting the sample of CCC absorbers to be at $z\le 0.28$, the  CCC and COS-Halos absorbers have similar metallicity PDFs. We argue that this implies that the CGM might be more homogeneous, with a smaller fraction of metal-poor ($\xh <-1$) gas at $z\la 0.2$ than at $0.2<z\la 1$. The smaller fraction of metal-poor absorbers is also tentatively observed in CCC  where we find only 1 out of 7 absorbers with $\xh <-1$ at $z\la 0.25$. 

\item We compare the CCC empirical metallicity PDFs to those of the FIRE and EAGLE simulations, finding similar issues highlighted in \citetalias{wotta19}, including a lack of the metallicity PDF evolution with \nhi, a lack of very metal-poor absorbers in the FIRE simulations, and a strong evolution of the metallicity with $z$ in the EAGLE simulations that is not observed in CCC.  The EAGLE simulations produce, however, a much broader range of metallicities than observed in FIRE at both low and high metallicities. 

\item Using the EAGLE HiRes cosmological simulations at $z\simeq 0.5$, we find that the majority of the absorbers with  $16.2< \mlnhi < 22$ arise in the CGM of galaxies with $M_{200} >10^{10.5}$\,M$_\sun$. In the SLFS regime, a large fraction still arises in the CGM of galaxies  with $M_{200} >10^{10.5}$\,M$_\sun$, but a substantial number of the SLFSs, especially at very low metallicity ($\xh <-1.4$), appears to originate in the IGM or dwarf galaxy halos with $M_{200} <10^{10.5}$\,M$_\sun$. At $\mlnhi < 15$, the great majority of the Ly$\alpha$ forest absorbers arise in the IGM as expected from the observations. 
\end{enumerate}

With this paper, we have completed the first phase of CCC. The unprecedented large scale  of this survey has shed new light on the gas around $z\la 1$ galaxies. We found in particular that low metallicity gas is not rare around $z\la 1$ galaxies, and, in fact, as important as metal-enriched gas, at least in the \hi\ column range $15<\mlnhi \la 18.5$.  Since the low metallicity gas is now found down to $\sim 10^{15.1}$ \cmm, this strongly suggests that the source of the low-metallicity gas is the IGM. This naturally  leads to conclude that the $z\la 1$ IGM has a metallicity that is below $\xh <-1.4$, although it would be difficult to demonstrate it without a new telescope that would provide a large sample of SNR\,$> 100$--200 ultraviolet $z \la 1.5$ QSO spectra (e.g., LUVOIR\footnote{See \url{https://asd.gsfc.nasa.gov/luvoir/}.}). According to EAGLE simulations shown in this paper, the metallicity of the IGM probed \lya-forest absorbers at $z\sim 0.5$  peaks around $\xh \simeq -2.3$ (see Fig.~\ref{f-HI-abs-eagle}), well below the metal-detection sensitivity for absorbers with $\mlnhi <15$. Observations and simulations have shown that gas with $15<\mlnhi <19$ is typically found around galaxies, and likely gravitationally bound in galaxy halos. Such gas is therefore a likely source of fuel for continued star formation within galaxies over billions of years.

This series of CCC papers also raises some new questions. Why  do the metallicity distributions of the pLLSs and LLSs evolve from a unimodal PDF at $z>2$ (\citealt{lehner16,fumagalli16}) to a bimodal PDF at $0.45 < z < 1.0$ to a unimodal PDF at $z<0.45$? Is the bimodal nature of the metallicity distribution a last imprint of the peak of star-formation rate of galaxies? Why does the metal-poor CGM gas appear to be scarcer at $z<0.2$ than at higher redshift? There are also several aspects still left to explore for a more complete picture of the gas surrounding $z<1$. In this paper we have touched on the multiple gas-phase nature of $15<\mlnhi <19$ absorbers, but that included only photoionization processes. Many absorbers have \ovi\ or other high ions that cannot be easily fit with a  photoionization model. In a pilot study using the \citetalias{lehner13} sample, \citet{fox13} find the detection rate of \ovi\ absorption associated with pLLSs at $z < 1$ is about 70\%--80\%, i.e., pLLSs probe multiple gas phases (and based on the spectra shown in \citetalias{lehner13}, we expect a similar frequency in the SLFSs and LLSs). With our new follow-up  \hst\ Legacy program of CCC, one of our goals will be to quantify the dichotomy in the multiphase gas properties between metal-poor and metal-rich absorbers to better understand the relationship of high ionization gas to accretion and feedback structures.  

\section*{Acknowledgements}
We thank the referee for useful recommendations, in particular for the suggestion to study the impact of the plausible  multiple gas-phase nature of the gas on the metallicity estimates using single-phase ionization modeling. We greatly appreciate help from and thank Xavier Prochaska and Michele Fumagalli in assisting implementing the Bayesian MCMC software. We are also grateful to Zach Hafen for sharing and helping with the data from the FIRE simulations and for providing useful comments. We thank the EAGLE consortium for making the simulations available. Support for this research was provided by NASA through grants HST-AR-12854 and HST-AR-15634 from the Space Telescope Science Institute, which is operated by the Association of Universities for Research in Astronomy, Incorporated, under NASA contract NAS5-26555. BDO acknowledges support by the \hst\ theory grant HST-AR-13262 and KLC acknowledges support from NSF grant AST-1615296. This material is also based upon work supported by the  NASA  Astrophysical Data Analysis Program (ADAP) grants NNX16AF52G under Grant No.\ AST-1212012.  Based on observations made with the NASA/ESA \hst, and obtained from the Hubble Legacy Archive, which is a collaboration between the Space Telescope Science Institute (STScI/NASA), the Space Telescope European Coordinating Facility (ST-ECF/ESA) and the Canadian Astronomy Data Centre (CADC/NRC/CSA). This work was also supported by a NASA Keck PI Data Award, administered by the NASA Exoplanet Science Institute. Some of the data presented in this work were obtained from the  Keck Observatory Database of Ionized Absorbers toward QSOs (KODIAQ), which was funded through NASA ADAP grants NNX10AE84G and NNX16AF52G. Some of the Keck data presented herein were obtained at the W. M. Keck Observatory from telescope time allocated to the National Aeronautics and Space Administration through the agency's scientific partnership with the California Institute of Technology and the University of California. The Observatory was made possible by the generous financial support of the W. M. Keck Foundation. The authors wish to acknowledge the significant cultural role that the summit of Maunakea has always had within the indigenous Hawaiian community. We are most fortunate to have the opportunity to conduct observations from this mountain. Based also on observations collected at the European Southern Observatory under ESO program 0100.A-0483(A,B) and archival programs 076.A-0860(A), 075.A-0841(A), 293.A-5038(A). This research was finally supported by the Notre Dame Center for Research Computing through the Grid Engine software and, together with the Notre Dame Cooperative Computing Lab, through the HTCondor software; we especially thank Dodi Heryadi and Scott Hampton for their assistance. 

\software{Astropy \citep{price-whelan18}, emcee \citep{foreman-mackey13}, Matplotlib \citep{hunter07}, PyIGM \citep{prochaska17a}}

\facilities{HST(COS), Keck(HIRES), LBT(MODS), VLT(UVES)}

\clearpage
\makeatletter
\renewcommand{\thefigure}{A\@arabic\c@figure}
\setcounter{figure}{0}

\appendix

In this Appendix, we  first provide information regarding the supplemental files [that available  on the AAS website or by a request to the first author of the paper]. Although for the pLLSs and LLSs, there is some redundancy for several absorbers from \citetalias{wotta19}, this new release takes into account the updates made in this paper (see \S\ref{s-met-det}). First, we provide the MCMC input files in a machine-readable format in Tables~\ref{t-mcmc-format} for all the absorbers in the CCC sample. In these tables (sorted by absorber category---SLFS, pLLS, LLS---and increasing right ascension), the first column provides the identification the absorber.  Columns 2 and 3 give the redshift of the absorber and its error; column 4 gives the ion or atom. Columns 5 and 6 report the column density of the ion and its $1\sigma$ error (for the purpose of the MCMC modeling, we have averaged the lower and upper error bars when they are not symmetric). Column 7 gives the flag indicating whether the measurement is a detection, an upper limit, or a lower limit (flag\,$ =0,-1,-2$, respectively). We only list in these tables the ions that were used in the MCMC photoionization modeling (for the full list of ions, the reader should refer to \citetalias{lehner18}). 

\begin{figure*}[tbp]
\plottwo{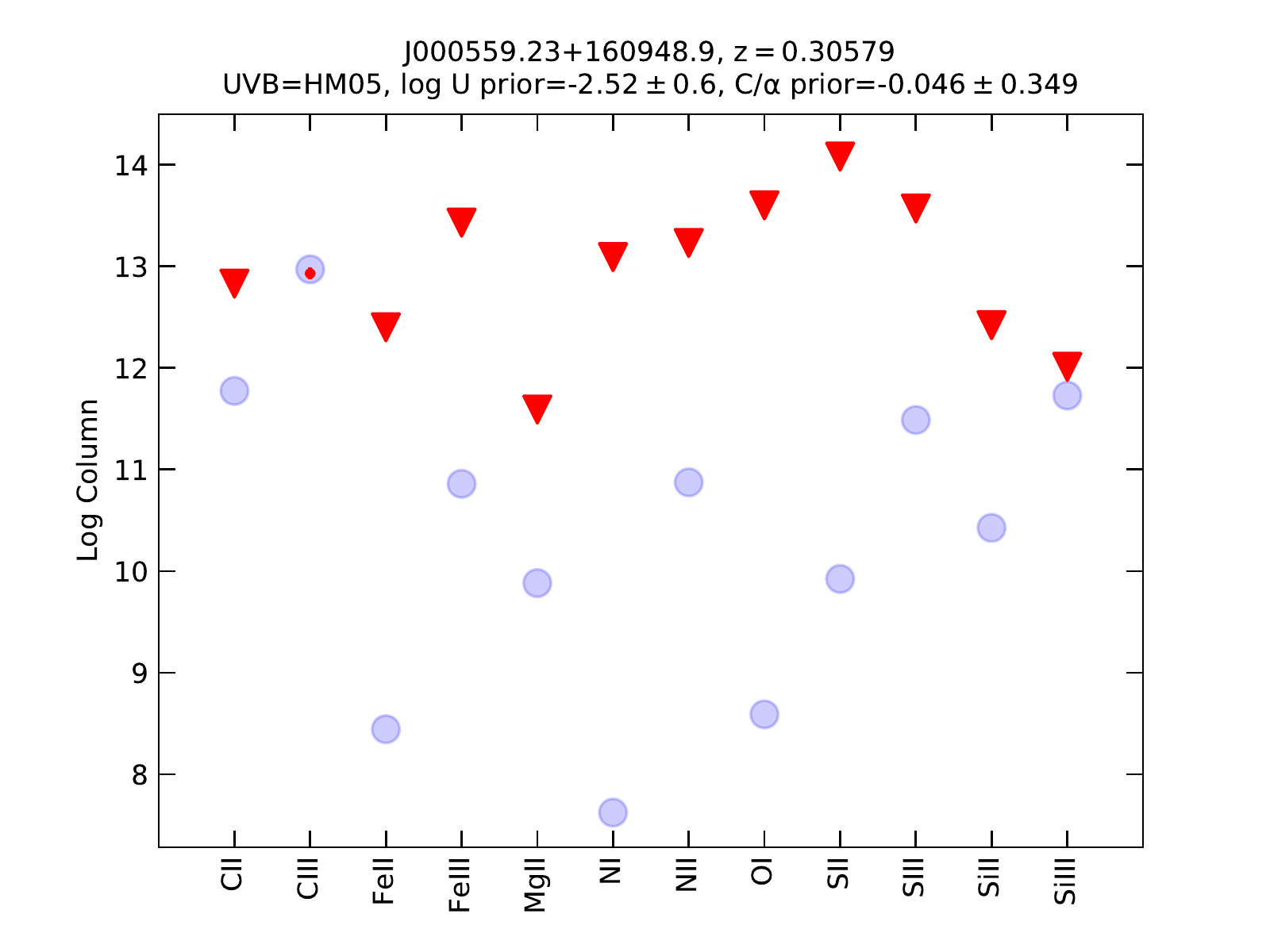}{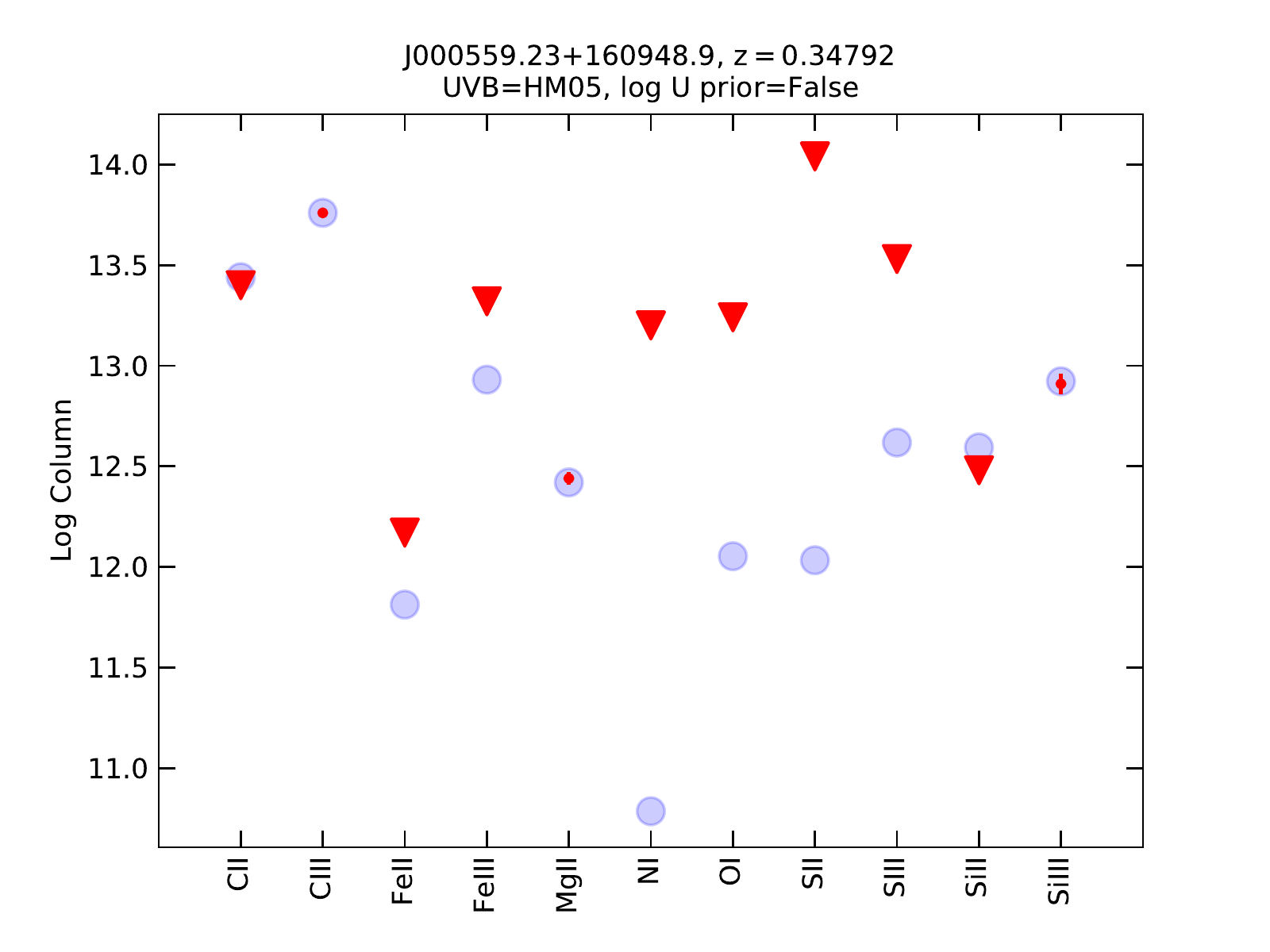}
\caption{Examples of MCMC comparison plots for the absorbers at $z=0.30579$  and $0.34792$ toward J000559.23+160948.9. It shows the measured column densities for each ion (red) and the predicted column densities from the median MCMC model (blue). Downward triangles show upper limits. 
\label{f-MCMC_output-residual}}
\end{figure*}

\begin{figure*}[tbp]
\plottwo{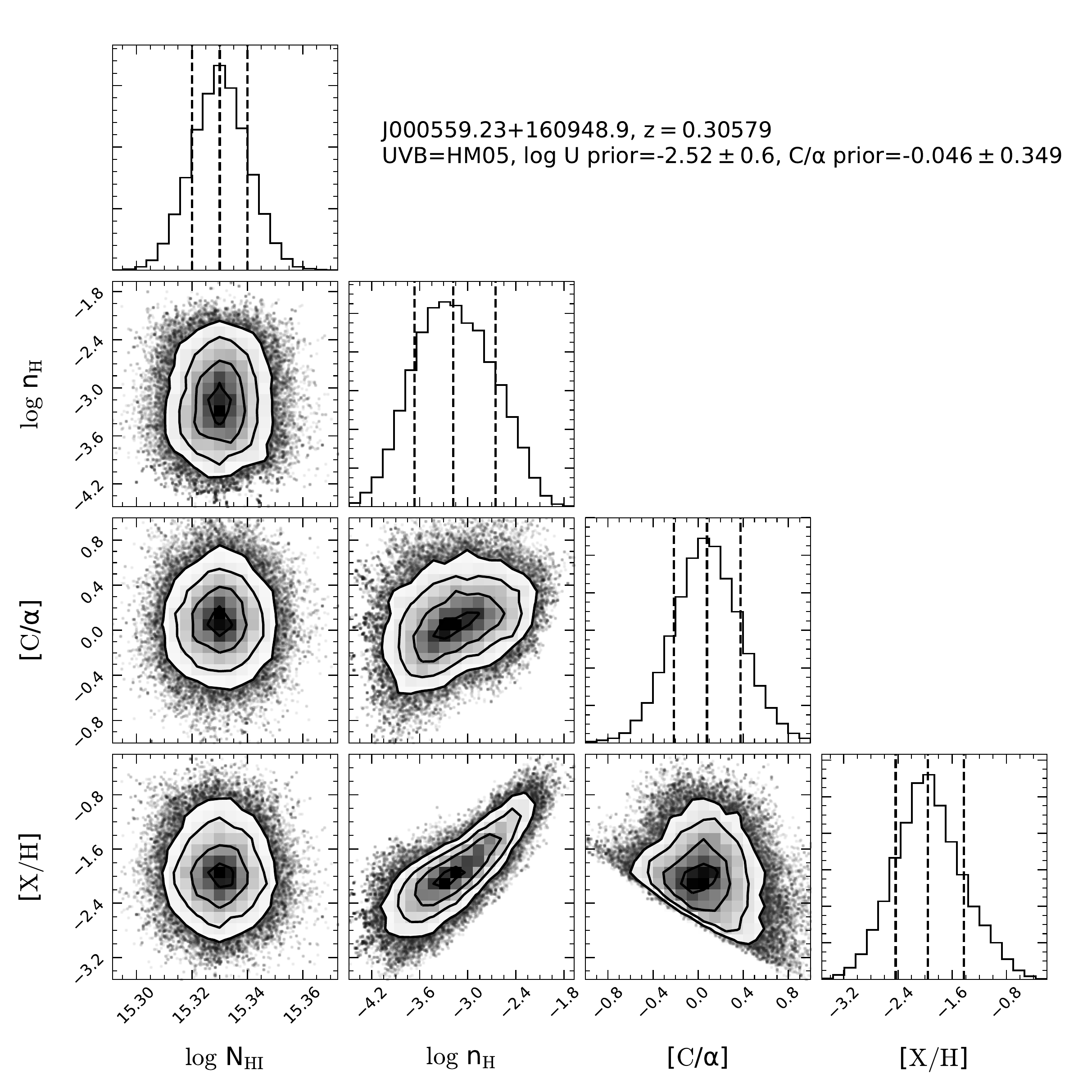}{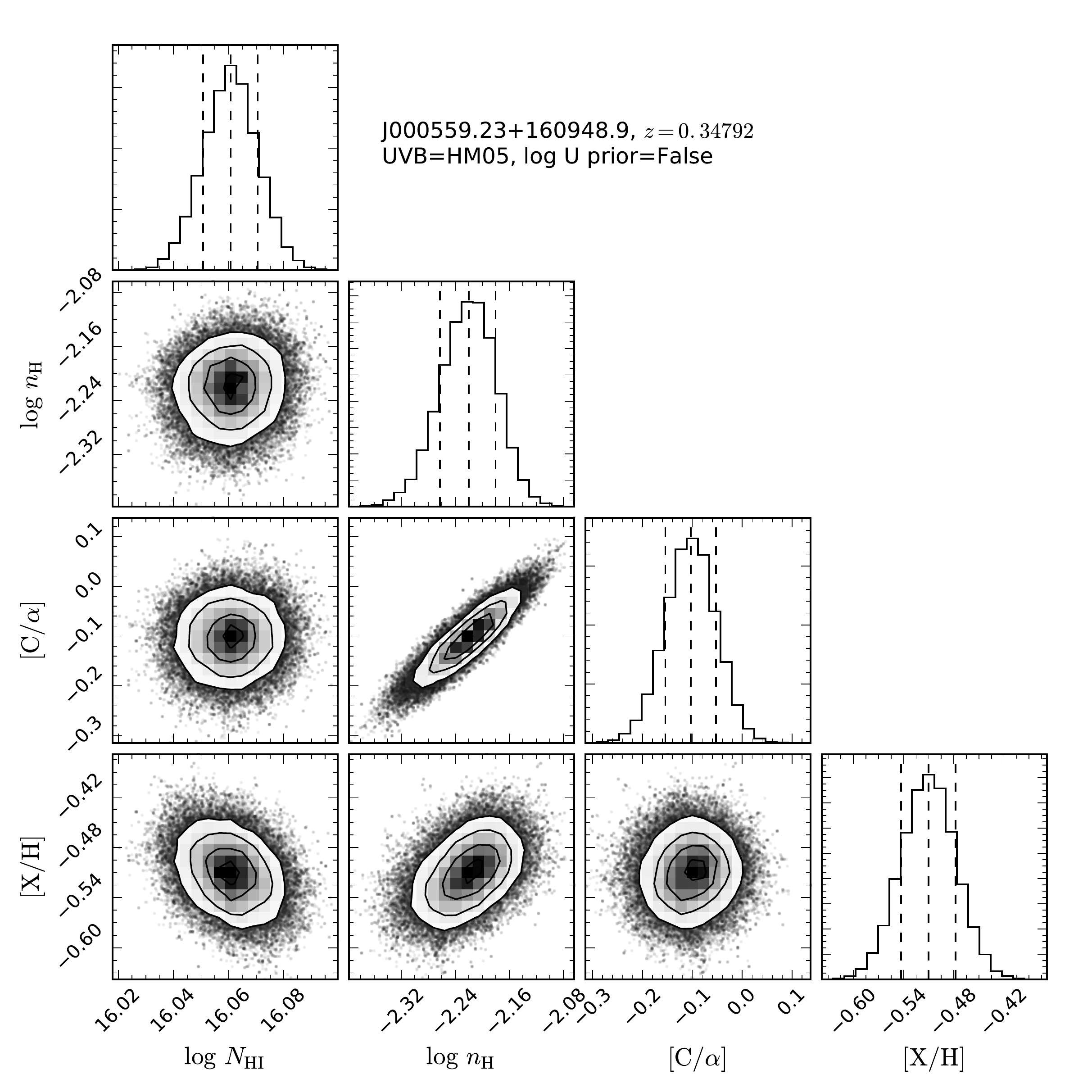}
\caption{Examples of MCMC corner plots for the absorbers at $z=0.30579$  and $0.34792$ toward J000559.23+160948.9. The histograms along the diagonal show the PDFs for \nhi, hydrogen number density ($n_{\rm H}$), \ca, and metallicity, respectively. The contour plots below the diagonal show the joint posterior PDFs of the given row and column. For the $z=0.30579$ absorber, there is not enough constraint on the metal column densities to derive the metallicity and \ca\ without using priors (see Fig.~\ref{f-MCMC_output-residual}).
\label{f-MCMC_output-corner}}
\end{figure*}

Second, for each CCC absorber, we provide the comparison plots between the observations and models and corner plots as shown in Figs.~\ref{f-MCMC_output-residual} and \ref{f-MCMC_output-corner}, respectively. Two examples of SLFSs are shown in these figures (corresponding to the first two absorbers listed in Table~\ref{t-mcmc-format}). In Fig.~\ref{f-MCMC_output-residual}, we compare the measured column densities for each ion (red) and the predicted column densities from the median MCMC model (blue). Triangles (when present) show lower limits (i.e., saturated transitions), while downward triangles show upper limits. Red data points with error bars (sometimes smaller than the circles) denote well-constrained column densities. The predicted and observed column densities are in good agreement for these two cases, but the $z=0.30579$ absorber has mostly upper limits on the column densities of the metal ions (all non-detections, except for \ciii). For that reason, we have to use a Gaussian prior on both $\log U$ and $\ca$ (see \S\ref{s-met-det} and \citetalias{wotta19} for more details). In the corner plots (Fig.~\ref{f-MCMC_output-corner}), the histograms along the diagonal show the posterior PDFs for \nhi, $n_{\rm H}$, $\ca$, and $\xh$ marginalized over the other parameters. The vertical dashed lines represent the median and 68\% CI of each PDF (80\% CI\ for upper and lower limits on \xh). The contour plots below the diagonal show the joint PDFs for the parameters of the given row and column (e.g., the topmost contour plot shows the joint posterior PDF of $n_{\rm H}$ and \nhi, the latter of which is an input parameter).  From the corner or comparison plots, one can determine readily which modeling was used: (1) if there is no entry for $\log U$ ($\log U\; {\rm prior} = {\rm False}$), then a flat prior on the ionization parameter was used; (2) if a value to $\log U$ is given, then a Gaussian prior on $\log U$ was used with the listed mean and dispersion values; (3) if $\ca$ is present, the absence of value indicates that a flat prior was used, otherwise a Gaussian prior was used on that ratio with the listed mean and dispersion values. The EUVB used in the modeling is also provided (in this case, HM05). 

\begin{figure}[tbp]
\epsscale{0.6}
\plotone{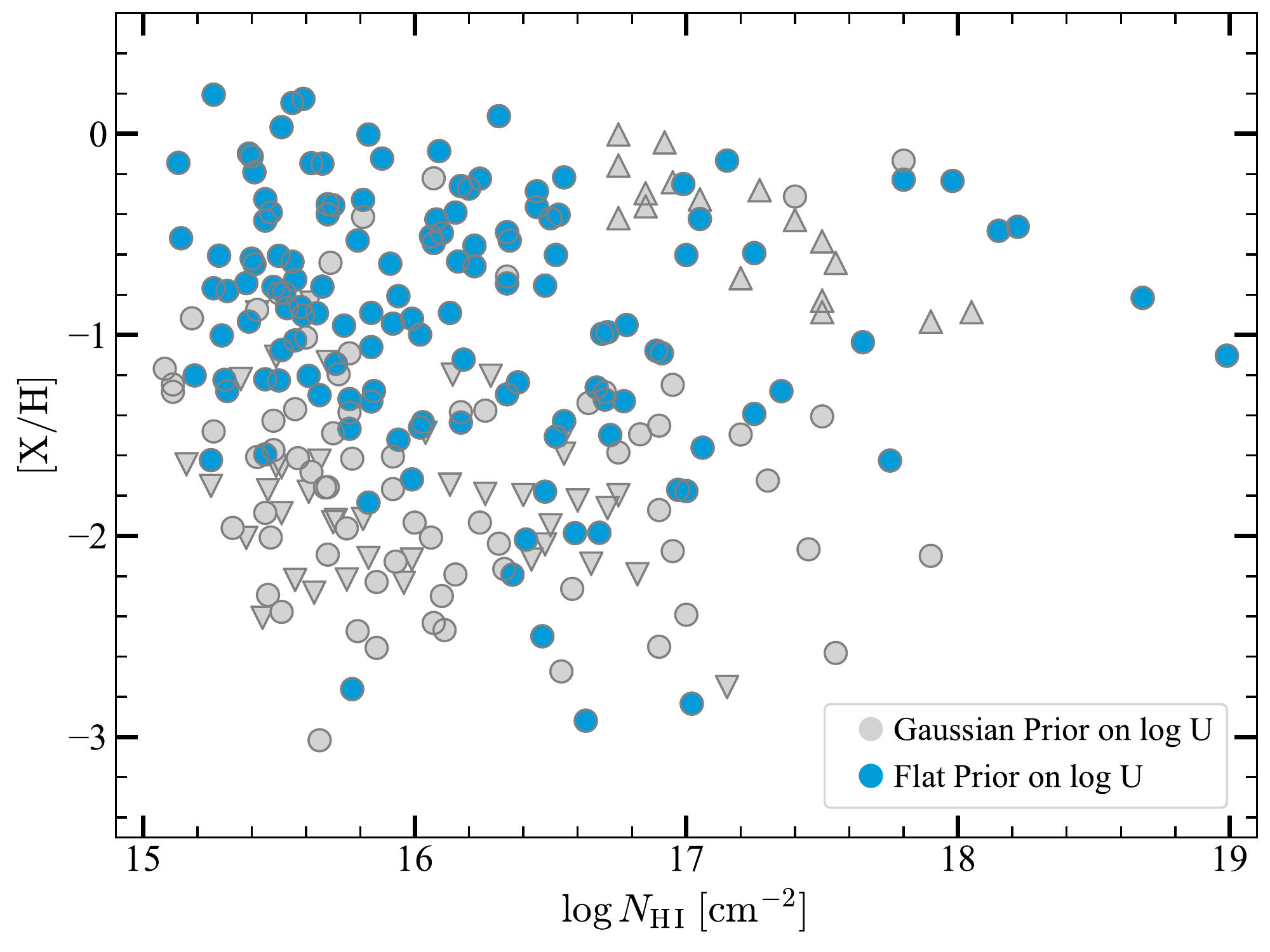}
\caption{Similar to Fig.~\ref{f-met_vs_nh1}, but only for the CCC absorbers and where we differentiate absorbers that were modeled with flat and Gaussian priors on \logU. The well-modeled absorbers are shown with circles where we adopt the median values. For lower (triangles) and upper (down triangles) limits where the values represent the $10^{\rm th}$ and $90^{\rm th}$ percentiles, respectively. For clarity, we do not show the error bars (see Fig.~\ref{f-met_vs_nh1} for these). 
\label{f-met-vs-nhi-logu}}
\end{figure}

Finally, in Fig.~\ref{f-met-vs-nhi-logu}, we show the metallicity of the CCC absorbers against \nhi\ where we differentiate the absorbers that were modeled using a Gaussian prior on \logU\ (gray symbols) from those that were modeled using a flat prior (blue symbols). Although this figure highlights specific \nhi\ and \xh\ ranges where the Gaussian and flat priors on \logU\ are not equally used, this is not unexpected since the Gaussian prior is only used to improve the estimates on the metallicities when constraints from metal ions are less than adequate. Hence, not surprisingly, the Gaussian prior is comparatively more used for absorbers with low metallicities with $\xh \la -1.7$ since the constraints are fewer in this regime (often a single ion detected with multiple non-detections or only non-detections). Furthermore as \nhi\ decreases, some of the upper limits (e.g., \mgii) also become less constraining, which is the reason why at $\mlnhi \la 16.2$ and  $\xh \la -1.7$, a Gaussian prior in this regime provides much better constraints on the metallicities. Thirty-three absorbers with $16.5 \la \mlnhi \la 18$  also come from the \citetalias{wotta16} COS G140L survey, which uses only \mgii\ as the metal to derive the metallicity, explaining the higher frequency at any metallicity in this \hi\ column density range where a Gaussian prior on \logU\ is used.  We emphasize again  this method has been tested on absorbers with sufficient constraints for the use of a flat prior on \logU, yielding similar results \citepalias{wotta16,wotta19}.  This  approach of using a Gaussian prior on \logU\ is powerful for estimating their metallicities when the available constraints are not sufficient to use a flat prior or to improve upper limits. For example, \citet{zahedy19} and \citet{chen19} derived a 95\% upper limit to the gas metallicity $\xh <-2.2$ of one of the absorbers in our sample. Using our approach with a Gaussian prior, we derive $\xh = -2.58 \pm 0.09$ (see \S\ref{s-disc-pristine}). It is fully consistent with the independently derived upper limit, but it is much better constrained thanks to the use of the Gaussian prior on \logU. 

\clearpage 
\begin{deluxetable}{lcc}
\tabcolsep=3pt
\tablecolumns{3}
\tablewidth{0pc}
\tablecaption{Updated Gaussian Prior Placed on \logU\ for the Low-Resolution MCMC Analysis\label{t-logu-constraint}}
\tabletypesize{\footnotesize}
\tablehead{\colhead{$\mlnhi$} & \colhead{$m$} & \colhead{$\left\langle\log U\right\rangle$}}
\startdata
$15.0 < \mlnhi < 15.6$       & 37  &   $-2.52 \pm 0.60$  \\
$15.6 \le \mlnhi < 16.2$     & 43   &  $-2.59 \pm 0.68$  \\
$16.2 \le \mlnhi < 16.8$     & 33   &  $-2.79 \pm 0.67$  \\
$16.8 \le \mlnhi \le 19.0$   & 20   &  $-3.03 \pm 0.23$  \\
\enddata
\tablecomments{$m$ is the number of absorbers in each \nhi\ interval for which we derive the mean and standard deviation of \logU.} 
\end{deluxetable}

\startlongtable
\begin{deluxetable*}{lccccc}
\tabcolsep=3pt
\tablecolumns{6}
\tablewidth{0pc}
\tablecaption{Metallicities of the Absorbers in CCC Sample \label{t-met-sum}}
\tabletypesize{\footnotesize}
\tablehead{\colhead{QSO Name}   & \colhead{\zabs}   & \colhead{\lnhi}   & \colhead{$\xh $}   & \colhead{$\log U$}   & \colhead{\ca}    \\
           \colhead{ }        & \colhead{ }               & \colhead{[\cmm]} & \colhead{ }               & \colhead{ }          & \colhead{ }                    }
\startdata
J035128.56$-$142908.0     & 0.328476       & $15.08 \pm 0.05$       & $-$1.67, $-$1.17, $-$0.64          & $-$2.89, $-$2.45:, $-$2.00         & $-$0.21, +0.08:, +0.39              \\ 
J011016.25$-$021851.0     & 0.535375       & $15.11 \pm 0.04$       & $-$1.89, $-$1.25, $-$0.63          & $-$3.00, $-$2.58:, $-$2.10         & $-$0.27, +0.02:, +0.35              \\ 
J124511.26+335610.1       & 0.587563       & $15.11 \pm 0.05$       & $-$1.72, $-$1.28, $-$0.83          & $-$2.73, $-$2.30:, $-$1.80         & $-$0.09, +0.16:, +0.45              \\ 
J000559.23+160948.9       & 0.366204       & $15.13 \pm 0.01$       & $-$0.19, $-$0.14, $-$0.10          & $-$2.40, $-$2.35, $-$2.31          & \nodata                             \\ 
J040748.42$-$121136.3     & 0.360742       & $15.14 \pm 0.02$       & $-$0.57, $-$0.52, $-$0.47          & $-$2.68, $-$2.63, $-$2.59          & $-$0.26, $-$0.23, $-$0.20           \\ 
J015513.20$-$450611.9     & 0.222165       & $15.16 \pm 0.04$       & $<$$-$4.63, $<$$-$3.12, $<$$-$1.64 & $-$3.17, $-$2.56:, $-$1.93         & $-$0.38, $-$0.04:, +0.30            \\ 
J121640.56+071224.3       & 0.382433       & $15.18 \pm 0.04$       & $-$1.47, $-$0.92, $-$0.33          & $-$2.77, $-$2.41:, $-$1.92         & $-$0.32, $-$0.03:, +0.30            \\ 
J115120.46+543733.0       & 0.252465       & $15.19 \pm 0.01$       & $-$1.39, $-$1.20, $-$0.91          & $-$2.91, $-$2.61, $-$2.27          & +0.15, +0.35, +0.49                 \\
\enddata
\tablecomments{
The lower to upper bounds for each quantity represent the 68\% CI for detections and the 80\% CI for the upper or lower limits; the middle number is the median metallicity. When a colon is present after the median value of \logU, this implies that a \logU\ Gaussian prior was used to determine the properties of the absorber. $^a$Proximate absorbers, i.e., absorbers with $\Delta v \equiv (z_{\rm em} - z_{\rm abs})/(1+z_{\rm abs})\, c < 3000$ \km\ (see \citetalias{lehner18}).
}
\end{deluxetable*}

\clearpage 
\startlongtable
\begin{deluxetable*}{lcccccc}
\tabcolsep=3pt
\tablecolumns{6}
\tablewidth{0pc}
\tablecaption{Closely Redshift Separated Absorbers \label{t-close}}
\tabletypesize{\footnotesize}
\tablehead{\colhead{Target}& \colhead{$[\mlnhi^1,\mlnhi^2]$}   & \colhead{$[z^1_{\rm abs},z^2_{\rm abs}]$} & \colhead{$\Delta v_c$}  & \colhead{Type} & \colhead{$([\xh^1),(\xh^2)]$}   \\
           \colhead{ }        & \colhead{[\cmm] } & \colhead{} & \colhead{(\km)} & \colhead{ }    & \colhead{ }                 }
\startdata
J004705.89+031954.9    & $[15.57 , 15.56]$ & $[0.313870 , 0.314257]$ &     88.3     & Int. & $[(-2.17, -1.61, -1.08), (-1.88, -1.37, -0.94)]$ \\
J011935.69-282131.4    & $[15.65 , 16.03]$ & $[0.348326 , 0.348731]$ &     90.1     & Prox & $[(-1.45, -1.30, -1.04), (-1.51, -1.43, -1.27)]$ \\
J020157.16-113233.1    & $[15.63 , 15.47]$ & $[0.322565 , 0.323073]$ &    115.2$^a$ & Int. & $[<(-4.75, -3.71, -2.28),(-0.50, -0.39, -0.28)]$ \\
J020157.16-113233.1    & $[15.47 , 16.09]$ & $[0.323073 , 0.323404]$ &     75.0$^a$ & Int. & $[(-0.50, -0.39, -0.28), (-0.45, -0.43, -0.40)]$ \\
J020157.16-113233.1    & $[16.09 , 15.84]$ & $[0.323404 , 0.324507]$ &    249.9$^a$ & Int. & $[(-0.45, -0.43, -0.40), (-1.24, -1.06, -0.77)]$ \\
J035128.56-142908.0    & $[16.38 , 16.55]$ & $[0.356924 , 0.357173]$ &     55.0     & Int. & $[(-1.31, -1.24, -1.15), (-0.31, -0.22, -0.12)]$ \\
J044011.90-524818.0    & $[16.20 , 15.66]$ & $[0.614962 , 0.615662]$ &    129.9     & Int. & $[(-0.29, -0.27, -0.25), (-0.18, -0.13, -0.11)]$ \\ 
J084349.47+411741.6    & $[17.00 , 16.17]$ & $[0.532565 , 0.533507]$ &    184.3     & Int. & $[(-0.77, -0.60, -0.39), (-0.31, -0.26, -0.21)]$ \\
\enddata
\tablecomments{Absorbers with $\Delta v \equiv (z^2_{\rm abs} - z^1_{\rm abs})/(1+z^1_{\rm abs})\, c < 500$ \km\ along the same sightline. For the metallicities, the lower to upper bounds for each quantity represent the 68\% CI for detections and the 80\% CI for the upper or lower limits; the middle number is the median metallicity.
}
\end{deluxetable*}

\begin{deluxetable}{lcc}
\tabcolsep=3pt
\tablecolumns{3}
\tablewidth{0pc}
\tablecaption{Metallicity Statistics \label{t-stat}}
\tabletypesize{\footnotesize}
\tablehead{\colhead{Abs. Type}        & \colhead{$\widetilde{\xh}$}  & \colhead{$\overline{\xh}$}  }
\startdata
\multicolumn{3}{c}{Entire sample}\\
\hline
SLFSs    		& $-1.18$ & $-1.47 \pm 1.18$   \\
pLLSs    		& $-1.34$ & $-1.41 \pm 1.38$   \\
LLSs     		& $-0.59$ & $-0.52 \pm 1.19$   \\
pLLSs+LLSs		& $-1.18$ & $-1.19 \pm 1.39$   \\
SLFSs+pLLSs+LLSs	& $-1.18$ & $-1.35 \pm 1.28$   \\
\hline
\multicolumn{3}{c}{Sample without limits} \\
\hline
SLFSs    		& $-1.00$ & $-1.11 \pm 0.81$   \\
pLLSs    		& $-1.28$ & $-1.29 \pm 0.81$   \\
LLSs     		& $-1.10$ & $-1.10 \pm 0.72$   \\
pLLSs+LLSs		& $-1.23$ & $-1.24 \pm 0.79$   \\
SLFSs+pLLSs+LLSs	& $-1.07$ & $-1.15 \pm 0.82$   \\
\hline
\enddata
\tablecomments{The tilde and bar above \xh\ represent the median and mean (with the standard deviation), respectively. To derive the mean, median, and standard deviation we directly use the walkers.}
\end{deluxetable}

\startlongtable
\begin{deluxetable*}{lcccccc}
\tabcolsep=3pt
\tablecolumns{7}
\tablewidth{0pc}
\tablecaption{Physical Properties of the CCC Sample\label{t-physical_properties}}
\tabletypesize{\footnotesize}
\tablehead{\colhead{QSO Name}   & \colhead{\zabs}   & \colhead{\lnhi}   & \colhead{$\mlnh$}  & \colhead{$\log n_{\rm H}$}  & \colhead{$\log T$}   & \colhead{$l$}    \\
           \colhead{ }        & \colhead{ }               & \colhead{[\cmm]} & \colhead{[\cmm]}       & \colhead{[\cmmm]}         & \colhead{[K]}          & \colhead{(kpc)}   }
\startdata
J035128.56$-$142908.0     & 0.328476       & $15.08 \pm 0.05$       & 17.55, 18.10, 18.63       & $-$3.50, $-$3.05:, $-$2.61         &  4.12,  4.28,  4.40       &    0.05,    0.45,    4.38           \\ 
J011016.25$-$021851.0     & 0.535375       & $15.11 \pm 0.04$       & 17.45, 17.98, 18.55       & $-$3.18, $-$2.70:, $-$2.28         &  4.11,  4.26,  4.39       &    0.02,    0.15,    1.72           \\ 
J124511.26+335610.1       & 0.587563       & $15.11 \pm 0.05$       & 17.77, 18.29, 18.88       & $-$3.43, $-$2.93:, $-$2.50         &  4.19,  4.32,  4.44       &    0.06,    0.54,    6.59           \\ 
J000559.23+160948.9       & 0.366204       & $15.13 \pm 0.01$       & 18.00, 18.06, 18.12       & $-$3.15, $-$3.10, $-$3.05          &  4.01,  4.03,  4.06       &    0.37,    0.47,    0.60           \\ 
J040748.42$-$121136.3     & 0.360742       & $15.14 \pm 0.02$       & 17.81, 17.87, 17.93       & $-$2.87, $-$2.83, $-$2.79          &  4.10,  4.13,  4.14       &    0.13,    0.16,    0.20           \\ 
J015513.20$-$450611.9     & 0.222165       & $15.16 \pm 0.04$       & 17.41, 18.10, 18.85       & $-$3.71, $-$3.07:, $-$2.46         &  4.18,  4.30,  4.45       &    0.02,    0.48,   11.69           \\ 
J121640.56+071224.3       & 0.382433       & $15.18 \pm 0.04$       & 17.70, 18.21, 18.80       & $-$3.51, $-$3.03:, $-$2.66         &  4.05,  4.25,  4.40       &    0.07,    0.56,    6.75           \\ 
J115120.46+543733.0       & 0.252465       & $15.19 \pm 0.01$       & 17.69, 18.05, 18.44       & $-$3.32, $-$2.98, $-$2.68          &  4.17,  4.25,  4.34       &    0.08,    0.35,    1.90           \\
\enddata
\tablecomments{
The lower  and upper  bounds for each quantity represent the 68\% CI, while the middle value is the median value. When a colon is present after the median value of $\mlnnh$, this implies that a \logU\ Gaussian prior was used to determine the properties of the absorber (see \S\ref{s-met-det}). }
\end{deluxetable*}

\begin{deluxetable*}{lcccccccccc}
\tabcolsep=3pt
\tablecolumns{11}
\tablewidth{0pc}
\tablecaption{Statistical Physical Properties of the Absorbers \label{t-phys-stat}}
\tabletypesize{\footnotesize}
\tablehead{\colhead{Absorber}   & \colhead{$\widetilde{\mlnhi}$}  & \colhead{$\overline{\mlnhi}$} & \colhead{$\widetilde{\log n_{\rm H}}$}  & \colhead{$\overline{\log n_{\rm H}}$}    & \colhead{$\widetilde{\log N_{\rm H}}$}  & \colhead{$\overline{\log N_{\rm H}}$}  & \colhead{$\widetilde{\log l}$}  & \colhead{$\overline{\log l}$}       & \colhead{$\widetilde{\log T}$}  & \colhead{$\overline{\log T}$}  	\\	     
           \colhead{Type}      	& \colhead{[\cmm]}  & \colhead{[\cmm]}   & \colhead{[\cmmm]}  & \colhead{[\cmmm]}    & \colhead{[\cmm]}  & \colhead{[\cmm]}  & \colhead{[kpc]}  & \colhead{[kpc]}       & \colhead{[K]}  & \colhead{[K]}  		     
}
\startdata
\multicolumn{11}{c}{Entire sample}\\
\hline
SLFSs+pLLSs+LLSs       & $16.04$ & $16.18 \pm 0.74$ &  $-2.64$ & $-2.65 \pm 0.48$  & $18.83$ & $18.84 \pm 0.67$ & $+0.00$ & $+0.00 \pm 0.98$ & $4.23$ & $4.12 \pm 0.36$   \\
SLFSs    	       & $15.65$ & $15.67 \pm 0.28$ &  $-2.89$ & $-2.84 \pm 0.48$  & $18.54$ & $18.56 \pm 0.59$ & $-0.11$ & $-0.09 \pm 1.01$ & $4.28$ & $4.23 \pm 0.15$   \\
pLLSs    	       & $16.63$ & $16.64 \pm 0.26$ &  $-2.46$ & $-2.47 \pm 0.46$  & $19.20$ & $19.12 \pm 0.60$ & $+0.15$ & $+0.10 \pm 1.00$ & $4.20$ & $4.05 \pm 0.46$   \\
LLSs     	       & $17.52$ & $17.68 \pm 0.43$ &  $-2.15$ & $-2.17 \pm 0.26$  & $19.57$ & $19.49 \pm 0.46$ & $+0.13$ & $+0.18 \pm 0.65$ & $4.03$ & $3.75 \pm 0.52$   \\
\hline
\multicolumn{11}{c}{Restricted sample without the \logU\ constraint}\\
\hline
SLFSs+pLLSs+LLSs       & $16.02$ & $16.16 \pm 0.75$ &  $-2.60$ & $-2.68 \pm 0.60$  & $18.87$ & $18.83 \pm 0.81$ & $+0.16$ & $+0.01 \pm 1.25$ & $4.16$ & $4.16 \pm 0.18$   \\
SLFSs    	       & $15.64$ & $15.67 \pm 0.28$ &  $-2.88$ & $-2.80 \pm 0.60$  & $18.53$ & $18.50 \pm 0.72$ & $-0.15$ & $-0.19 \pm 1.27$ & $4.18$ & $4.17 \pm 0.18$   \\
\enddata
\tablecomments{The tilde and bar above each parameter represent the median and mean (with the standard deviation), respectively, of the logarithmic values of the parameter. The entire sample consists of 262 absorbers: 152 SLFSs, 82 pLLSs, and 28 LLSs. The restricted sample without the \logU\ constraint  contains 134 absorbers (i.e., absorbers that were modeled with a flat prior on \logU): 80 SLFSs, 43 pLLSs, and 11 LLSs. Note that \nhi\ is estimated directly from the data, while all other quantities are estimated from our photoionization modeling (see \citetalias{wotta19} and \S\ref{s-met-det}).}
\end{deluxetable*}

\begin{deluxetable*}{lcccccc}
\tabcolsep=3pt
\tablecolumns{7}
\tablewidth{0pc}
\tablecaption{Comparison of the Metallicities Between CCC and COS-Halos\label{t-compcos}}
\tabletypesize{\footnotesize}
\tablehead{\colhead{Name} & \colhead{$z^{\rm CCC}_{\rm abs}$} & \colhead{$z^{\rm CH}_{\rm abs}$} & \colhead{$\mlnhi^{\rm CCC}$} & \colhead{$\mlnhi^{\rm CH}$} & \colhead{$\xh_{\rm CCC}$} & \colhead{$\xh_{\rm CH}$}\\ 
\colhead{ } & \colhead{ } & \colhead{ } & \colhead{[\cmm]} & \colhead{[\cmm]} & \colhead{ } & \colhead{ }}
\startdata
J040148.98-054056.5 & 0.21945 & 0.21969 & $15.64 \pm 0.04$ & $15.45 \pm 0.03$ & $-0.89 \,^{+0.16}_{-0.11}$  & $-0.71 \,^{+0.32}_{-0.18}$ \\
J091440.38+282330.6 & 0.24425 & 0.24431 & $15.56 \pm 0.03$ & $15.45 \pm 0.03$ & $-1.03 \pm 0.12$ 	    & $-0.90 \,^{+0.38}_{-0.29}$ \\
J094331.61+053131.4 & 0.22902 & 0.22839 & $15.59 \pm 0.09$ & $16.04 \pm 0.53$ & $-0.90 \,^{+0.33}_{-0.24}$  & $-1.24 \,^{+0.73}_{-0.77}$ \\
J100902.06+071343.8 & 0.35597 & 0.35569 & $18.68 \pm 0.18$ & $18.50 \pm 0.36$ & $-0.82 \pm 0.10$ 	    & $-0.61 \pm 0.14$ 		 \\
J101622.60+470643.3 & 0.25233 & 0.25195 & $17.25 \pm 0.10$ & $17.09 \pm 0.02$ & $-0.59 \pm 0.10$ 	    & $-0.47 \pm 0.05$ 		 \\
J111239.11+353928.2 & 0.24650 & 0.24670 & $15.58 \pm 0.06$ & $16.70 \pm 0.64$ & $-0.86 \,^{+0.24}_{-0.19}$  & $-1.29 \,^{+0.56}_{-0.52}$ \\
J123304.05-003134.1 & 0.31876 & 0.31850 & $15.56 \pm 0.01$ & $15.56 \pm 0.03$ & $-0.73 \pm 0.11$ 	    & $-0.65 \,^{+0.35}_{-0.24}$ \\
J124154.02+572107.3 & 0.21795 & 0.21780 & $15.40 \pm 0.04$ & $15.30 \pm 0.06$ & $-0.11 \pm 0.09$ 	    & $+0.08 \pm 0.11$ 		 \\
\enddata
\tablecomments{CH stands for COS-Halos. The COS-Halos \hi\ and metallicities are from \citet{prochaska17}, but we use HM05 to derive the metallicities. For J040148.98-054056.5, our results are fully consistent with the original estimate of \nhi\ by \citet{tumlinson13} ($\mlnhi = 15.63 \pm 0.05$); it is unclear why this value was changed in \citet{prochaska17} since no new COS spectrum was obtained for that QSO. For J094331.61+053131.4, our result overlaps within 1$\sigma$ with the original COS-Halos measurement from \citet{tumlinson13} that did not use the low resolution COS G140L spectrum (they derived $\mlnhi = 15.43 \pm 0.08$; also we note that the break at the Lyman limit only provides an upper limit on \nhi, see Fig.~3 in \citealt{prochaska17}). For J111239.11+353928.2, we use in CCC only the strongest component of the absorber; the other component is not in CCC because its \hi\ column density combined with the signal-to-noise of these data  are too low to be part of our survey with $\mlnhi  \simeq 15.25$. The total column density from our measurements is, however, consistent with the original value derived by \citet{tumlinson13} without using the break at the Lyman limit ($\mlnhi = 15.79 \pm 0.13$ compared to $\mlnhi \simeq 15.75$). In our opinion, the lack of absorption in \hi\ $\lambda$930 and well-separated components in \hi\ $\lambda$1025 both favor a low \nhi\ value for this absorber.}
\end{deluxetable*}

\clearpage
\makeatletter
\renewcommand{\thetable}{A\@arabic\c@table}
\setcounter{table}{0}
\startlongtable
\begin{deluxetable*}{lcclccr}
\tabletypesize{\scriptsize}
\tablecaption{Summary of the MCMC Inputs for the Absorbers in CCC \label{t-mcmc-format}}
\tablehead{
\colhead{ID}  & \colhead{$z_{\rm abs}$} & \colhead{$\sigma_{z_{\rm abs}}$}& \colhead{Ion} & \colhead{$\log N$} & \colhead{$\sigma_{\log N}$}  & \colhead{Flag} \\
\colhead{} & \colhead{} & \colhead{} & \colhead{} & \colhead{[cm$^{-2} $]}& \colhead{[cm$^{-2} $]}& \colhead{}\\
\colhead{(1)} & \colhead{(2)} & \colhead{(3)} & \colhead{(4)} & \colhead{(5)} & \colhead{(6)}& \colhead{(7)} 
}
\startdata
\hline
\multicolumn{7}{c}{SLFSs}\\
\hline
J000559.23+160948.9\_z0.305789 &   0.305789 &   0.000040 & \ion{H}{1}   &   15.33 &    0.01 & $  0$ \\
J000559.23+160948.9\_z0.305789 &   0.305789 &   0.000040 & \ion{C}{2}   &   12.83 &    0.24 & $ -1$ \\
J000559.23+160948.9\_z0.305789 &   0.305789 &   0.000040 & \ion{C}{3}   &   12.93 &    0.06 & $  0$ \\
J000559.23+160948.9\_z0.305789 &   0.305789 &   0.000040 & \ion{Fe}{2}  &   12.40 &    0.24 & $ -1$ \\
J000559.23+160948.9\_z0.305789 &   0.305789 &   0.000040 & \ion{Fe}{3}  &   13.43 &    0.24 & $ -1$ \\
J000559.23+160948.9\_z0.305789 &   0.305789 &   0.000040 & \ion{Mg}{2}  &   11.59 &    0.24 & $ -1$ \\
J000559.23+160948.9\_z0.305789 &   0.305789 &   0.000040 & \ion{N}{1}   &   13.09 &    0.24 & $ -1$ \\
\enddata
\tablecomments{The flag column has the following definition: $0=$ detection (not saturated or contaminated); $-1 =$ upper limit; $-2 =$ lower limit (due to saturation of the line).}
\end{deluxetable*}

\phantom{This is needed for the table to fully resolve.}


\begin{thebibliography}{}
\expandafter\ifx\csname natexlab\endcsname\relax\def\natexlab#1{#1}\fi

\bibitem[{{Akerman} {et~al.}(2004){Akerman}, {Carigi}, {Nissen}, {Pettini}, \&
  {Asplund}}]{akerman04}
{Akerman}, C.~J., {Carigi}, L., {Nissen}, P.~E., {Pettini}, M., \& {Asplund},
  M. 2004, \aap, 414, 931

\bibitem[{{Altay} {et~al.}(2011){Altay}, {Theuns}, {Schaye}, {Crighton}, \&
  {Dalla Vecchia}}]{altay11}
{Altay}, G., {Theuns}, T., {Schaye}, J., {Crighton}, N.~H.~M., \& {Dalla
  Vecchia}, C. 2011, \apjl, 737, L37

\bibitem[{{Asplund} {et~al.}(2009){Asplund}, {Grevesse}, {Sauval}, \&
  {Scott}}]{asplund09}
{Asplund}, M., {Grevesse}, N., {Sauval}, A.~J., \& {Scott}, P. 2009, \araa, 47,
  481

\bibitem[{{Berg} {et~al.}(2016){Berg}, {Skillman}, {Henry}, {Erb}, \&
  {Carigi}}]{berg16}
{Berg}, D.~A., {Skillman}, E.~D., {Henry}, R.~B.~C., {Erb}, D.~K., \& {Carigi},
  L. 2016, \apj, 827, 126

\bibitem[{{Berg} {et~al.}(2019){Berg}, {Howk}, {Lehner}, {Wotta}, {O'Meara},
  {Bowen}, {Burchett}, {Peeples}, \& {Tejos}}]{berg19}
{Berg}, M.~A., {Howk}, J.~C., {Lehner}, N., {et~al.} 2019, \apj,
  arXiv:1811.10717, in press

\bibitem[{{Bordoloi} {et~al.}(2014){Bordoloi}, {Tumlinson}, {Werk},
  {Oppenheimer}, {Peeples}, {Prochaska}, {Tripp}, {Katz}, {Dav{\'e}}, {Fox},
  {Thom}, {Ford}, {Weinberg}, {Burchett}, \& {Kollmeier}}]{bordoloi14}
{Bordoloi}, R., {Tumlinson}, J., {Werk}, J.~K., {et~al.} 2014, \apj, 796, 136

\bibitem[{{Borthakur} {et~al.}(2013){Borthakur}, {Heckman}, {Strickland},
  {Wild}, \& {Schiminovich}}]{borthakur13}
{Borthakur}, S., {Heckman}, T., {Strickland}, D., {Wild}, V., \&
  {Schiminovich}, D. 2013, \apj, 768, 18

\bibitem[{{Borthakur} {et~al.}(2016){Borthakur}, {Heckman}, {Tumlinson},
  {Bordoloi}, {Kauffmann}, {Catinella}, {Schiminovich}, {Dav{\'e}}, {Moran}, \&
  {Saintonge}}]{borthakur16}
{Borthakur}, S., {Heckman}, T., {Tumlinson}, J., {et~al.} 2016, \apj, 833, 259

\bibitem[{{Bowen} {et~al.}(2002){Bowen}, {Pettini}, \& {Blades}}]{bowen02}
{Bowen}, D.~V., {Pettini}, M., \& {Blades}, J.~C. 2002, \apj, 580, 169

\bibitem[{{Bregman} {et~al.}(2018){Bregman}, {Anderson}, {Miller},
  {Hodges-Kluck}, {Dai}, {Li}, {Li}, \& {Qu}}]{bregman18}
{Bregman}, J.~N., {Anderson}, M.~E., {Miller}, M.~J., {et~al.} 2018, \apj, 862,
  3

\bibitem[{{Cescutti} {et~al.}(2009){Cescutti}, {Matteucci}, {McWilliam}, \&
  {Chiappini}}]{cescutti09}
{Cescutti}, G., {Matteucci}, F., {McWilliam}, A., \& {Chiappini}, C. 2009,
  \aap, 505, 605

\bibitem[{{Chen} {et~al.}(2000){Chen}, {Lanzetta}, \&
  {Fern{\'a}ndez-Soto}}]{chen00}
{Chen}, H.-W., {Lanzetta}, K.~M., \& {Fern{\'a}ndez-Soto}, A. 2000, \apj, 533,
  120

\bibitem[{{Chen} {et~al.}(2001){Chen}, {Lanzetta}, {Webb}, \&
  {Barcons}}]{chen01a}
{Chen}, H.-W., {Lanzetta}, K.~M., {Webb}, J.~K., \& {Barcons}, X. 2001, \apj,
  559, 654

\bibitem[{{Chen} {et~al.}(2018){Chen}, {Zahedy}, {Johnson}, {Pierce}, {Huang},
  {Weiner}, \& {Gauthier}}]{chen18}
{Chen}, H.-W., {Zahedy}, F.~S., {Johnson}, S.~D., {et~al.} 2018, \mnras, 479,
  2547

\bibitem[{{Chen} {et~al.}(2019){Chen}, {Johnson}, {Straka}, {Zahedy}, {Schaye},
  {Muzahid}, {Bouch{\'e}}, {Cantalupo}, {Marino}, \& {Wendt}}]{chen19}
{Chen}, H.-W., {Johnson}, S.~D., {Straka}, L.~A., {et~al.} 2019, \mnras, 484,
  431

\bibitem[{{Cooke} {et~al.}(2011){Cooke}, {Pettini}, {Steidel}, {Rudie}, \&
  {Nissen}}]{cooke11a}
{Cooke}, R., {Pettini}, M., {Steidel}, C.~C., {Rudie}, G.~C., \& {Nissen},
  P.~E. 2011, \mnras, 417, 1534

\bibitem[{{Cooksey} {et~al.}(2008){Cooksey}, {Prochaska}, {Chen}, {Mulchaey},
  \& {Weiner}}]{cooksey08}
{Cooksey}, K.~L., {Prochaska}, J.~X., {Chen}, H.-W., {Mulchaey}, J.~S., \&
  {Weiner}, B.~J. 2008, \apj, 676, 262

\bibitem[{{Cooper} {et~al.}(2015){Cooper}, {Simcoe}, {Cooksey}, {O'Meara}, \&
  {Torrey}}]{cooper15}
{Cooper}, T.~J., {Simcoe}, R.~A., {Cooksey}, K.~L., {O'Meara}, J.~M., \&
  {Torrey}, P. 2015, \apj, 812, 58

\bibitem[{{Corlies} {et~al.}(2018){Corlies}, {Peeples}, {Tumlinson}, {O'Shea},
  {Lehner}, {Howk}, \& {O'Meara}}]{corlies19}
{Corlies}, L., {Peeples}, M.~S., {Tumlinson}, J., {et~al.} 2018, arXiv
  e-prints, arXiv:1811.05060, submitted to the ApJ

\bibitem[{{Crain} {et~al.}(2015){Crain}, {Schaye}, {Bower}, {Furlong},
  {Schaller}, {Theuns}, {Dalla Vecchia}, {Frenk}, {McCarthy}, {Helly},
  {Jenkins}, {Rosas-Guevara}, {White}, \& {Trayford}}]{crain15}
{Crain}, R.~A., {Schaye}, J., {Bower}, R.~G., {et~al.} 2015, \mnras, 450, 1937

\bibitem[{{Crighton} {et~al.}(2013){Crighton}, {Hennawi}, \&
  {Prochaska}}]{crighton13a}
{Crighton}, N.~H.~M., {Hennawi}, J.~F., \& {Prochaska}, J.~X. 2013, \apjl, 776,
  L18

\bibitem[{{Crighton} {et~al.}(2015){Crighton}, {Hennawi}, {Simcoe}, {Cooksey},
  {Murphy}, {Fumagalli}, {Prochaska}, \& {Shanks}}]{crighton15}
{Crighton}, N.~H.~M., {Hennawi}, J.~F., {Simcoe}, R.~A., {et~al.} 2015, \mnras,
  446, 18

\bibitem[{{Danforth} {et~al.}(2016){Danforth}, {Keeney}, {Tilton}, {Shull},
  {Stocke}, {Stevans}, {Pieri}, {Savage}, {France}, {Syphers}, {Smith},
  {Green}, {Froning}, {Penton}, \& {Osterman}}]{danforth16}
{Danforth}, C.~W., {Keeney}, B.~A., {Tilton}, E.~M., {et~al.} 2016, \apj, 817,
  111

\bibitem[{{Fabbian} {et~al.}(2010){Fabbian}, {Khomenko}, {Moreno-Insertis}, \&
  {Nordlund}}]{fabbian09}
{Fabbian}, D., {Khomenko}, E., {Moreno-Insertis}, F., \& {Nordlund}, {\AA}.
  2010, \apj, 724, 1536

\bibitem[{{Feigelson} \& {Nelson}(1985)}]{feigelson85}
{Feigelson}, E.~D., \& {Nelson}, P.~I. 1985, \apj, 293, 192

\bibitem[{{Ferland} {et~al.}(2013){Ferland}, {Porter}, {van Hoof}, {Williams},
  {Abel}, {Lykins}, {Shaw}, {Henney}, \& {Stancil}}]{ferland13}
{Ferland}, G.~J., {Porter}, R.~L., {van Hoof}, P.~A.~M., {et~al.} 2013, \rmxaa,
  49, 137

\bibitem[{{Foreman-Mackey} {et~al.}(2013){Foreman-Mackey}, {Hogg}, {Lang}, \&
  {Goodman}}]{foreman-mackey13}
{Foreman-Mackey}, D., {Hogg}, D.~W., {Lang}, D., \& {Goodman}, J. 2013, \pasp,
  125, 306

\bibitem[{{Fox} {et~al.}(2005){Fox}, {Wakker}, {Savage}, {Tripp}, {Sembach}, \&
  {Bland-Hawthorn}}]{fox05}
{Fox}, A.~J., {Wakker}, B.~P., {Savage}, B.~D., {et~al.} 2005, \apj, 630, 332

\bibitem[{{Fox} {et~al.}(2013){Fox}, {Lehner}, {Tumlinson}, {Howk}, {Tripp},
  {Prochaska}, {O'Meara}, {Werk}, {Bordoloi}, {Katz}, {Oppenheimer}, \&
  {Dav{\'e}}}]{fox13}
{Fox}, A.~J., {Lehner}, N., {Tumlinson}, J., {et~al.} 2013, \apj, 778, 187

\bibitem[{{Fumagalli} {et~al.}(2011){Fumagalli}, {O'Meara}, \&
  {Prochaska}}]{fumagalli11b}
{Fumagalli}, M., {O'Meara}, J.~M., \& {Prochaska}, J.~X. 2011, Science, 334,
  1245

\bibitem[{{Fumagalli} {et~al.}(2016){Fumagalli}, {O'Meara}, \&
  {Prochaska}}]{fumagalli16}
---. 2016, \mnras, 455, 4100

\bibitem[{{Glidden} {et~al.}(2016){Glidden}, {Cooper}, {Cooksey}, {Simcoe}, \&
  {O'Meara}}]{glidden16}
{Glidden}, A., {Cooper}, T.~J., {Cooksey}, K.~L., {Simcoe}, R.~A., \&
  {O'Meara}, J.~M. 2016, \apj, 833, 270

\bibitem[{{Haardt} \& {Madau}(1996)}]{haardt96}
{Haardt}, F., \& {Madau}, P. 1996, \apj, 461, 20

\bibitem[{{Haardt} \& {Madau}(2001)}]{haardt01}
{Haardt}, F., \& {Madau}, P. 2001, in Clusters of Galaxies and the High
  Redshift Universe Observed in X-rays, ed. D.~M. {Neumann} \& J.~T.~V. {Tran},
  64

\bibitem[{{Haardt} \& {Madau}(2012)}]{haardt12}
---. 2012, \apj, 746, 125

\bibitem[{{Hafen} {et~al.}(2017){Hafen}, {Faucher-Gigu{\`e}re},
  {Angl{\'e}s-Alc{\'a}zar}, {Kere{\v s}}, {Feldmann}, {Chan}, {Quataert},
  {Murray}, \& {Hopkins}}]{hafen17}
{Hafen}, Z., {Faucher-Gigu{\`e}re}, C.-A., {Angl{\'e}s-Alc{\'a}zar}, D.,
  {et~al.} 2017, \mnras, 469, 2292

\bibitem[{{Hopkins} {et~al.}(2018){Hopkins}, {Wetzel}, {Kere{\v s}},
  {Faucher-Gigu{\`e}re}, {Quataert}, {Boylan-Kolchin}, {Murray}, {Hayward},
  {Garrison-Kimmel}, {Hummels}, {Feldmann}, {Torrey}, {Ma},
  {Angl{\'e}s-Alc{\'a}zar}, {Su}, {Orr}, {Schmitz}, {Escala}, {Sanderson},
  {Grudi{\'c}}, {Hafen}, {Kim}, {Fitts}, {Bullock}, {Wheeler}, {Chan},
  {Elbert}, \& {Narayanan}}]{hopkins18}
{Hopkins}, P.~F., {Wetzel}, A., {Kere{\v s}}, D., {et~al.} 2018, \mnras, 480,
  800

\bibitem[{{Hummels} {et~al.}(2019){Hummels}, {Smith}, {Hopkins}, {O'Shea},
  {Silvia}, {Werk}, {Lehner}, {Wise}, {Collins}, \& {Butsky}}]{hummels19}
{Hummels}, C.~B., {Smith}, B.~D., {Hopkins}, P.~F., {et~al.} 2019, \apj,
  arXiv:1811.12410, in press

\bibitem[{Hunter(2007)}]{hunter07}
Hunter, J.~D. 2007, Computing In Science \& Engineering, 9, 90

\bibitem[{{Isobe} {et~al.}(1986){Isobe}, {Feigelson}, \& {Nelson}}]{isobe86}
{Isobe}, T., {Feigelson}, E.~D., \& {Nelson}, P.~I. 1986, \apj, 306, 490

\bibitem[{{Jenkins} {et~al.}(2005){Jenkins}, {Bowen}, {Tripp}, \&
  {Sembach}}]{jenkins05}
{Jenkins}, E.~B., {Bowen}, D.~V., {Tripp}, T.~M., \& {Sembach}, K.~R. 2005,
  \apj, 623, 767

\bibitem[{{Jenkins} \& {Wallerstein}(2017)}]{jenkins17}
{Jenkins}, E.~B., \& {Wallerstein}, G. 2017, \apj, 838, 85

\bibitem[{{Kacprzak} {et~al.}(2012){Kacprzak}, {Churchill}, {Steidel},
  {Spitler}, \& {Holtzman}}]{kacprzak12}
{Kacprzak}, G.~G., {Churchill}, C.~W., {Steidel}, C.~C., {Spitler}, L.~R., \&
  {Holtzman}, J.~A. 2012, \mnras, 427, 3029

\bibitem[{{Keeney} {et~al.}(2017){Keeney}, {Stocke}, {Danforth}, {Shull},
  {Pratt}, {Froning}, {Green}, {Penton}, \& {Savage}}]{keeney17}
{Keeney}, B.~A., {Stocke}, J.~T., {Danforth}, C.~W., {et~al.} 2017, \apjs, 230,
  6

\bibitem[{{Kirkman} {et~al.}(2007){Kirkman}, {Tytler}, {Lubin}, \&
  {Charlton}}]{kirkman07}
{Kirkman}, D., {Tytler}, D., {Lubin}, D., \& {Charlton}, J. 2007, \mnras, 376,
  1227

\bibitem[{{Kulkarni} {et~al.}(2005){Kulkarni}, {Fall}, {Lauroesch}, {York},
  {Welty}, {Khare}, \& {Truran}}]{kulkarni05}
{Kulkarni}, V.~P., {Fall}, S.~M., {Lauroesch}, J.~T., {et~al.} 2005, \apj, 618,
  68

\bibitem[{{Lanzetta} {et~al.}(1995){Lanzetta}, {Bowen}, {Tytler}, \&
  {Webb}}]{lanzetta95}
{Lanzetta}, K.~M., {Bowen}, D.~V., {Tytler}, D., \& {Webb}, J.~K. 1995, \apj,
  442, 538

\bibitem[{{Lehner}(2017)}]{lehner17}
{Lehner}, N. 2017, in Astrophysics and Space Science Library, Vol. 430, Gas
  Accretion onto Galaxies, ed. A.~{Fox} \& R.~{Dav{\'e}}, 117

\bibitem[{{Lehner} {et~al.}(2016){Lehner}, {O'Meara}, {Howk}, {Prochaska}, \&
  {Fumagalli}}]{lehner16}
{Lehner}, N., {O'Meara}, J.~M., {Howk}, J.~C., {Prochaska}, J.~X., \&
  {Fumagalli}, M. 2016, \apj, 833, 283

\bibitem[{{Lehner} {et~al.}(2009){Lehner}, {Prochaska}, {Kobulnicky},
  {Cooksey}, {Howk}, {Williger}, \& {Cales}}]{lehner09}
{Lehner}, N., {Prochaska}, J.~X., {Kobulnicky}, H.~A., {et~al.} 2009, \apj,
  694, 734

\bibitem[{{Lehner} {et~al.}(2007){Lehner}, {Savage}, {Richter}, {Sembach},
  {Tripp}, \& {Wakker}}]{lehner07}
{Lehner}, N., {Savage}, B.~D., {Richter}, P., {et~al.} 2007, \apj, 658, 680

\bibitem[{{Lehner} {et~al.}(2001){Lehner}, {Sembach}, {Dufton}, {Rolleston}, \&
  {Keenan}}]{lehner01a}
{Lehner}, N., {Sembach}, K.~R., {Dufton}, P.~L., {Rolleston}, W.~R.~J., \&
  {Keenan}, F.~P. 2001, \apj, 551, 781

\bibitem[{{Lehner} {et~al.}(2018){Lehner}, {Wotta}, {Howk}, {O'Meara},
  {Oppenheimer}, \& {Cooksey}}]{lehner18}
{Lehner}, N., {Wotta}, C.~B., {Howk}, J.~C., {et~al.} 2018, \apj, 866, 33

\bibitem[{{Lehner} {et~al.}(2013){Lehner}, {Howk}, {Tripp}, {Tumlinson},
  {Prochaska}, {O'Meara}, {Thom}, {Werk}, {Fox}, \& {Ribaudo}}]{lehner13}
{Lehner}, N., {Howk}, J.~C., {Tripp}, T.~M., {et~al.} 2013, \apj, 770, 138

\bibitem[{{Liang} \& {Chen}(2014)}]{liang14}
{Liang}, C.~J., \& {Chen}, H.-W. 2014, \mnras, 445, 2061

\bibitem[{{Maiolino} \& {Mannucci}(2019)}]{maiolino19}
{Maiolino}, R., \& {Mannucci}, F. 2019, Astronomy and Astrophysics Review, 27,
  3

\bibitem[{{Mattsson}(2010)}]{mattsson10}
{Mattsson}, L. 2010, \aap, 515, A68

\bibitem[{{McAlpine} {et~al.}(2016){McAlpine}, {Helly}, {Schaller}, {Trayford},
  {Qu}, {Furlong}, {Bower}, {Crain}, {Schaye}, {Theuns}, {Dalla Vecchia},
  {Frenk}, {McCarthy}, {Jenkins}, {Rosas-Guevara}, {White}, {Baes}, {Camps}, \&
  {Lemson}}]{mcalpine16}
{McAlpine}, S., {Helly}, J.~C., {Schaller}, M., {et~al.} 2016, Astronomy and
  Computing, 15, 72

\bibitem[{{M{\'e}nard} \& {Fukugita}(2012)}]{menard12}
{M{\'e}nard}, B., \& {Fukugita}, M. 2012, \apj, 754, 116

\bibitem[{{M{\'e}nard} {et~al.}(2010){M{\'e}nard}, {Scranton}, {Fukugita}, \&
  {Richards}}]{menard10}
{M{\'e}nard}, B., {Scranton}, R., {Fukugita}, M., \& {Richards}, G. 2010,
  \mnras, 405, 1025

\bibitem[{{Muzahid} {et~al.}(2015){Muzahid}, {Kacprzak}, {Churchill},
  {Charlton}, {Nielsen}, {Mathes}, \& {Trujillo-Gomez}}]{muzahid15}
{Muzahid}, S., {Kacprzak}, G.~G., {Churchill}, C.~W., {et~al.} 2015, \apj, 811,
  132

\bibitem[{{O'Meara} {et~al.}(2007){O'Meara}, {Prochaska}, {Burles}, {Prochter},
  {Bernstein}, \& {Burgess}}]{omeara07}
{O'Meara}, J.~M., {Prochaska}, J.~X., {Burles}, S., {et~al.} 2007, \apj, 656,
  666

\bibitem[{{O'Meara} {et~al.}(2011){O'Meara}, {Prochaska}, {Chen}, \&
  {Madau}}]{omeara11}
{O'Meara}, J.~M., {Prochaska}, J.~X., {Chen}, H.-W., \& {Madau}, P. 2011,
  \apjs, 195, 16

\bibitem[{{O'Meara} {et~al.}(2013){O'Meara}, {Prochaska}, {Worseck}, {Chen}, \&
  {Madau}}]{omeara13}
{O'Meara}, J.~M., {Prochaska}, J.~X., {Worseck}, G., {Chen}, H.-W., \& {Madau},
  P. 2013, \apj, 765, 137

\bibitem[{{Peeples} {et~al.}(2017){Peeples}, {Tumlinson}, {Fox}, {Aloisi},
  {Fleming}, {Jedrzejewski}, {Oliveira}, {Ayres}, {Danforth}, {Keeney}, \&
  {Jenkins}}]{peeples17}
{Peeples}, M., {Tumlinson}, J., {Fox}, A., {et~al.} 2017, {The Hubble
  Spectroscopic Legacy Archive}, Tech. rep., {STScI}

\bibitem[{{Peeples} {et~al.}(2014){Peeples}, {Werk}, {Tumlinson},
  {Oppenheimer}, {Prochaska}, {Katz}, \& {Weinberg}}]{peeples14}
{Peeples}, M.~S., {Werk}, J.~K., {Tumlinson}, J., {et~al.} 2014, \apj, 786, 54

\bibitem[{{Peeples} {et~al.}(2019){Peeples}, {Corlies}, {Tumlinson}, {O'Shea},
  {Lehner}, {O'Meara}, {Howk}, {Earl}, {Smith}, {Wise}, \&
  {Hummels}}]{peeples19}
{Peeples}, M.~S., {Corlies}, L., {Tumlinson}, J., {et~al.} 2019, \apj, 873, 129

\bibitem[{{Penprase} {et~al.}(2010){Penprase}, {Prochaska}, {Sargent},
  {Toro-Martinez}, \& {Beeler}}]{penprase10}
{Penprase}, B.~E., {Prochaska}, J.~X., {Sargent}, W.~L.~W., {Toro-Martinez},
  I., \& {Beeler}, D.~J. 2010, \apj, 721, 1

\bibitem[{{Penton} {et~al.}(2002){Penton}, {Stocke}, \& {Shull}}]{penton02}
{Penton}, S.~V., {Stocke}, J.~T., \& {Shull}, J.~M. 2002, \apj, 565, 720

\bibitem[{{P{\'e}roux} {et~al.}(2006){P{\'e}roux}, {Kulkarni}, {Meiring},
  {Ferlet}, {Khare}, {Lauroesch}, {Vladilo}, \& {York}}]{peroux06}
{P{\'e}roux}, C., {Kulkarni}, V.~P., {Meiring}, J., {et~al.} 2006, \aap, 450,
  53

\bibitem[{{P{\'e}roux} {et~al.}(2008){P{\'e}roux}, {Meiring}, {Kulkarni},
  {Khare}, {Lauroesch}, {Vladilo}, \& {York}}]{peroux08}
{P{\'e}roux}, C., {Meiring}, J.~D., {Kulkarni}, V.~P., {et~al.} 2008, \mnras,
  386, 2209

\bibitem[{{Pettini} {et~al.}(2008){Pettini}, {Zych}, {Steidel}, \&
  {Chaffee}}]{pettini08}
{Pettini}, M., {Zych}, B.~J., {Steidel}, C.~C., \& {Chaffee}, F.~H. 2008,
  \mnras, 385, 2011

\bibitem[{{Planck Collaboration} {et~al.}(2016){Planck Collaboration}, {Ade},
  {Aghanim}, {Arnaud}, {Ashdown}, {Aumont}, {Baccigalupi}, {Banday},
  {Barreiro}, {Bartlett}, \& et~al.}]{planck16}
{Planck Collaboration}, {Ade}, P.~A.~R., {Aghanim}, N., {et~al.} 2016, \aap,
  594, A13

\bibitem[{{Prochaska} {et~al.}(2003){Prochaska}, {Gawiser}, {Wolfe}, {Castro},
  \& {Djorgovski}}]{prochaska03}
{Prochaska}, J.~X., {Gawiser}, E., {Wolfe}, A.~M., {Castro}, S., \&
  {Djorgovski}, S.~G. 2003, \apjl, 595, L9

\bibitem[{{Prochaska} {et~al.}(2005){Prochaska}, {Herbert-Fort}, \&
  {Wolfe}}]{prochaska05}
{Prochaska}, J.~X., {Herbert-Fort}, S., \& {Wolfe}, A.~M. 2005, \apj, 635, 123

\bibitem[{{Prochaska} {et~al.}(2010){Prochaska}, {O'Meara}, \&
  {Worseck}}]{prochaska10}
{Prochaska}, J.~X., {O'Meara}, J.~M., \& {Worseck}, G. 2010, \apj, 718, 392

\bibitem[{{Prochaska} {et~al.}(2011){Prochaska}, {Weiner}, {Chen}, {Mulchaey},
  \& {Cooksey}}]{prochaska11c}
{Prochaska}, J.~X., {Weiner}, B., {Chen}, H.-W., {Mulchaey}, J., \& {Cooksey},
  K. 2011, \apj, 740, 91

\bibitem[{{Prochaska} \& {Wolfe}(2000)}]{prochaska00}
{Prochaska}, J.~X., \& {Wolfe}, A.~M. 2000, \apjl, 533, L5

\bibitem[{{Prochaska} {et~al.}(2017{\natexlab{a}}){Prochaska}, {Tejos},
  {Wotta}, {Burchett}, {Fumagalli}, marijana777, {O'Meara}, {Werk}, {Rafelski},
  {Neeleman}, \& {Lee}}]{prochaska17a}
{Prochaska}, J.~X., {Tejos}, N., {Wotta}, C.~B., {et~al.} 2017{\natexlab{a}},
  pyigm/pyigm: Initial release for publications,  {UCSC},
  doi:10.5281/zenodo.1045480

\bibitem[{{Prochaska} {et~al.}(2017{\natexlab{b}}){Prochaska}, {Werk},
  {Worseck}, {Tripp}, {Tumlinson}, {Burchett}, {Fox}, {Fumagalli}, {Lehner},
  {Peeples}, \& {Tejos}}]{prochaska17}
{Prochaska}, J.~X., {Werk}, J.~K., {Worseck}, G., {et~al.} 2017{\natexlab{b}},
  \apj, 837, 169

\bibitem[{{Prochter} {et~al.}(2010){Prochter}, {Prochaska}, {O'Meara},
  {Burles}, \& {Bernstein}}]{prochter10}
{Prochter}, G.~E., {Prochaska}, J.~X., {O'Meara}, J.~M., {Burles}, S., \&
  {Bernstein}, R.~A. 2010, \apj, 708, 1221

\bibitem[{{Quiret} {et~al.}(2016){Quiret}, {P{\'e}roux}, {Zafar}, {Kulkarni},
  {Jenkins}, {Milliard}, {Rahmani}, {Popping}, {Rao}, {Turnshek}, \&
  {Monier}}]{quiret16}
{Quiret}, S., {P{\'e}roux}, C., {Zafar}, T., {et~al.} 2016, \mnras, 458, 4074

\bibitem[{{Rafelski} {et~al.}(2012){Rafelski}, {Wolfe}, {Prochaska},
  {Neeleman}, \& {Mendez}}]{rafelski12}
{Rafelski}, M., {Wolfe}, A.~M., {Prochaska}, J.~X., {Neeleman}, M., \&
  {Mendez}, A.~J. 2012, \apj, 755, 89

\bibitem[{{Rahmati} \& {Oppenheimer}(2018)}]{rahmati18}
{Rahmati}, A., \& {Oppenheimer}, B.~D. 2018, \mnras, arXiv:1712.03988

\bibitem[{{Rahmati} {et~al.}(2015){Rahmati}, {Schaye}, {Bower}, {Crain},
  {Furlong}, {Schaller}, \& {Theuns}}]{rahmati15}
{Rahmati}, A., {Schaye}, J., {Bower}, R.~G., {et~al.} 2015, \mnras, 452, 2034

\bibitem[{{Rahmati} {et~al.}(2016){Rahmati}, {Schaye}, {Crain}, {Oppenheimer},
  {Schaller}, \& {Theuns}}]{rahmati16}
{Rahmati}, A., {Schaye}, J., {Crain}, R.~A., {et~al.} 2016, \mnras, 459, 310

\bibitem[{{Rahmati} {et~al.}(2013){Rahmati}, {Schaye}, {Pawlik}, \& {Rai{\v
  c}evi{\'c}}}]{rahmati13}
{Rahmati}, A., {Schaye}, J., {Pawlik}, A.~H., \& {Rai{\v c}evi{\'c}}, M. 2013,
  \mnras, 431, 2261

\bibitem[{{Ribaudo} {et~al.}(2011){Ribaudo}, {Lehner}, {Howk}, {Werk}, {Tripp},
  {Prochaska}, {Meiring}, \& {Tumlinson}}]{ribaudo11}
{Ribaudo}, J., {Lehner}, N., {Howk}, J.~C., {et~al.} 2011, \apj, 743, 207

\bibitem[{{Robert} {et~al.}(2019){Robert}, {Murphy}, {O'Meara}, {Crighton}, \&
  {Fumagalli}}]{robert19}
{Robert}, P.~F., {Murphy}, M.~T., {O'Meara}, J.~M., {Crighton}, N.~H.~M., \&
  {Fumagalli}, M. 2019, \mnras, 483, 2736

\bibitem[{{Rosenwasser} {et~al.}(2018){Rosenwasser}, {Muzahid}, {Charlton},
  {Kacprzak}, {Wakker}, \& {Churchill}}]{rosenwasser18}
{Rosenwasser}, B., {Muzahid}, S., {Charlton}, J.~C., {et~al.} 2018, \mnras,
  476, 2258

\bibitem[{{Schaye} {et~al.}(2015){Schaye}, {Crain}, {Bower}, {Furlong},
  {Schaller}, {Theuns}, {Dalla Vecchia}, {Frenk}, {McCarthy}, {Helly},
  {Jenkins}, {Rosas-Guevara}, {White}, {Baes}, {Booth}, {Camps}, {Navarro},
  {Qu}, {Rahmati}, {Sawala}, {Thomas}, \& {Trayford}}]{schaye15}
{Schaye}, J., {Crain}, R.~A., {Bower}, R.~G., {et~al.} 2015, \mnras, 446, 521

\bibitem[{{Shull} {et~al.}(2017){Shull}, {Danforth}, {Tilton}, {Moloney}, \&
  {Stevans}}]{shull17}
{Shull}, J.~M., {Danforth}, C.~W., {Tilton}, E.~M., {Moloney}, J., \&
  {Stevans}, M.~L. 2017, \apj, 849, 106

\bibitem[{{Shull} {et~al.}(2012){Shull}, {Smith}, \& {Danforth}}]{shull12}
{Shull}, J.~M., {Smith}, B.~D., \& {Danforth}, C.~W. 2012, \apj, 759, 23

\bibitem[{{Som} {et~al.}(2015){Som}, {Kulkarni}, {Meiring}, {York},
  {P{\'e}roux}, {Lauroesch}, {Aller}, \& {Khare}}]{som15}
{Som}, D., {Kulkarni}, V.~P., {Meiring}, J., {et~al.} 2015, \apj, 806, 25

\bibitem[{{Stocke} {et~al.}(2013){Stocke}, {Keeney}, {Danforth}, {Shull},
  {Froning}, {Green}, {Penton}, \& {Savage}}]{stocke13}
{Stocke}, J.~T., {Keeney}, B.~A., {Danforth}, C.~W., {et~al.} 2013, \apj, 763,
  148

\bibitem[{{Suresh} {et~al.}(2019){Suresh}, {Nelson}, {Genel}, {Rubin}, \&
  {Hernquist}}]{suresh19}
{Suresh}, J., {Nelson}, D., {Genel}, S., {Rubin}, K.~H.~R., \& {Hernquist}, L.
  2019, \mnras, 483, 4040

\bibitem[{{Tejos} {et~al.}(2014){Tejos}, {Morris}, {Finn}, {Crighton},
  {Bechtold}, {Jannuzi}, {Schaye}, {Theuns}, {Altay}, {Le F{\`e}vre},
  {Ryan-Weber}, \& {Dav{\'e}}}]{tejos14}
{Tejos}, N., {Morris}, S.~L., {Finn}, C.~W., {et~al.} 2014, \mnras, 437, 2017

\bibitem[{{The Astropy Collaboration} {et~al.}(2018){The Astropy
  Collaboration}, {Price-Whelan}, {Sip{\H o}cz}, {G{\"u}nther}, {Lim},
  {Crawford}, {Conseil}, {Shupe}, {Craig}, {Dencheva}, {Ginsburg},
  {VanderPlas}, {Bradley}, {P{\'e}rez-Su{\'a}rez}, {de Val-Borro}, {Aldcroft},
  {Cruz}, {Robitaille}, {Tollerud}, {Ardelean}, {Babej}, {Bachetti}, {Bakanov},
  {Bamford}, {Barentsen}, {Barmby}, {Baumbach}, {Berry}, {Biscani}, {Boquien},
  {Bostroem}, {Bouma}, {Brammer}, {Bray}, {Breytenbach}, {Buddelmeijer},
  {Burke}, {Calderone}, {Cano Rodr{\'{\i}}guez}, {Cara}, {Cardoso},
  {Cheedella}, {Copin}, {Crichton}, {D{\'A}vella}, {Deil}, {Depagne},
  {Dietrich}, {Donath}, {Droettboom}, {Earl}, {Erben}, {Fabbro}, {Ferreira},
  {Finethy}, {Fox}, {Garrison}, {Gibbons}, {Goldstein}, {Gommers}, {Greco},
  {Greenfield}, {Groener}, {Grollier}, {Hagen}, {Hirst}, {Homeier}, {Horton},
  {Hosseinzadeh}, {Hu}, {Hunkeler}, {Ivezi{\'c}}, {Jain}, {Jenness}, {Kanarek},
  {Kendrew}, {Kern}, {Kerzendorf}, {Khvalko}, {King}, {Kirkby}, {Kulkarni},
  {Kumar}, {Lee}, {Lenz}, {Littlefair}, {Ma}, {Macleod}, {Mastropietro},
  {McCully}, {Montagnac}, {Morris}, {Mueller}, {Mumford}, {Muna}, {Murphy},
  {Nelson}, {Nguyen}, {Ninan}, {N{\"o}the}, {Ogaz}, {Oh}, {Parejko}, {Parley},
  {Pascual}, {Patil}, {Patil}, {Plunkett}, {Prochaska}, {Rastogi}, {Reddy
  Janga}, {Sabater}, {Sakurikar}, {Seifert}, {Sherbert}, {Sherwood-Taylor},
  {Shih}, {Sick}, {Silbiger}, {Singanamalla}, {Singer}, {Sladen}, {Sooley},
  {Sornarajah}, {Streicher}, {Teuben}, {Thomas}, {Tremblay}, {Turner},
  {Terr{\'o}n}, {van Kerkwijk}, {de la Vega}, {Watkins}, {Weaver}, {Whitmore},
  {Woillez}, \& {Zabalza}}]{price-whelan18}
{The Astropy Collaboration}, {Price-Whelan}, A.~M., {Sip{\H o}cz}, B.~M.,
  {et~al.} 2018, ArXiv e-prints, arXiv:1801.02634

\bibitem[{{Tripp} {et~al.}(2011){Tripp}, {Meiring}, {Prochaska}, {Willmer},
  {Howk}, {Werk}, {Jenkins}, {Bowen}, {Lehner}, {Sembach}, {Thom}, \&
  {Tumlinson}}]{tripp11}
{Tripp}, T.~M., {Meiring}, J.~D., {Prochaska}, J.~X., {et~al.} 2011, Science,
  334, 952

\bibitem[{{Tumlinson} {et~al.}(2017){Tumlinson}, {Peeples}, \&
  {Werk}}]{tumlinson17}
{Tumlinson}, J., {Peeples}, M.~S., \& {Werk}, J.~K. 2017, \araa, 55, 389

\bibitem[{{Tumlinson} {et~al.}(2011){Tumlinson}, {Werk}, {Thom}, {Meiring},
  {Prochaska}, {Tripp}, {O'Meara}, {Okrochkov}, \& {Sembach}}]{tumlinson11}
{Tumlinson}, J., {Werk}, J.~K., {Thom}, C., {et~al.} 2011, \apj, 733, 111

\bibitem[{{Tumlinson} {et~al.}(2013){Tumlinson}, {Thom}, {Werk}, {Prochaska},
  {Tripp}, {Katz}, {Dav{\'e}}, {Oppenheimer}, {Meiring}, {Ford}, {O'Meara},
  {Peeples}, {Sembach}, \& {Weinberg}}]{tumlinson13}
{Tumlinson}, J., {Thom}, C., {Werk}, J.~K., {et~al.} 2013, \apj, 777, 59

\bibitem[{{Tytler}(1987)}]{tytler87}
{Tytler}, D. 1987, \apj, 321, 49

\bibitem[{{van de Voort} {et~al.}(2018){van de Voort}, {Springel}, {Mandelker},
  {van den Bosch}, \& {Pakmor}}]{vandevoort18}
{van de Voort}, F., {Springel}, V., {Mandelker}, N., {van den Bosch}, F.~C., \&
  {Pakmor}, R. 2018, ArXiv e-prints, arXiv:1808.04369

\bibitem[{{Welty} {et~al.}(1997){Welty}, {Lauroesch}, {Blades}, {Hobbs}, \&
  {York}}]{welty97}
{Welty}, D.~E., {Lauroesch}, J.~T., {Blades}, J.~C., {Hobbs}, L.~M., \& {York},
  D.~G. 1997, \apj, 489, 672

\bibitem[{{Werk} {et~al.}(2013){Werk}, {Prochaska}, {Thom}, {Tumlinson},
  {Tripp}, {O'Meara}, \& {Peeples}}]{werk13}
{Werk}, J.~K., {Prochaska}, J.~X., {Thom}, C., {et~al.} 2013, \apjs, 204, 17

\bibitem[{{Werk} {et~al.}(2014){Werk}, {Prochaska}, {Tumlinson}, {Peeples},
  {Tripp}, {Fox}, {Lehner}, {Thom}, {O'Meara}, {Ford}, {Bordoloi}, {Katz},
  {Tejos}, {Oppenheimer}, {Dav{\'e}}, \& {Weinberg}}]{werk14}
{Werk}, J.~K., {Prochaska}, J.~X., {Tumlinson}, J., {et~al.} 2014, \apj, 792, 8

\bibitem[{{Wotta} {et~al.}(2019){Wotta}, {Lehner}, {Howk}, {O'Meara},
  {Oppenheimer}, \& {Cooksey}}]{wotta19}
{Wotta}, C.~B., {Lehner}, N., {Howk}, J.~C., {et~al.} 2019, \apj, 872, 81

\bibitem[{{Wotta} {et~al.}(2016){Wotta}, {Lehner}, {Howk}, {O'Meara}, \&
  {Prochaska}}]{wotta16}
{Wotta}, C.~B., {Lehner}, N., {Howk}, J.~C., {O'Meara}, J.~M., \& {Prochaska},
  J.~X. 2016, \apj, 831, 95

\bibitem[{{Zahedy} {et~al.}(2019){Zahedy}, {Chen}, {Johnson}, {Pierce},
  {Rauch}, {Huang}, {Weiner}, \& {Gauthier}}]{zahedy19}
{Zahedy}, F.~S., {Chen}, H.-W., {Johnson}, S.~D., {et~al.} 2019, \mnras, 484,
  2257

\bibitem[{{Zonak} {et~al.}(2004){Zonak}, {Charlton}, {Ding}, \&
  {Churchill}}]{zonak04}
{Zonak}, S.~G., {Charlton}, J.~C., {Ding}, J., \& {Churchill}, C.~W. 2004,
  \apj, 606, 196

\end{thebibliography}
\end{document}